\documentclass[12pt]{article}

\usepackage{graphicx,psfrag}
\usepackage{float}
\usepackage{amsmath, amssymb, graphics, setspace, relsize}
\usepackage{wrapfig}
\usepackage{multirow}

\newcommand \bra[1]{\left< {#1} \,\right\vert}
\newcommand \ket[1]{\left\vert\, {#1} \, \right>}
\newcommand \braket[2]{\hbox{$\left< {#1} \,\vrule\, {#2} \right>$}}
\newcommand{\bea}{\begin{eqnarray}}
\newcommand{\eea}{\end{eqnarray}}
\newcommand{\simgt}{\hbox{ \raise3pt\hbox to 0pt{$>$}\raise-3pt\hbox{$\sim$} }}
\newcommand{\simlt}{\hbox{ \raise3pt\hbox to 0pt{$<$}\raise-3pt\hbox{$\sim$} }}

\newcommand{\clfn}{\setcounter{footnote}{0}}

\newcommand{\LQ}{\Lambda_{\rm QCD}}
\newcommand{\alfs}{\alpha_{s}}
\newcommand{\msbar}{$\overline{\rm MS}$}

\begin{document}

\begin{titlepage}

    \begin{flushright}
      \normalsize TU--986\\
      \today
    \end{flushright}

\vskip1.5cm
\begin{center}
\Large\bf\boldmath
Understanding Interquark Force and 
Quark Masses\\ 
in Perturbative
QCD\footnote{
Based on the lecture courses given at Rikkyo Univ., Kyoto Univ., Karlsruhe
Univ.\ and 
Nagoya Univ., during 2012--2014.
}
\unboldmath
\end{center}

\vspace*{0.8cm}
\begin{center}

{\sc Y. Sumino}\\[5mm]
  {\small\it Department of Physics, Tohoku University}\\[0.1cm]
  {\small\it Sendai, 980-8578 Japan}

\end{center}

\vspace*{0.8cm}
\begin{abstract}
\noindent
This lecture note presents a self-contained 
introduction to the theory of a heavy quark-antiquark 
($Q\bar{Q}$) system
in terms of perturbative QCD.
The lecture is intended for non-experts, such as
graduate course students.
The heavy $Q\bar{Q}$ system serves as an
ideal laboratory for testing various aspects
of QCD:
We can examine the nature of renormalons in perturbative
series;
an effective field theory Potential-NRQCD
is constructed, whose derivation from full
QCD can be traced stepwise;
we see
absorption of renormalons by non-perturbative matrix
elements in OPE clearly;
a systematic 
short-distance expansion of UV contributions can be performed,
which predicts a ``Coulomb+linear'' potential
in perturbative QCD;
we can test these theoretical
formulations by comparison to lattice computations, where we observe
a significant overlap with perturbative regime;
finally we can test our microscopic understanding
by comparing to experimental data for
the bottomonium states.
These subjects are covered
in a concise and elementary manner.
Overall,
we provide a microscopic description of the main dynamics of
a heavy $Q\bar{Q}$ system, as an example for which
theoretical framework, practical computations and qualitative
understanding have been most advanced.
\vspace*{0.8cm}
\noindent

\end{abstract}

\vfil
\end{titlepage}

\tableofcontents

\section{Introduction}

Studying properties of various hadrons has long been 
one of the standard analysis methods to elucidate 
the dynamics of the strong interaction.
Among various observed hadrons,
the heavy quarkonium states
are unique, which are the
bound states of a heavy quark-antiquark ($Q\bar{Q}$)
pair.
This is because they are the
only known individual hadronic states whose
properties can be predicted in a self-contained manner
within perturbative QCD.
Namely, we can compute
several observables associated with individual
heavy quarkonium states
(such as energy levels,  leptonic decay widths  and
transition rates) systematically in expansions in 
the strong coupling constant $\alpha_s$.
Such series expansions make sense,
since the large mass of
the heavy quarks,  $m_Q(\gg \LQ$),
and the color-singlet nature of the bound states restrict the relevant
dynamical degrees of freedom to be in
a short-distance region
and the asymptotic freedom of QCD designates
expansions in a small coupling constant.

The purpose of this lecture  is to provide a
theoretical basis to study properties of
the heavy $Q\bar{Q}$ system in perturbative QCD.
In particular, we describe the nature of the force
between $Q$ and $\bar{Q}$ and 
of their masses inside the bound states.
Indeed, these ingredients determine the main
dynamics inside
the heavy quarkonium states.
For pedagogical reasons,
in this lecture
we restrict our discussion to the leading-order contributions
in the heavy mass limit
$m_Q\to\infty$ and also in the leading logarithmic (LL) order
of the perturbative series.
With these tools, we provide a microscopic description of the
interquark force and quark masses.

Some key aspects are as follows.
First, higher-order terms of perturbative expansions play
crucial roles
to study properties of the heavy $Q\bar{Q}$ bound states
quantitatively as well as qualitatively.
In fact, their properties often turn out to be quite far from
those of the Coulomb bound states such as positronium
states.
Secondly, it is important to separate systematically
ultra-violet (UV) and infra-red (IR)
energy scales involved.
For this purpose it is essential to use a low-energy
effective field theory (EFT) for 
a heavy $Q\bar{Q}$
system, such as potential-NRQCD (pNRQCD).

An interesting feature is that different theoretical frameworks
and various concepts are mutually linked and converge towards
a consistent picture.
Namely, IR renormalons in the purely perturbative computation,
OPE in an EFT, and extraction of UV contributions after
log resummations, all point to a consistent result, which
also agrees with results of  lattice computations.
At the same time we find an interrelation between
the concepts of the running coupling constant,
linear potential and quark self-energies (which resemble
constituent quark masses) from a
microscopic viewpoint, even though
the validity range of the theory
is  restricted to a short-distance
region $r< \LQ^{-1}$.
These subjects are covered in the main body of the lecture.

Let us sketch the outline of the
lecture.
(See also the table of contents in page 1.)
\begin{description}
\item[Sec.~2:]
After explaining the setup, we provide a microscopic picture
of spontaneous breakdown of chiral symmetry and introduce the concept of
constituent quark mass.
\item[Sec.~3:]
To facilitate the reading for beginners, we explain the basic
tools such as renormalization in the 
$\overline{\rm MS}$ scheme, renormalization-group (RG) equation
and running coupling constant.
\item[Sec.~4:]
In order to study the interquark force, we define the static QCD
potential from a Wilson loop and relate it to the energy of
a heavy $Q\bar{Q}$
system.
\item[Sec.~5:]
We explain the notion of IR renormalons as uncertainties characteristic to
perturbative series.
\item[Sec.~6:]
It is shown that the most dominant renormalon uncertainty cancels
in a color-singlet heavy $Q\bar{Q}$ system, which leads to
a dramatic improvement of convergence of perturbative series.
\item[Sec.~7:]
We present a systemtic method for a short-distance expansion of
the UV contributions to the QCD potential and 
show that it predicts
a ``Coulomb+linear'' potential in perturbative QCD.
\item[Sec.~8:]
We provide a microscopic picture and interpretation of
the heavy $Q\bar{Q}$ system, which
can be obtained by the preceding analyses.
\item[Sec.~9:]
We construct pNRQCD EFT from full QCD and explain the relation
between an operator-product expansion (OPE) and renormalons.
A solid theoretical framework to analyze the interquark force
is provided.
\item[Sec.~10:]
We list up references for each of Secs.~2--9
and for further applications,
to assist the readers who want to learn the subjects more deeply 
or those who
are interested in computations of various
observables of heavy quarkonium states.
\end{description}
We collect technical details and necessary knowledge
from related fields
in appendices.
To keep simplicity of explanations,
in the main body of the lecture we do not quote
related papers or describe how the relevant researches were carried out
historically,
apart from some monumental papers.
The refereces are collected in the final section.
Even there, we tend to refer to 
review-like recent papers, from which interested readers
can trace original papers and details of researches at the frontiers.

\begin{figure}[h]
\vspace*{-15mm}
\begin{center}
\includegraphics[width=12cm]{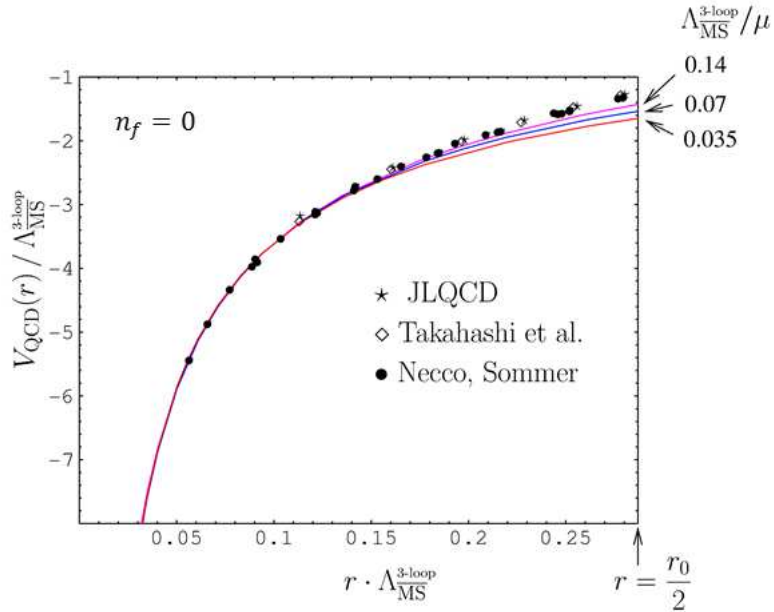}
\end{center}
\vspace*{-13mm}
\caption{\footnotesize
Static QCD potential as a function of the distance between the static
charges $r$.
Both axes are scaled by powers of
$\Lambda_{\overline{\rm MS}}^{\mbox{\scriptsize 3-loop}}$.
Solid lines represent NNNLO perturbative QCD predictions
with different scale choices.
Data points represent lattice computations by three
different groups.
\label{3LoopQCDPot}}
\vspace*{-2mm}
\end{figure}

Before starting the whole discussion,
let us quote the current status of the static QCD potential.
Fig.~\ref{3LoopQCDPot} shows the potential
energy between two static color charges as a function of
the distance
$r$ between the charges.
The scales are measured in units of the QCD scale in the
{\msbar} scheme at
three-loop order, 
$\Lambda_{\overline{\rm MS}}^{\mbox{\scriptsize 3-loop}}$.
The next-to-next-to-next-to-leading order
(NNNLO) perturbative QCD prediction and lattice computations are
compared.
The three solid lines correspond to the perturbative predictions
with different scale choices.\footnote{
It is customary to vary the renormalization scale $\mu$ by
a factor 2 or 1/2 in estimating uncertainties of 
perturbative QCD predictions.
}
The data points represent lattice results by three
different groups.
The number of quark flavors $n_f$ is set to zero in both computations.
$r_0$ denotes the Sommer scale, which is interpreted as about 0.5~fm.
Hence, the largest $r$ in this figure is about 0.25~fm
[$\approx (0.8~{\rm GeV})^{-1}$].
Since the relation between the lattice scale ($r_0$) and
$\Lambda_{\overline{\rm MS}}^{\mbox{\scriptsize 3-loop}}$ 
is taken from other source,
the only adjustable parameter in this comparison is
an $r$--independent constant added to each potential,
whose value is chosen such that
all the potentials coincide at
$r \Lambda_{\overline{\rm MS}}^{\mbox{\scriptsize 3-loop}} = 0.1$.
We see a good agreement between
the perturbative and lattice predictions in the displayed
range.

\section{QCD Lagrangian, Chiral Symmetry, Quark Masses}
\clfn

\subsection{Setup}
\label{sec:setup}

The chiral quark fields are defined as eigenstates
of $\gamma_5$ as
\bea
\gamma_5 \psi_L(x)=-\psi_L(x)
,~~~~~
\gamma_5 \psi_R(x)=+\psi_R(x),
\eea
where $\psi_L$ and $\psi_R$ represent the left-handed and right-handed
quark fields, respectively.
The QCD Lagrangian is given by
\bea
&&
{\cal L}_{\rm QCD}=\sum_{q=u,d,s,c,b,t}
\left[
\overline{\psi}_L^{(q)}i\!\not{\!\! D}\,\psi_L^{(q)}
+
\overline{\psi}_R^{(q)}i\!\not{\!\! D}\,\psi_R^{(q)}
-m_q\left(
\overline{\psi}_L^{(q)}\psi_R^{(q)}
+
\overline{\psi}_R^{(q)}\psi_L^{(q)}
\right)
\right]
\nonumber\\ &&
~~~~~~~~~~
-\frac{1}{4}G_{\mu\nu}^aG^{\mu\nu a} ,
\eea
where the covariant derivative is given by $D_\mu=\partial_\mu-igA_\mu$.

\begin{figure}[t]
\begin{center}
\includegraphics[width=9cm]{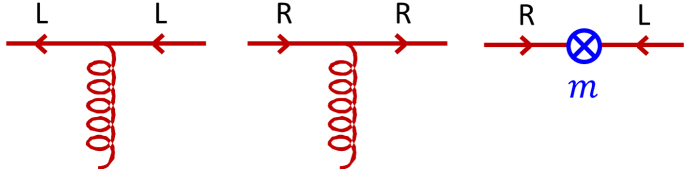}
\end{center}
\vspace*{-.5cm}
\caption{\footnotesize
Chirality-conserving and chirality-violating vertices,
which follow from eq.~(\ref{QuarkLagrangian}).
\label{chirality-diagrams}}
\end{figure}

Chiral transformation rotates $\psi_L$ and $\psi_R$ independently as
\bea
\psi_L \to e^{i\theta_L}\,\psi_L
, ~~~~~
\psi_R \to e^{i\theta_R}\,\psi_R .
\eea
The quark part of the Lagrangian consists of invariant and
non-invariant terms under this transformation:
\bea
\overline{\psi}(i\!\not{\!\! D}-m)\,\psi
= \underbrace{
\overline{\psi}_Li\!\not{\!\! D}\,\psi_L
+ \overline{\psi}_Ri\!\not{\!\! D}\,\psi_R
}_\text{chiral inv.}
-m
\underbrace{
\left(
\overline{\psi}_L\psi_R
+
\overline{\psi}_R\psi_L
\right)}_\text{chiral non-inv.},
\label{QuarkLagrangian}
\eea
where we omit the flavor indices.
Chiral invariance is synonymous to chirality conservation,
which means that a left-handed (right-handed) quark
remains to be left-handed (right-handed) through interactions;
see Fig.~\ref{chirality-diagrams}.

In QCD it is believed that, even if all the masses of quarks vanish
($m_q\to 0$),
hadrons would still remain massive, and
the order parameter of the chiral symmetry  (chiral quark condensate)
would remain non-zero:
\bea
\bra{0}
\overline{\psi}_L\psi_R
+
\overline{\psi}_R\psi_L
\ket{0}
\ne 0 .
\eea
In fact, in Nature, the $\overline{\rm MS}$ masses of the
$u$ and $d$ quarks are much smaller than the proton
and neutron masses:
\bea
m_u\approx 2~{\rm MeV},~~
m_d\approx 5~{\rm MeV}~~
\ll ~~~
m_p\approx m_n \approx 1~{\rm GeV} .
\eea
Conventionally this observation has led to the notion of the
``constituent quark mass'' of order
300~MeV, possessed by each quark inside a nucleon.
It is considered as indicating
spontaneous breakdown of chiral symmetry in the ideal limit $m_q\to 0$,
and that chirality is not conserved by the QCD vacuum.

\subsection{Picture of spontaneous chiral symmetry breakdown}

Let us present a qualitative picture at a microscopic
level of the spontaneous
breakdown of chiral symmetry in the QCD vacuum.

Suppose that initially the state is equal to the perturbative vacuum
state,
namely the ground
state of the free field theory
(the state without any quarks and gluons).
As time
\begin{wrapfigure}{l}{30mm}
\vspace*{-10mm}
\begin{center}
\includegraphics[width=2cm]{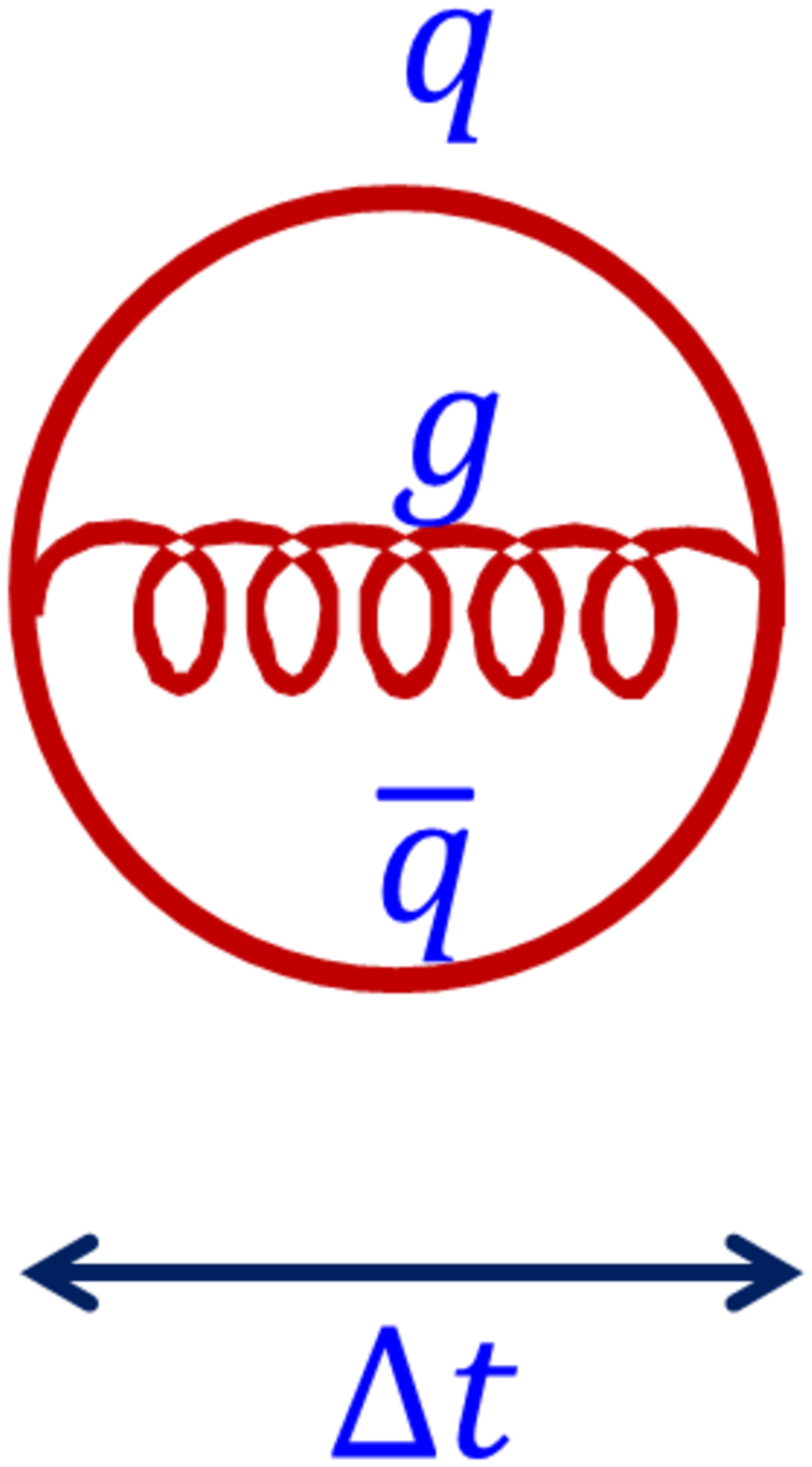}
\end{center}
\vspace*{0mm}
\end{wrapfigure}
evolves, the QCD interaction generates quantum fluctuations.
As indicated by
the left diagram, 
a quark-antiquark
pair and gluon can be created from the perturbative vacuum,
for a short time interval determined by the uncertainty principle, 
\bea
\Delta t \cdot \Delta E \sim 1 .
\eea
$\Delta E$ denotes the energy required to create the three particles.
As $\Delta E$ 

\noindent
becomes smaller, the lifetime of $q\bar{q}g$ 
becomes longer.
$\Delta E$ is given by the sum of the 
energies of the individual particles,
minus the binding energy $E_{\rm bin}$.
The largest
\begin{wrapfigure}{l}{33mm}
\vspace*{-5mm}
\begin{center}
\hspace*{-5mm}
\includegraphics[width=2.5cm]{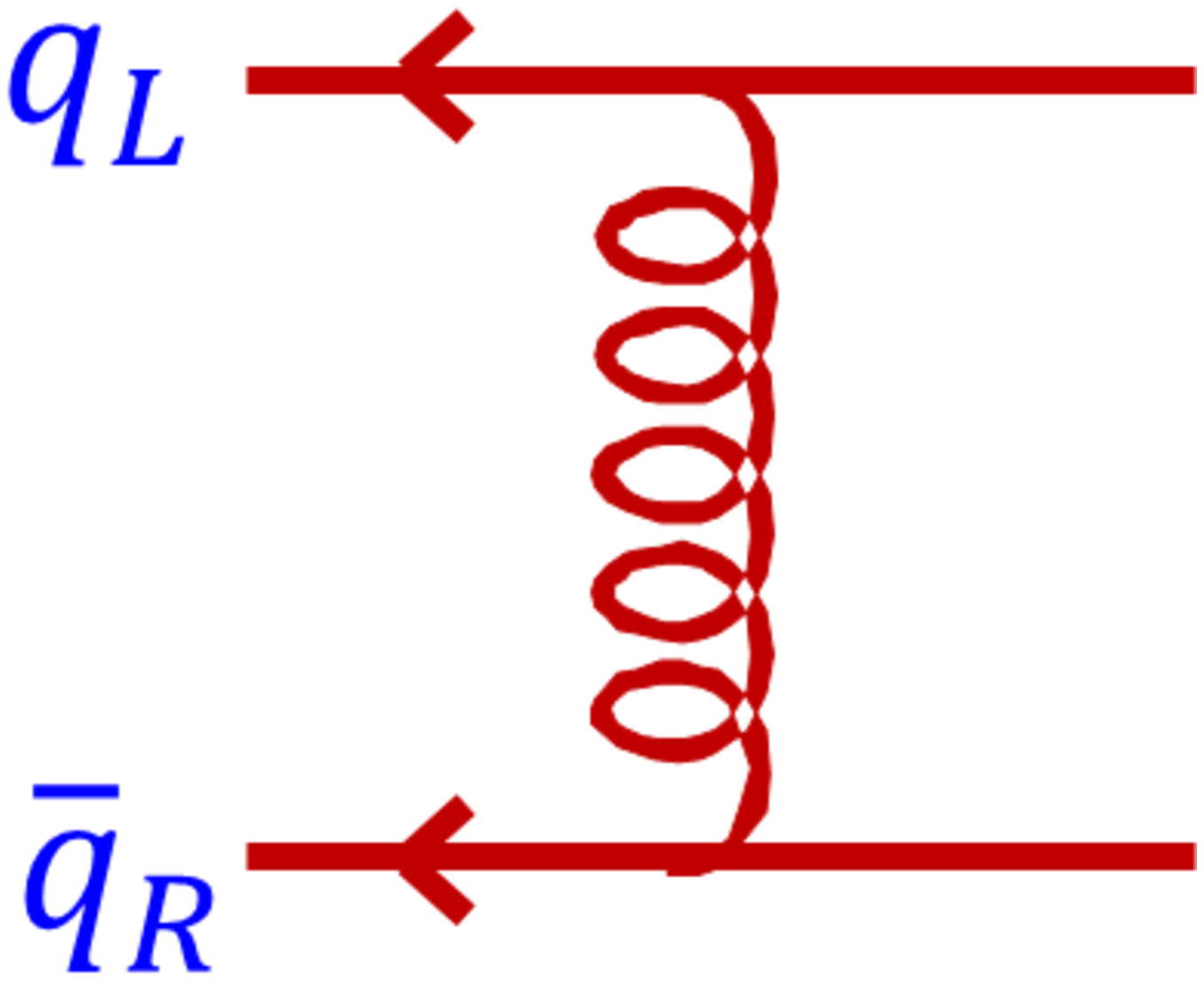}
\end{center}
\vspace*{-5mm}
\end{wrapfigure}
binding energy is expected to be
induced by an attractive force between
the left-handed quark and 
right-handed antiquark\footnote{
More precisely, it is an anti-particle of a
right-handed quark, $\overline{(q_R)}$.
Nevertheless we use the short-hand terminology and notation, for brevity.
} 
($\bar{q}_R$) by exchange of gluons.
Since the energy of each particle is not smaller than its
rest energy (mass), the energy of the system is bounded 
from below as
\bea
\Delta E \geq 2 m_q -E_{\rm bin} .
\eea
$E_{\rm bin}$ is expected to be large, since the attractive force
by QCD interaction is strong.
If the quark masses are small, the binding energy may exceed
$2m_q$.
It is conjectured that this is the case for 
$q=u,d$, such that $\Delta E$ turns negative.

In the case $\Delta E > 0$, the lifetime of $q\bar{q}g$
becomes longer as $\Delta E$ decreases.
If $\Delta E$ decreases further and turns negative,
it becomes energetically more favorable 
(compared to the
perturbative vacuum without any particles)
to create quarks and gluons
and lower the energy of the whole system by the large binding energy.
Hence, $\bar{q}_Lq_R$ and $\bar{q}_R q_L$ bound states are
created everywhere, a large energy is emitted, and
the total energy of the whole system gets lowered down.
This is the condensation of $\bar{q}q$, by which essentially
the space is covered with $\bar{q}q$ bound states, whose energies
are lower than that of the perturbative vacuum.
This state constitutes the true vacuum of QCD.

\begin{wrapfigure}{l}{70mm}
\vspace*{-5mm}
\begin{center}
\includegraphics[width=7.5cm]{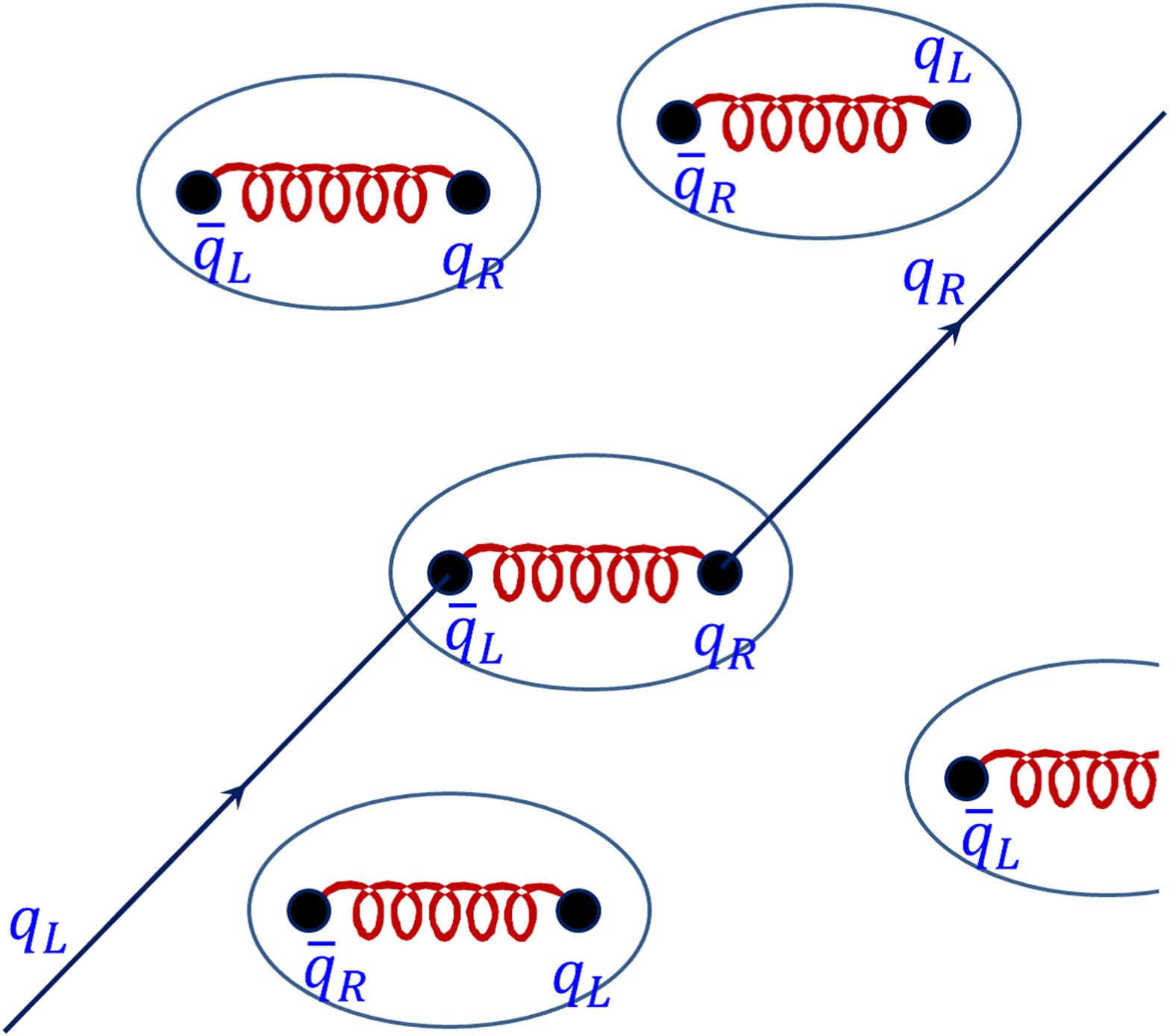}
\end{center}
\vspace*{-5mm}
\end{wrapfigure}
Since there are $\bar{q}_Lq_R$ everywhere in the
true vacuum, 
if a $q_L$ propagates through this 
vacuum, it sometimes
pair-annihilates with $\bar{q}_L$ in a $\bar{q}_Lq_R$ bound state.
Then, $q_R$ loses its partner and starts propagating in place 
of $q_L$.
(See the left figure.)
Thus, it looks as if $q_L$ has turned into $q_R$ while propagating
through the vacuum.

\begin{figure}[h]
\begin{center}
\begin{tabular}{cc}
\includegraphics[width=8cm]{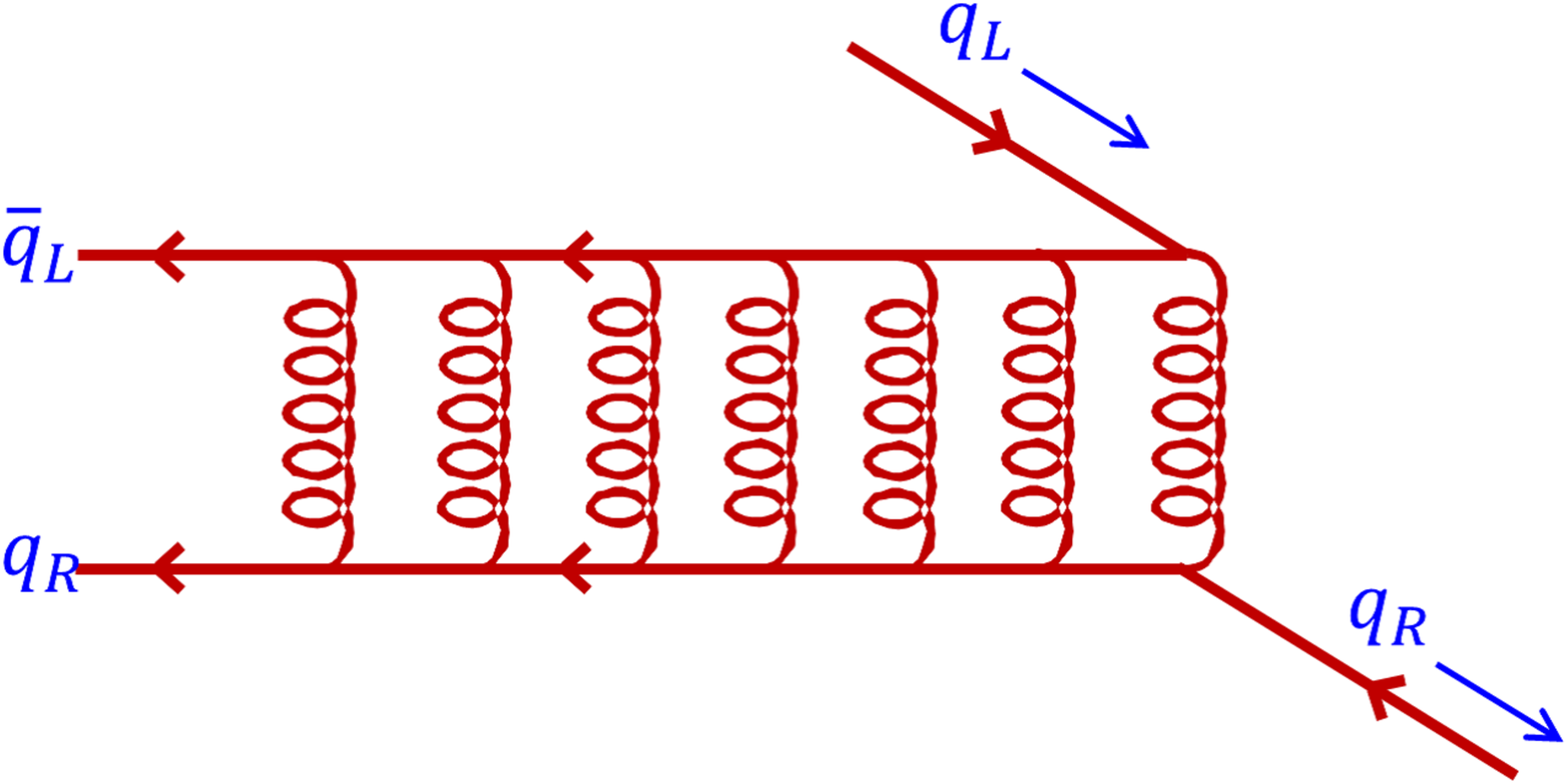}
&~~~~~
\includegraphics[width=3.7cm]{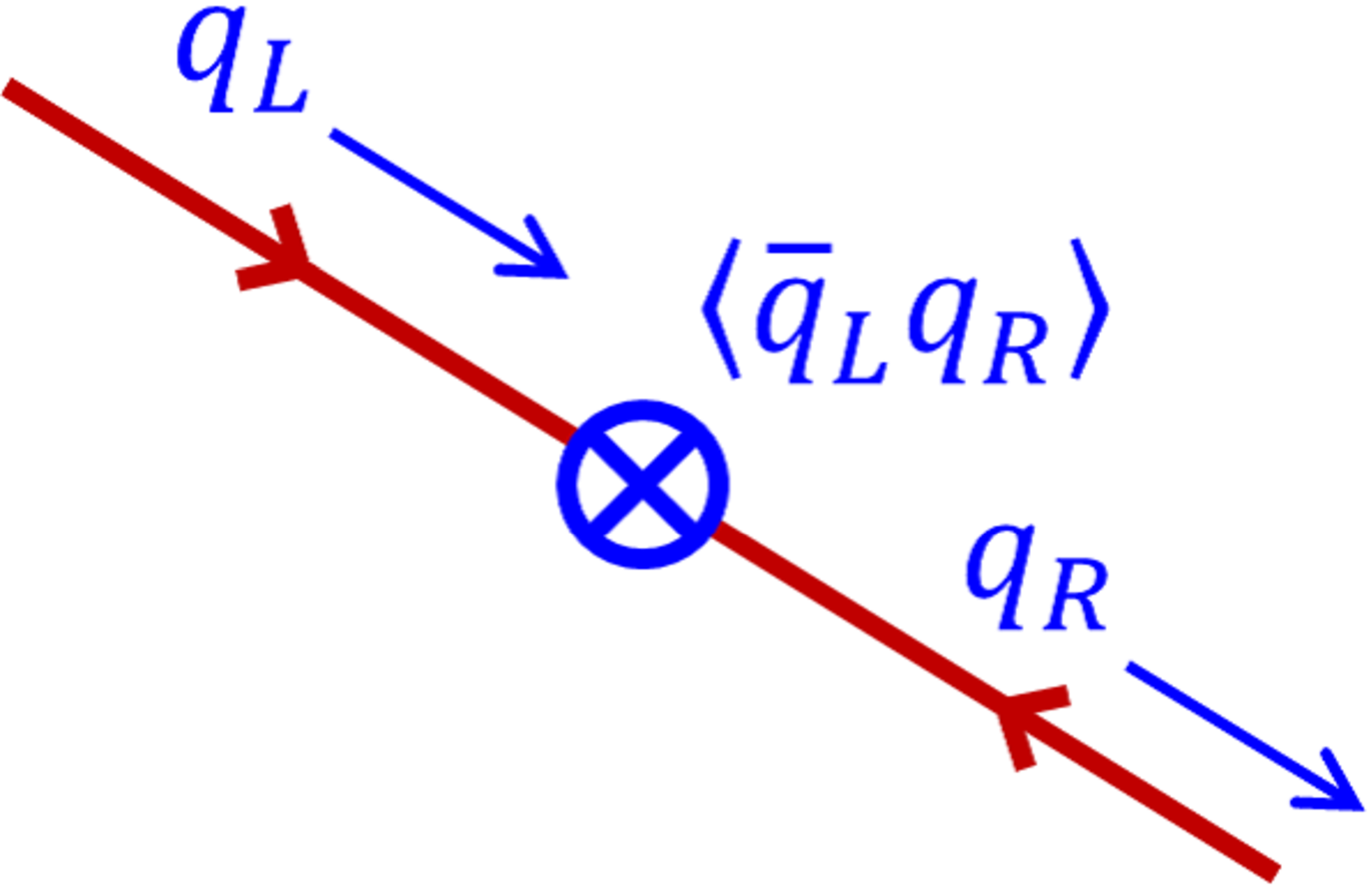}
\vspace*{-2mm}\\
(a)&~~~~~~~(b)
\end{tabular}
\end{center}
\vspace*{-.5cm}
\caption{\footnotesize
(a) Schematic picture of chirality transition when
a quark propagates through the vacuum, depicted in a
Feynman diagrammatic manner.
(Time evolves from left to right.)
The ladder-type multiple gluon exchanges symbolize
a bound state of $\bar{q}_Lq_R$. 
Note that chirality is preserved at the level of fundamental
interaction.
(b) Chirality violation by quark condensate in the vacuum with
spontaneous symmetry breaking,
when one does not see the structure of the vacuum.
\label{FeynDiag-ChiralSSB}}
\end{figure}

Let us depict the same process in a Feynman diagrammatic manner.
As can be seen in Fig.~\ref{FeynDiag-ChiralSSB}(a), 
the fundamental interaction
always preserves chirality, in the limit $m_q\to 0$.
Nevertheless, propagation of a quark in the vacuum looks as if
chirality is non-conserving, if one does not see
the structure of the vacuum; see Fig.~\ref{FeynDiag-ChiralSSB}(b). 

To quantify the above picture, we need to perform a non-perturbative
analysis, e.g., to use Nambu-Jona-Lasino model, Schwinger-Dyson equation,
or lattice QCD simulation.
Nevertheless, some aspects can be elucidated even within 
perturbative QCD, as we will see in the subsequent sections.
It is not possible to describe spontaneous-symmetry-breaking
phenomenon in perturbative QCD.
It is, however, possible to access the ``constituent quark
mass picture,'' which used to be considered also impossible.
We will quantify to which extent this can be handled in perturbative QCD.

At this stage let us ask a paradoxical question.
On the one hand, the constituent quark mass is known to be
larger than the $\overline{\rm MS}$ mass inside a nucleon.
Hence, QCD interaction would increase the mass of a quark.
On the other hand, just like the hydrogen atom, the mass of 
a heavy quarkonium is expected to be 
lighter than the sum of the masses of
the individual heavy quarks, due to a negative binding energy.
Hence, in this case, the masses of the quarks seem to be reduced
by QCD interaction.
The question is, whether QCD interaction increases or decreases
the masses of quarks inside hadrons? 
It is one of the goals of this lecture 
that at the end the readers can answer to
this question from a microscopic viewpoint.

\section{\boldmath $\overline{\rm MS}$ scheme and Running Coupling Constant}

In the following sections we perform analyses of logarithms
in the perturbative prediction for the interquark force.
They are based on the renormalization group (RG) equation, and
we will prepare necessary theoretical basis in this section.
Those who are familiar with this subject
 may as well skip this section.

\subsection{\boldmath Renormalization and $\overline{\text{MS}}$ scheme}

Let us explain the renormalization procedure
in perturbative QCD.
In particular we explain the
$\overline{\rm MS}$ scheme, which is the most frequently used
renormalization prescription in contemporary computations
in perturbative QCD.

For simplicity we focus only on the gauge field,
namely we ignore the quark and ghost fields in this subsection
(which can be treated in a similar manner).
The Lagrangian is given by
\bea
{\cal L}_A=
-\frac{1}{4}G_{\mu\nu}^aG^{\mu\nu a} ;
~~~~~~
G_{\mu\nu}^a=\partial_\mu A_\nu^a - \partial A_\nu^a+gf^{abc}A_\mu^b
A_\nu^c .
\eea
${\cal L}_A$ is expressed in terms of the bare field and bare coupling
constant.
We work
with dimensional regularization and set the dimensions of the space-time
as
$D=4-2\epsilon$ (one temporal and $(3-2\epsilon)$ spatial dimensions).
Here, $\epsilon$ is treated as a general complex variable.
In $D$ dimensions, the (bare) gauge coupling constant
is a dimensionful parameter.
The gauge coupling and gauge field are  rewritten in terms of
the renormalized quantities as
\bea
g=\bar{\mu}^\epsilon Z_g \, g_R ,
~~~~~~~
A_\mu = \sqrt{Z_A}\, A_\mu^R ,
\label{renormalization}
\eea
where $\bar{\mu}$ is a parameter with the dimension of mass,
while the renormalization constants $Z_i$ and 
renormalized coupling constant $g_R$ are defined to be dimensionless;
$A_\mu^R$ denotes the renormalized gauge field.
If we choose $Z_i=Z_i(\epsilon ,g_R)$ appropriately such that
they diverge as $\epsilon \to 0$, perturbative series of
all the physical quantities
($S$-matrix elements, cross sections, spectrum, etc.)
can be made UV finite ('t Hooft).

In the first equation of
(\ref{renormalization}), we separate the single parameter $g$ 
of the theory into the parameters
$\bar{\mu}$ and $g_R$ (and $Z_g$).
Since the number of parameters in the theory should not change,
$\bar{\mu}$ (which will be rewritten by $\mu$ later) and
$\displaystyle \alpha_s=\frac{g_R^2}{4\pi}$ are related if we fix the
bare coupling.
Therefore, we obtain the renormalized coupling constant
as a function of the renormalization scale $\mu$:
\bea
\alpha_s=\alpha_s(\mu)
.
\eea
Theoretical predictions are unchanged if $\alpha_s$ and $\mu$
are varied satisfying this relation.

In the $\overline{\rm MS}$ scheme,
we rewrite 
\bea
\bar{\mu}= \frac{\mu}{\sqrt{4\pi}}\, e^{\gamma_E/2},
\label{mubar}
\eea
where $\gamma_E=0.5772\cdots$ denotes the Euler constant,
and the renormalization constant is taken in the form
\bea
&&
Z_g=1+\frac{\alpha_s(\mu)}{4\pi}\,\frac{Z_{11}}{\epsilon}
+\biggl(\frac{\alpha_s(\mu)}{4\pi}\biggr)^2
\biggl(\frac{Z_{22}}{\epsilon^2}+\frac{Z_{21}}{\epsilon}\biggr)
\nonumber\\ &&
~~~~~~~~~
+\biggl(\frac{\alpha_s(\mu)}{4\pi}\biggr)^3
\biggl(\frac{Z_{33}}{\epsilon^3}+
\frac{Z_{32}}{\epsilon^2}+\frac{Z_{31}}{\epsilon}\biggr)
+ \cdots .
\eea
We take $Z_A$ also in the same form.
Namely, only the poles of $\epsilon$ are included in
$Z_i$, whereas order $\epsilon^0,\epsilon^1,\epsilon^2,\cdots$ terms
are not included.\footnote{
Physical quantities can be made finite even if we include terms of
order $\epsilon^0,\epsilon^1,\epsilon^2,\cdots$ in $Z_i$.
This corresponds to taking other renormalization prescription.
}
In this way
$Z_i$'s are defined uniquely and
the renormalization prescription is fixed ($\overline{\rm MS}$ scheme).

The reason why the $\overline{\rm MS}$ scheme
is used more often than other schemes is that empirically
perturbative series for various physical quantities
exhibit good convergence behaviors.

\subsection{Renormalization group and running coupling constant}
\label{sec:RG}

Let us consider an observable $A=A(\alpha_s(\mu); \mu/Q)$,
which includes only one scale $Q$.
We normalize $A$ by powers of $Q$ such that it becomes dimensionless.
A typical example of $A$ is the $R$-ratio in the
case that quark masses are neglected;
in this case $Q=\sqrt{s}$ represents the c.m.\ energy.

If we compute $A$ in perturbative expansion, we obtain
a polynomial of $\log(\mu/Q)$ in the form
\bea
&&
A(\alpha_s(\mu); \mu/Q)= a_0 + a_1\alpha_s(\mu)
+\alpha_s(\mu)^2\Bigl[
a_2^{(1)}\log \Bigl(\frac{\mu}{Q} \Bigr) + a_2^{(0)} \Bigr]
\nonumber\\ &&
~~~~~~~~~~~~~~~~~~~~~
+\alpha_s(\mu)^3\Bigl[
a_3^{(2)}\log^2 \Bigl(\frac{\mu}{Q} \Bigr) 
+ a_3^{(1)}\log \Bigl(\frac{\mu}{Q} \Bigr) + 
a_3^{(0)} \Bigr]
+ \cdots .
\label{PertExpOfA}
\eea
Dependence on $\mu/Q$ of a single-scale observable
emerges only through $\log(\mu/Q)$, due to the following reason.
Since $g$ is proportional to $\mu^\epsilon$,\footnote{
This is the only source of $\mu$.
} 
at each order of perturbative
expansion, $\mu/Q$ appears in the form
$(\mu/Q)^\epsilon = 1+\epsilon \log(\mu/Q)+\frac{1}{2}\epsilon^2
\log^2(\mu/Q)+\cdots$.
We expand in $\epsilon$ and send $\epsilon\to 0$
after subtracting poles in $\epsilon$, hence we obtain
powers of $\log(\mu/Q)$.

According to the previous subsection, physical quantities
are independent of $\mu$ if we fix the bare coupling constant,
or if $\alpha_s(\mu)$ is varied appropriately.
This leads to the RG equation:
\bea
&&
0=\mu \frac{d}{d\mu}\, A(\alpha_s(\mu);\mu/Q)
\nonumber\\ &&
~~
= \left[
\mu \frac{\partial}{\partial\mu}
+ \mu \frac{d\alpha_s(\mu)}{d\mu}\, 
\frac{\partial}{\partial\alpha_s(\mu)}
\right] \, A(\alpha_s(\mu);\mu/Q) ,
\label{RGeqOfA}
\eea
where we take into account 
the fact that $\mu$ dependence enters directly as well as
indirectly through $\alpha_s(\mu)$.
From the above equation, the beta function is defined by
\bea
\beta(\alpha_s(\mu))=\mu \frac{d\alpha_s(\mu)}{d\mu},
\label{defbetafn}
\eea
which is independent of the observable.

A formal solution to the RG equation (\ref{RGeqOfA}) can be
derived as follows.
Let $t=\log(\mu/Q)$, then $\displaystyle \frac{\partial}{\partial t}
= \mu \frac{\partial}{\partial \mu}$.
Hence, the RG equation has a similar form to the Schr\"odinger equation
in quantum mechanics:
\bea
&&
\biggl( \frac{\partial}{\partial t} - \hat{H} \biggr)
A(\alpha_s;t)=0
 ~~~~~~~\mbox{with}~~~
 \hat{H}= -\beta(\alpha_s)\frac{\partial}{\partial \alpha_s}
 .
 \eea
Its formal solution is given by
\bea
&&
 A(\alpha_s;t)=e^{\hat{H}t}\,A(\alpha_s;t=0)=
 \sum_{n=0}^{\infty}\frac{t^n}{n!}\,\hat{H}^n 
A(\alpha_s;t=0)
\nonumber\\ &&
~~~~~
=
 \sum_{n=0}^{\infty}\frac{\log^n(\mu/Q)}{n!}
\biggl[-\beta(\alpha_s)\frac{\partial}{\partial \alpha_s}\biggr]^n 
\Bigl(
 a_0 + a_1\alpha_s
+a_2^{(0)} \alpha_s^2
++a_3^{(0)} \alpha_s^3
+ \cdots \Bigr)
,
\label{FormalSolToRG}
\eea
where eq.~(\ref{PertExpOfA}) is used to rewrite $A(\alpha_s;t=0)$.
This means that, if we know $\beta(\alpha_s)$
and the log-independent terms of
$A(\alpha_s;\log(\mu/Q))$, 
all the log-dependent terms are determined
by the RG equation.

The beta function can be determined from eqs.~(\ref{RGeqOfA})
and (\ref{defbetafn})
using any observable $A$.
By explicit perturbative computations, it is known to have a form
\bea
\beta(\alpha_s)=-b_0\alpha_s^2
-b_1\alpha_s^3-b_2\alpha_s^4-\cdots .
\label{betafn}
\eea
So far, the expansion coefficients have been computed up to $b_3$.

Let us insert eq.~(\ref{betafn}) to
eq.~(\ref{FormalSolToRG}) and compute the first few terms.
\\ \\
$\bullet~~n=1$\vspace*{-2mm}
\bea
-\beta(\alpha_s)\frac{\partial}{\partial \alpha_s}\alpha_s
=-\beta(\alpha_s)
=
b_0\alpha_s^2
+b_1\alpha_s^3+b_2\alpha_s^4+\cdots ,
\label{An=1}
\eea
$\bullet~~n=2$\vspace*{-2mm}
\bea
&&
\biggl[-\beta(\alpha_s)\frac{\partial}{\partial \alpha_s}\biggl]^2
\alpha_s
=(b_0\alpha_s^2
+b_1\alpha_s^3+b_2\alpha_s^4+\cdots )
\frac{\partial}{\partial \alpha_s}\,
[\mbox{RHS of eq.~(\ref{An=1})}]
\nonumber\\ &&
~~~~~~~~~~~~~~~~~~~~~~~
=
2b_0^2\alpha_s^3
+5b_0b_1\alpha_s^4+\cdots ,
\eea
$\bullet~~n=3$\vspace*{-2mm}
\bea
&&
\biggl[-\beta(\alpha_s)\frac{\partial}{\partial \alpha_s}\biggl]^3
\alpha_s
=
3!\,b_0^3\alpha_s^4
+26b_0^2b_1\alpha_s^5+\cdots ,
\eea
\vspace*{-5mm}\\
\hspace*{10mm}
$\vdots$\hspace*{70mm}$\vdots$
\\ \\
Repeating the same procedure, 
and noting that differentiation by $\alpha_s$ reduces
the power of $\alpha_s$ by one while $\beta(\alpha_s)$
raises the power by at least two,
we find that the coefficients of logarithms in
eq.~(\ref{PertExpOfA}) are determined
by the beta function and $a_1,a_i^{(0)}$'s to have the following form:
\bea
&&
A(\alpha_s;\mu/Q)=
a_0+
\sum_{n=0}^{\infty}\Biggl\{
\underbrace{
a_1\alpha_s\Bigl[b_0\alpha_s\log\Bigl(\frac{\mu}{Q}\Bigr)\Bigr]^n
}_{\rm LL} 
+\underbrace{
C^{\rm NLL}_n\alpha_s^2\Bigl[b_0\alpha_s\log\Bigl(\frac{\mu}{Q}\Bigr)\Bigr]^n
}_{\rm NLL}
+\cdots\Biggr\} ,
\nonumber\\ &&
\eea
where $C^{\rm NLL}_n$ is given by a linear combination of
$a_2^{(0)}/b_0$ and $b_1/b_0$.
The terms, which have $n$th power of $\log(\mu/Q)$ and $(n+1)$th power
of $\alpha_s$, are called the leading logarithmic (LL) terms;
the terms, which have one additional power of $\alpha_s$ compared
to the LL terms, are called the  next-to-leading logarithmic (NLL) terms,
etc.

In the case that $\alpha_s$ is small but $b_0\alpha_s\log(\mu/Q)$
is not small, it is necessary to resum the LL terms, since
terms with higher $n$ are not suppressed.
It is expected that NLL terms and beyond are comparatively smaller,
since $\alpha_s$ is small.
It is possible to resum the LL terms if one knows $b_0$ and $a_1$.
Likewise, it is possible to resum the NLL terms if one knows
$b_1$ and $a_2^{(0)}$ in addition, and so on.
The necessity of resummation occurs when one wants to predict the observable
in a wide range of $Q$, so that $\mu/Q$ varies substantially.

As we have seen, the leading (one-loop)
coefficient of the beta function, $b_0$, in
\bea
\mu \frac{d\alpha_s(\mu)}{d\mu}=-b_0\alpha_s(\mu)^2-b_1\alpha_s(\mu)^3
-\cdots 
\label{DefRunningCoupling}
\eea
dictates the LL terms.
For pedagogical reasons,
in the following we mostly ignore $b_1,b_2,\cdots$
and constrain the argument to the LL case.
Furthermore, we write
\bea
b_0 =\frac{\beta_0}{2\pi}
\label{RedefBeta0}
\eea
in the rest of this lecture.

The solution to the differential equation
(\ref{DefRunningCoupling}) in the
case $b_1=b_2=\cdots=0$ and the replacement
(\ref{RedefBeta0}) defines the one-loop running coupling constant
\bea
\alpha_s(\mu)=\frac{2\pi}{\beta_0\log(\mu/\Lambda_{\rm QCD})} ,
\eea
where $\Lambda_{\rm QCD}$ is a constant of integration.
The inverse relation is given by
\hspace*{20mm}
\begin{wrapfigure}{l}{73mm}
\vspace*{-7mm}
\begin{center}
\includegraphics[width=7cm]{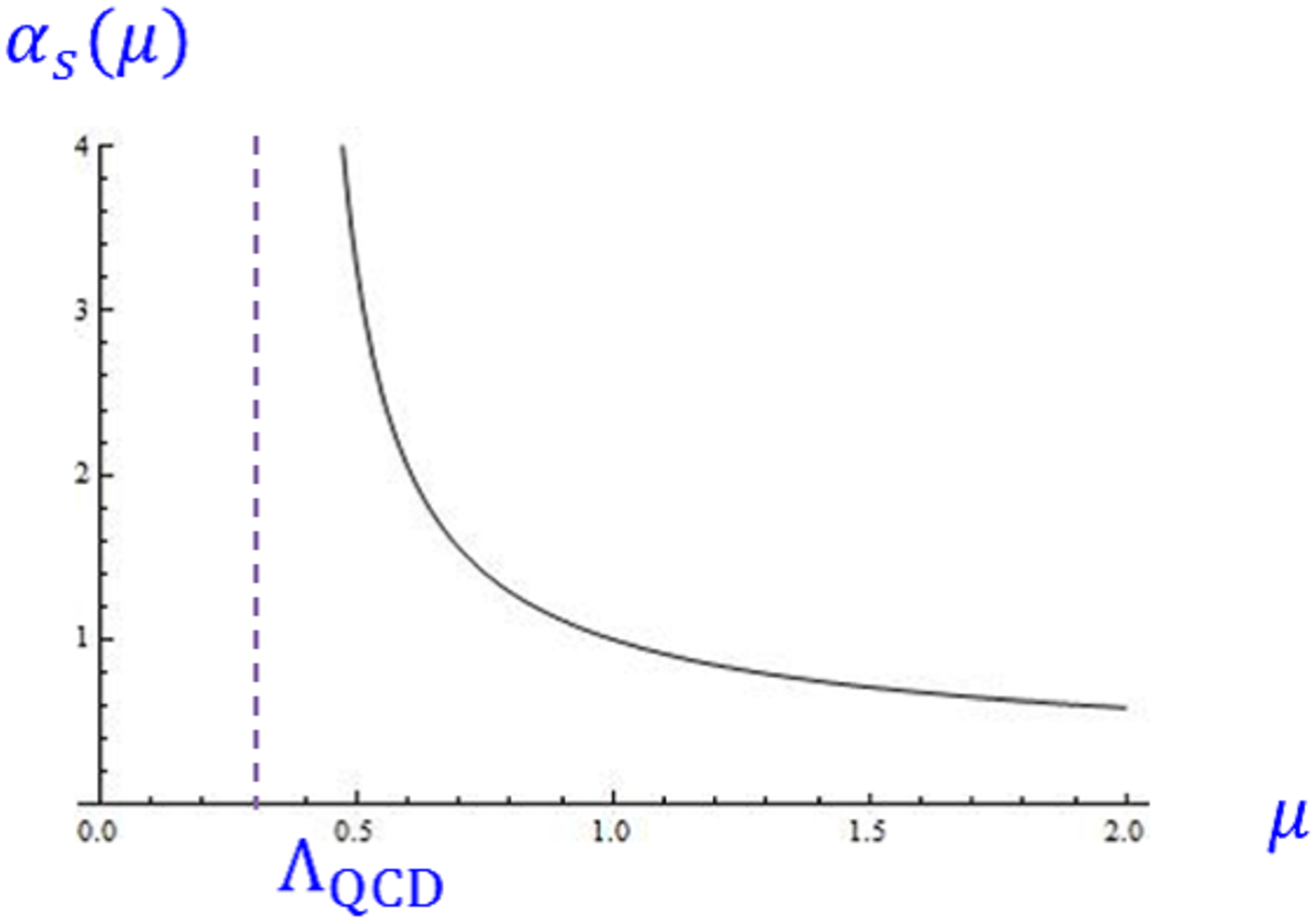}
\end{center}
\vspace*{-15mm}
\end{wrapfigure}
\bea
\Lambda_{\rm QCD}=\mu \,\exp \left[ -\frac{2\pi}{\beta_0\alpha_s(\mu)}
\right] .
\label{1LoopLambda}
\eea
The running coupling constant $\alpha_s(\mu)$
is shown in the left figure.
$\Lambda_{\rm QCD}$ is the scale 
where the 
coupling beomes large,
$\alpha_s(\mu)\simgt {\cal O}(1)$, and the
perturbation theory breaks down.
By comparing to experimental values, it is known that roughly
$\Lambda_{\rm QCD}\sim 200$--300~MeV.

\section{Interquark Force and QCD Potential}
\clfn

In this section we introduce
the static QCD potential, which has long been analyzed
to study the nature of the  force between an infinitely heavy (static) 
quark-antiquark pair.

\subsection{QCD potential from Wilson loop}
\label{sec:IntroQCDpot}

\begin{wrapfigure}{l}{63mm}
\vspace*{-5mm}
\begin{center}
\includegraphics[width=6cm]{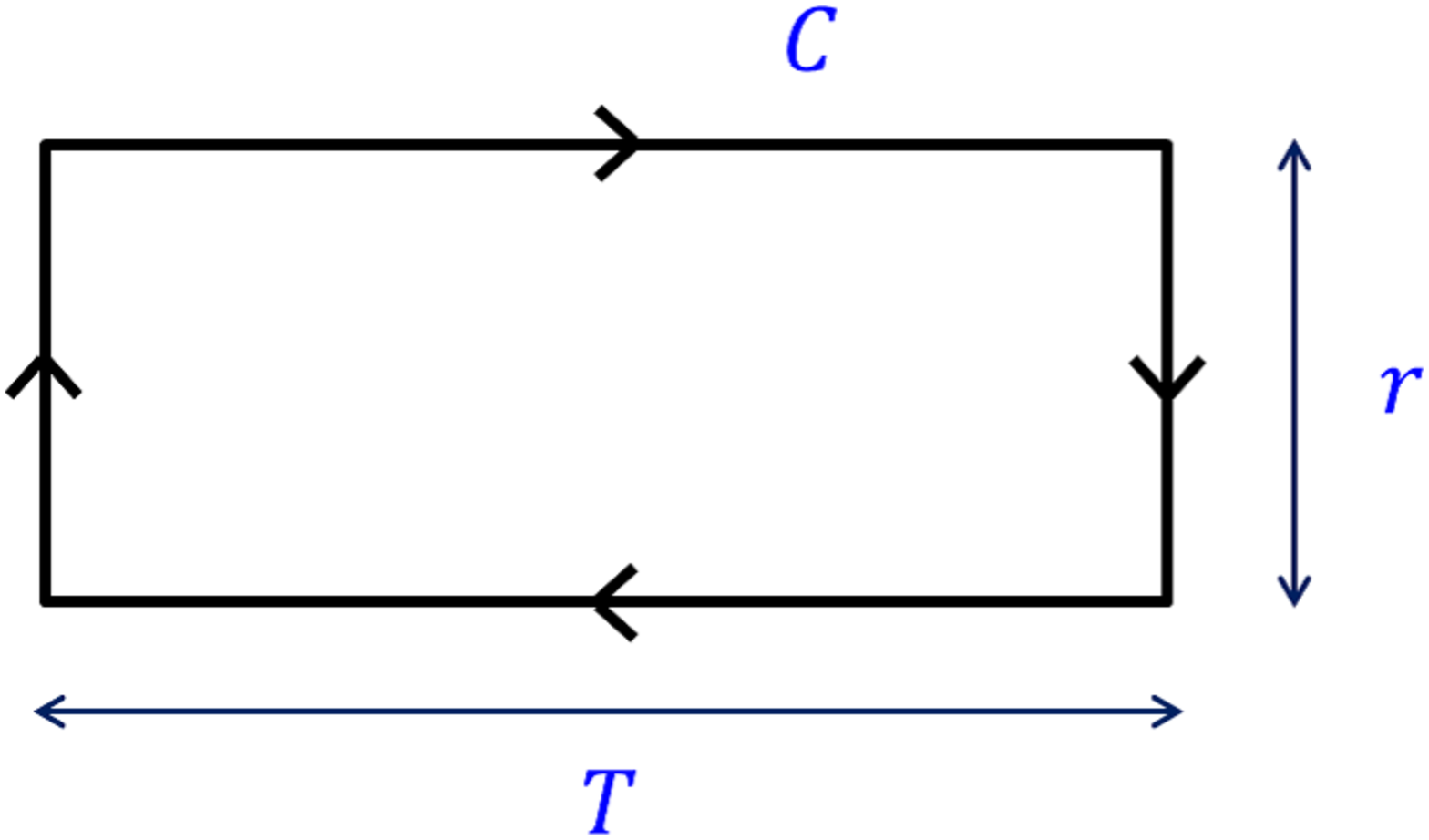}
\end{center}
\vspace*{-10mm}
\end{wrapfigure}
Consider a Wilson loop
\bea
W[A_\mu]=
{\rm Tr}\,\biggl[ {\rm P}
\exp\biggl\{ i g \oint_{C} dx^\mu A_\mu (x) \biggr\}
\biggr] .
\label{WilsonLoop}
\eea
Here, the integral
contour $C$ is taken to be rectangular, 
of spatial extent $r$ and
time extent $T$; see the left figure.
P stands for the path-ordered product along the contour $C$.
(Note that  $A_\mu (x)$
at different $x$ do not generally commute.)
We define the static QCD potential $V_{\rm QCD}(r)$,
from the expectation value of the Wilson loop 
in the large $T$ limit, by
\bea
&&
\langle W[A_\mu] \rangle =
\int {\cal D}A_\mu {\cal D}\psi_q{\cal D}\bar{\psi}_q \,\, W[A_\mu]\,\,
\exp\left(i\int d^4 x \, {\cal L}_{\rm QCD}\right)
\nonumber\\ &&
~~~~~~~~~~~
\approx \mbox{const}.\times
\exp\left[-iTV_{\rm QCD}(r)\right]
~~~\mbox{as}~~~
T \to \infty
.
\label{QCDPotFromWilsonLoop}
\eea
As will be explained in the next subsection, $V_{\rm QCD}(r)$
represents the energy between heavy quarks (static color charges).

\begin{wrapfigure}{l}{75mm}
\vspace*{-5mm}
\begin{center}
\includegraphics[width=7cm]{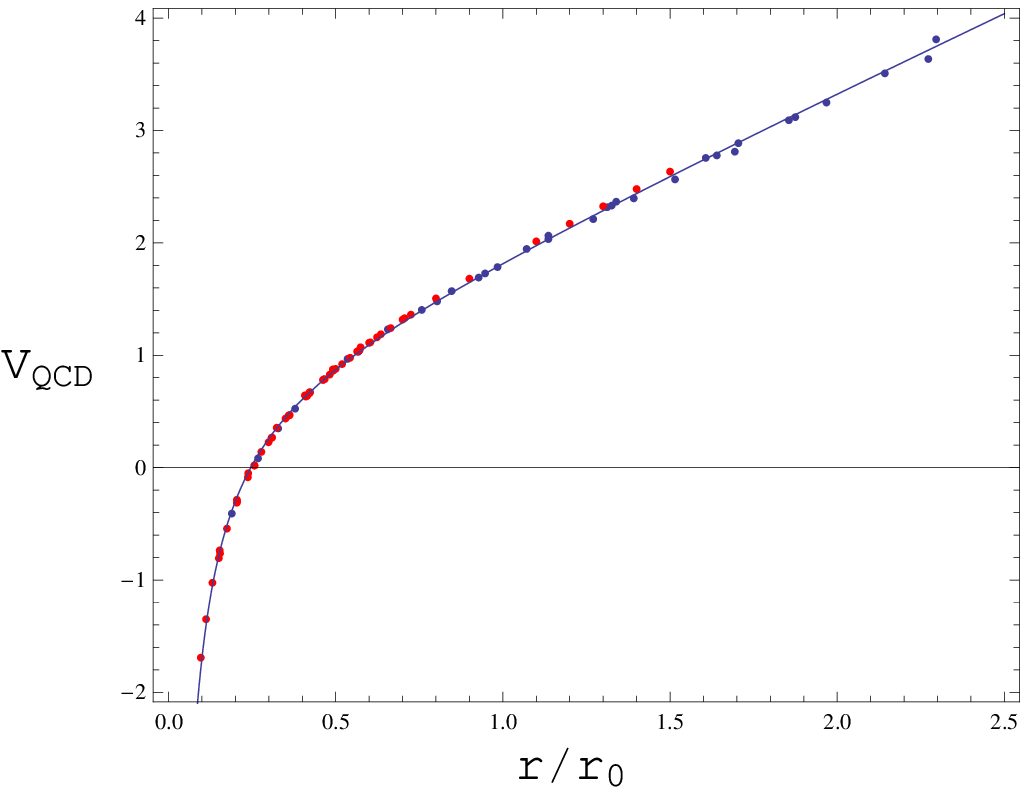}
\end{center}
\vspace*{-7mm}
\end{wrapfigure}
Known from numerical computations by lattice simulations,
$V_{\rm QCD}(r)$ can be fitted well by a
``Coulomb+linear'' form:\footnote{
It has also been known that
phenomenological potentials, which reproduce the
measured bottomonium and
charmonium spectra, have shapes close to $V_{\rm QCD}(r)$
computed by lattice simulations.
}
\bea
V_{\rm QCD}(r) \approx -\frac{a}{r} + Kr + \mbox{const}.
\label{Coulomb+LinearForm}
\eea
We show in the left figure lattice computations of 
$V_{\rm QCD}(r)$ by
two different groups and a fit of these numerical data
by the above ``Coulomb+linear'' form.
In general, lattice simulations can compute $V_{\rm QCD}(r)$
accurately at large $r$, whereas perturbative QCD
can predict $V_{\rm QCD}(r)$ more accurately at small $r$.

There has been a kind of folklore that
in eq.~(\ref{Coulomb+LinearForm}) the perturbative
prediction gives the Coulomb part $-a/r$,
while non-perturbative contribution gives the
linear part $Kr$, 
and both contributions add up.\footnote{
There is an indication in this
direction.
In lattice simulations with gauge fixing
in maximally abelian gauge, it is known that $V_{\rm QCD}(r)$
can be approximated well by the sum of the contribution
from `abelian gluons,' which gives a Coulomb-like potential,
and the contribution from monopoles, which gives rise
to a linear potential.
This feature seems to be valid down to fairly short distances.
}
We will show that at least at short distances,
$r \simlt 0.5$~fm, this statement is incorrect,
in that the linear part is included in perturbative
prediction.

\subsection{Energy of a static quark pair}
\label{Sec:MeaningV_QCD}

\begin{wrapfigure}{l}{40mm}
\vspace*{-10mm}
\begin{center}
\includegraphics[width=3.5cm]{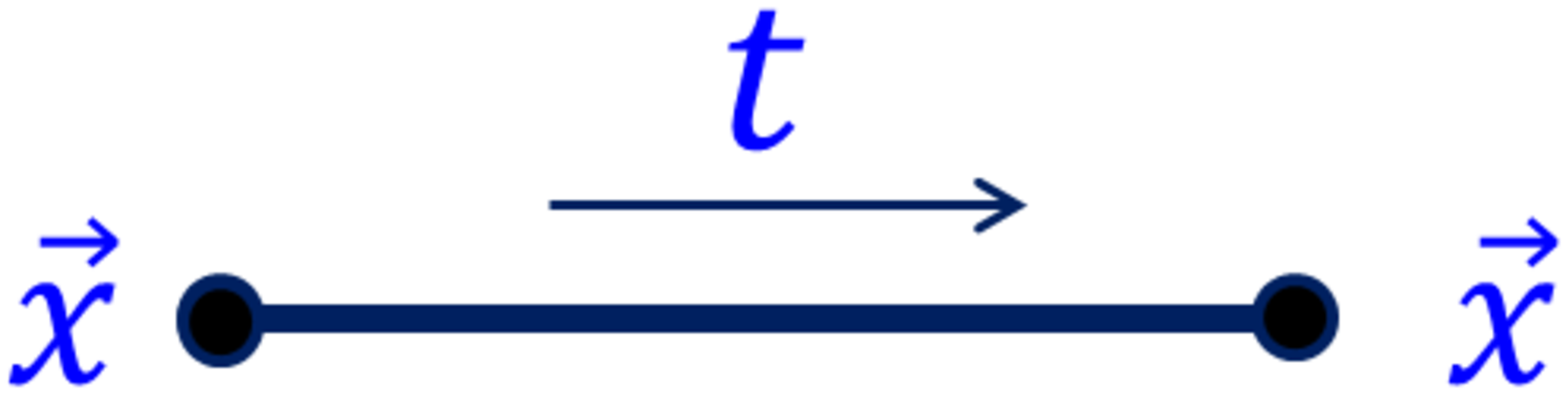}
\end{center}
\vspace*{-10mm}
\end{wrapfigure}
In order to clarify the meaning of the static QCD potential,
we rewrite the Wilson line by path integral.
The Wilson line in the time direction (see left)
can be written as\footnote{
The delta function on the left-hand side follows from the
usual equal-time commutation relation
of $\psi(t,\vec{x})$ and $\psi^\dagger(t,\vec{y})$.
}
\bea
{\rm P}\, \exp \!\left[ ig\!\int_0^T\!\!dt\, A_0(t,\vec{x}) \right]
\delta^3(\vec{x}-\vec{y})= \int\!
{\cal D}\psi{\cal D}\psi^\dagger\,
\psi(T,\vec{x})\psi^\dagger(0,\vec{y})\,
\exp\!\left[ i\!\int \! d^4x\,{\cal L}_{\rm HQET} \right] ,
\nonumber\\
(T\ge 0)~~~~~~
\label{WilsonLineByPI}
\eea
where $\psi(x)$ is a complex scalar field,\footnote{
Since we deal only with zero- or one-particle states,
it does not matter whether $\psi(x)$ is a boson or fermion.
The point here is that we do not consider the spin degrees of freedom
of the color charge.
}
\!which belongs to the
$N$ representation of $SU(N)$ gauge group.
\!($N=3$ for QCD.)\,
${\cal L}_{\rm HQET}$ denotes the Lagrangian of the 
Heavy Quark Effective Theory (HQET), given by
\bea
{\cal L}_{\rm HQET}=\psi(x)^\dagger iD_t \, \psi(x),
~~~~~~~~~
D_t=\partial_t-igA_0(x) .
\eea
The Feynman rules of HQET are shown in Fig.~\ref{FeynRule-HQET} below.
\begin{figure}[h]
\begin{center}
\includegraphics[width=6.3cm]{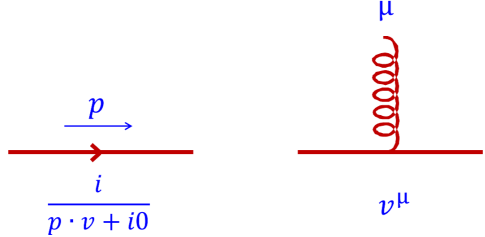}
\end{center}
\vspace*{-5mm}
\caption{\footnotesize
Feynman rules of HQET.
The left diagram represents the heavy quark propagator,
and the right diagram represents the gluon vertex.
$v^\mu=(1,\vec{0})$ denotes the unit vector in the
time direction.
\label{FeynRule-HQET}}
\end{figure}

The equality (\ref{WilsonLineByPI}) can be shown in the following
manner.
Since ${\cal L}_{\rm HQET}$ is bilinear in $\psi$, the path
integral on the right-hand side can be performed to yield
\bea
[\mbox{RHS of eq.~(\ref{WilsonLineByPI})}]=
D_t^{-1}(T,\vec{x};0,\vec{y}) ,
\eea
which means that it is the kernel of the
covariant derivative $D_t$:
\bea
\Bigl[ \partial_t - ig A_0(T,\vec{x}) \Bigr]
[\mbox{RHS of eq.~(\ref{WilsonLineByPI})}]=
\delta(T)\delta^3(\vec{x}-\vec{y}) .
\eea
On the other hand, at $T>0$, we find
\bea
&&
\Bigl[ \partial_t - ig A_0(T,\vec{x}) \Bigr]
[\mbox{LHS of eq.~(\ref{WilsonLineByPI})}]=
\Bigl[ ig A_0(T,\vec{x}) - ig A_0(T,\vec{x}) \Bigr]
[\mbox{LHS of eq.~(\ref{WilsonLineByPI})}]
\nonumber\\&&
~~~~~~~~~~~~~~~~
~~~~~~~~~~~~~~~~
~~~~~~~~~~~~~
=
0
\eea
by differentiation.
Furthermore, if we take the limit $T\to 0$, we find
\bea
[\mbox{LHS, RHS of eq.~(\ref{WilsonLineByPI})}]
\to \delta^3(\vec{x}-\vec{y}) .
\eea
Thus, both sides satisfy the same first-order differential
equation with the same initial condition for $T\ge 0$.

Using the above relation, we can express the Wilson loop
(\ref{WilsonLoop})
as a path integral:
\bea
&&
W[A_\mu]\, \delta^3(\vec{x}-\vec{x}')\delta^3(\vec{y}-\vec{y}^{\,\prime})
\nonumber\\&&
=\int\!
{\cal D}\psi{\cal D}\psi^\dagger\,
{\cal D}\chi{\cal D}\chi^\dagger\,
\exp\!\left[ i\!\int \! d^4x\,
(\psi^\dagger iD_t \, \psi+\chi^\dagger iD_t \, \chi)
 \right]
\nonumber\\&&
~~~~~~~~~
\times 
\underbrace{\left[ 
\psi^\dagger(0,\vec{x})\,\phi(\vec{x},\vec{y};0)\,
\chi(0,\vec{y})\right]}_{\text{(i)}} 
\, 
\underbrace{\left[ 
\psi(T,\vec{x}')\,\phi^\dagger(\vec{x}',\vec{y}^{\,\prime};T)\,
\chi^\dagger(T,\vec{y}^{\,\prime})\right]}_{\text{(ii)}}
,
\label{WilsonLoopByPI}
\eea
where $\psi(x)$ and $\chi(x)$ are both complex scalar fields in the
$N$ representation, and 
\bea
\phi(\vec{x},\vec{y};t)={\rm P}\,\exp\left[ig\!\int_{\vec{y}}^{\vec{x}}\!
d\vec{x}\cdot\vec{A}(t,\vec{x})
\right]
\eea
denotes the color string spanned between $\vec{x}$ and $\vec{y}$.
Here, $\chi(x)$ represents the field, which {\it creates}
the anti-color charge and belongs to the $N$ representation (rather
than $\overline{N}$ representation).
Namely, we adopt the definition, which corresponds to
the anti-particle component of the Dirac field
(i.e., $\chi \sim d^\dagger$ of
$b_{\vec{p},s}e^{-i\vec{p}\cdot\vec{x}}+d_{\vec{p},-s}^\dagger
e^{i\vec{p}\cdot\vec{x}}$).
Eq.~(\ref{WilsonLoopByPI}) is depicted in Fig.~\ref{MeaningWilsonLoop} below.
\begin{figure}[h]
\begin{center}
\includegraphics[width=9cm]{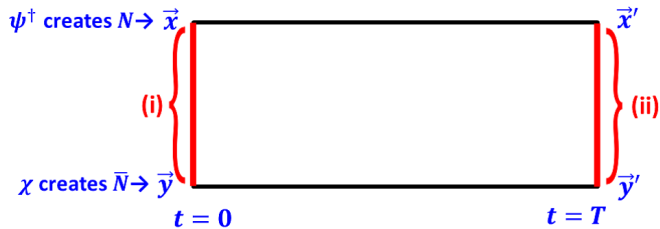}
\end{center}
\vspace*{-5mm}
\caption{\footnotesize
Correspondence between eq.~(\ref{WilsonLoopByPI}) and the
Wilson loop. 
Only the configuration $\vec{x}=\vec{x}'$ and $\vec{y}=\vec{y}^{\,\prime}$
survives, since $\psi$ and $\chi$ cannot propagate in the spatial
direction.
\label{MeaningWilsonLoop}}
\end{figure}

We compute $V_{\rm QCD}(r)$ from $\langle W[A_\mu] \rangle$
as given by eq.~(\ref{QCDPotFromWilsonLoop}).
Substituting the expression (\ref{WilsonLoopByPI}) [after
dividing by 
$\delta^3(\vec{x}-\vec{x}')\delta^3(\vec{y}-\vec{y}^{\,\prime})$], 
the Lagrangian of eq.~(\ref{QCDPotFromWilsonLoop}) is replaced
as
\bea
{\cal L}_{\rm QCD} \to
{\cal L}_{\rm QCD} +
\psi^\dagger iD_t \, \psi+\chi^\dagger iD_t \, \chi
,
\label{QCD+HQET-Lagrangian}
\eea
and there are insertions of the color source
operators (i) and (ii).
Starting from this expression we can compute the
Wilson loop [and therefore $V_{\rm QCD}(r)$]
in perturbative QCD, using the Feynman rules in Fig.~\ref{FeynRule-HQET}
in combination with those of QCD.
The leading contribution to
$V_{\rm QCD}(r)$ is given by the diagram with one-gluon exchange
between the two color charges
at $\vec{x}$ and $\vec{y}$, which simply gives
a Coulomb potential.

Equivalently, we can express $\langle W[A_\mu] \rangle$
in canonical formulation.
Let\footnote{
Since the HQET Lagrangian is first-order in time
derivative, the Hamiltonian is given by dropping the kinetic
term of the Lagrangian, similarly to the Dirac theory.
(It can be derived from the path integral in  holomorphic representation.)
}
\bea
&&
\ket{\alpha}=\psi^\dagger(\vec{x})\, \phi(\vec{x},\vec{y};0)
\,\chi(\vec{y})\ket{0} ,
\\ &&
\ket{\beta}=\psi^\dagger(\vec{x}')\, \phi(\vec{x}',\vec{y}^{\,\prime};T)
\,\chi(\vec{y}^{\,\prime})\ket{0} ,
\\ &&
H=\int d^3 \vec{x}\, ({\cal H}_{\rm QCD} +
ig\,\psi^\dagger \!A_0 \psi + ig \,\chi^\dagger\! A_0 \chi ) ,
\label{Hamiltonian}
\eea
then we can express
\bea
&&
\langle W[A_\mu] \rangle=\frac{
\bra{\beta}e^{-iHT}\ket{\alpha}
}{
\delta^3(\vec{x}-\vec{x}')\delta^3(\vec{y}-\vec{y}^{\,\prime})
}
\nonumber\\ &&
~~~~~~~~~~~
\approx \mbox{const}.\times
\exp\left[-iTV_{\rm QCD}(r)\right]
~~~\mbox{as}~~~
T \to \infty
.
\eea
This shows that $V_{\rm QCD}(r)$ is the lowest energy eigenvalue
of the energy eigenstates of $H$, which have overlaps with $\ket{\alpha}$
and $\ket{\beta}$.
Indeed, by inserting completeness relation in terms of the
eigenstates of $H$, we find
\bea
&&
\bra{\beta}e^{-iHT}\ket{\alpha}
=\sum_n\braket{\beta}{n}\braket{n}{\alpha}\,e^{-iE_nT}
\nonumber\\ &&
~~~~~~~~~~~~~~~~~~
\longrightarrow 
\braket{\beta}{n_0}\braket{n_0}{\alpha}\,e^{-iE_{n_0}T}
~~~\mbox{as}~~~
T \to \infty,
\eea
where the lowest energy state $\ket{n_0}$ is selected by
$+i0$ prescription, as usual.

One way to see
that the Lagrangian (\ref{QCD+HQET-Lagrangian})
[or the Hamiltonian (\ref{Hamiltonian})]
corresponds to inclusion of infinitely heavy colored particles
may be as follows.
The Lagrangian of the non-relativisitic Schr\"odinger
equation in quantum mechanics is given by
\bea
{\cal L}_{\rm QM}=\psi^\dagger \biggl(
iD_t-\frac{\vec{D}^2}{2m} \biggr)\, \psi .
\eea
If we send $m\to\infty$, it reduces to ${\cal L}_{\rm HQET}$.

Thus, $V_{\rm QCD}(r)$ can be interpreted as
the energy between infinitely heavy (static) color charges,
which are separated by distance $r=|\vec{x}-\vec{y}|$ and in a color-singlet
state
(since $\ket{\alpha},\ket{\beta}$ are color singlet).
The force between the static charges is obtained by differentiating 
$V_{\rm QCD}(r)$:
\bea
F(r)=-\frac{d}{dr}V_{\rm QCD}(r) .
\eea

\section{Renormalons in QCD Potential}
\label{sec:RenormalonsInV_QCD}
\clfn

In this section we explain renormalons in 
the perturbative prediction of $V_{\rm QCD}(r)$,
which cause large theoretical uncertainties.

\subsection{Theoretical background: asymptotic series}

First we explain the notion of asymptotic series.
As an example, we consider a one-dimensional integral
\bea
F(\lambda)=\int_{-\infty}^\infty\! dx\,
e^{-\frac{1}{2}x^2-\frac{\lambda}{4}x^4} ,
\eea
which is a toy model imitating the partition function of
the $\lambda \phi^4$ theory.
This integral has the following properties:
\begin{itemize}
\item
The integral is well defined for ${\rm Re}\,\lambda \ge 0$.
\item
The integral is divergent for ${\rm Re}\,\lambda < 0$.
\item
$\lambda=0$ is an essential singularity.
\end{itemize}
As a result, the convergence radius of the 
series expansion of $F$ in $\lambda$ is zero.\footnote{
$F(\lambda)$ can be expressed analytically using
the Bessel function, and its analytic properties can be
studied in full detail.
}
Namely, the series expansion does not converge for whatever small
value of $|\lambda|$.
In fact, the perturbative series can be computed as 
\bea
&&
F(\lambda)=\int_{-\infty}^\infty\! dx\,
e^{-\frac{1}{2}x^2}
\sum_{n=0}^\infty \frac{\Bigl(-\frac{1}{4}\lambda x^4\Bigr)^n}{n!}
=\sum_{n=0}^\infty\lambda^nB_n
 ,
 \\ &&
B_n=\frac{\sqrt{2}\,(-1)^n}{n!}
\,\Gamma\!\Bigl(2n+{\textstyle\frac{1}{2}}\Bigr)
.
\eea

\begin{wrapfigure}{l}{70mm}
\vspace*{-10mm}
\begin{center}
\includegraphics[width=6.5cm]{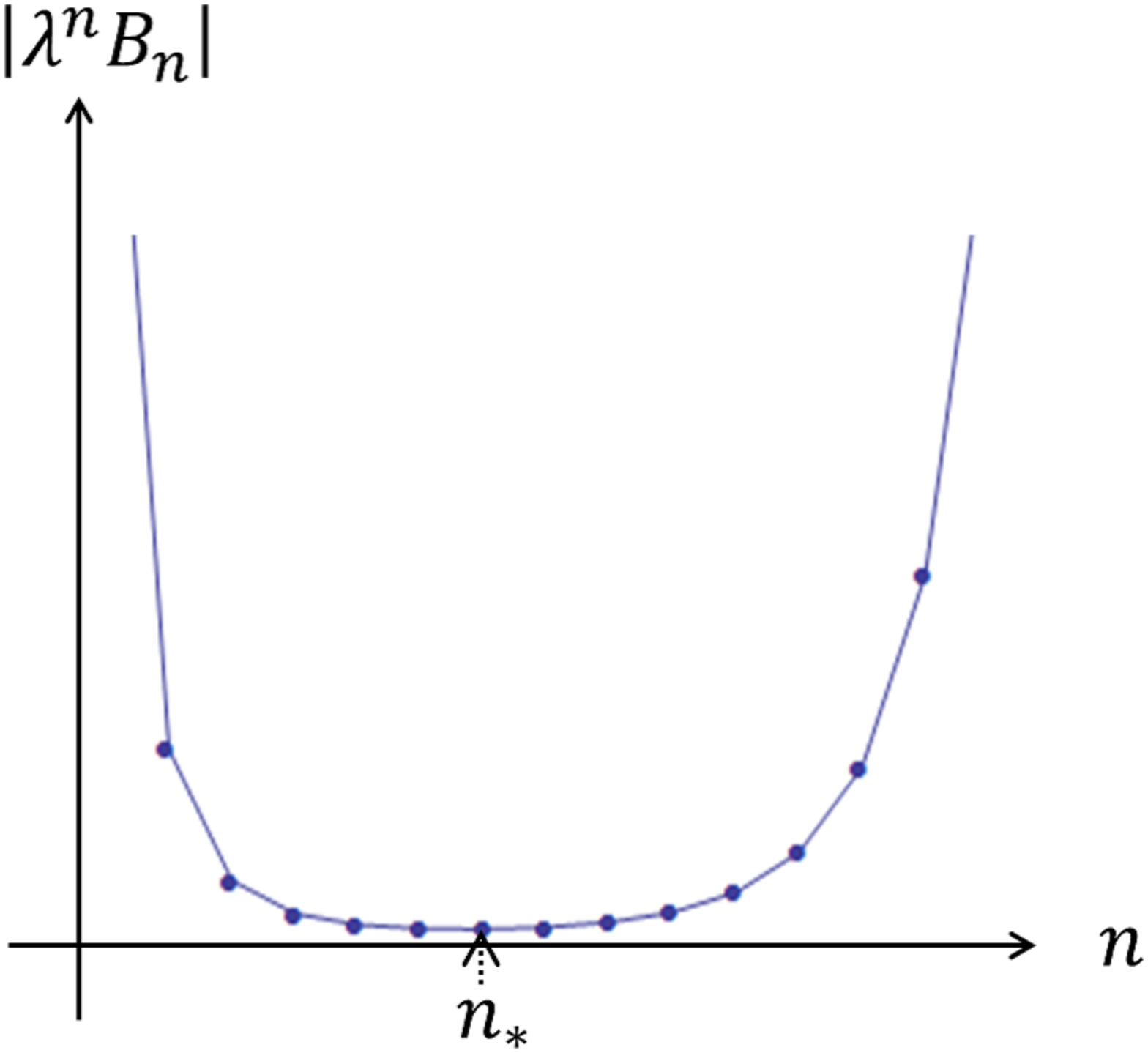}
\end{center}
\vspace*{-15mm}
\end{wrapfigure}
\noindent
$B_n$ grows rapidly for large $n$.
Hence, for $|\lambda| \ll 1$, the series first
converges apparently but
starts diverging from some $n=n_*$;
see the left 
figure, 

Let us define the difference between the true
value $F(\lambda)$ and the sum of a truncated
series by
\bea
\delta F_N \equiv \left|
F(\lambda)-\sum_{n=0}^N \lambda^n B_n
\right|
\eea
for Re\,$\lambda\ge 0$ 
[such  that $F(\lambda)$ is well defined].
It can be shown that $\delta F_N\sim {\cal O}(|\lambda^N B_N|)$. 
Namely, for $N\simlt n_*$, the truncated series
approaches the true value as we include more terms.
The best approximation is obtained if we truncate at
$N\approx n_*$, where $\delta F_N \sim |\lambda^{n_*} B_{n_*}|$.

\begin{figure}[h]
\begin{center}
\includegraphics[width=11cm]{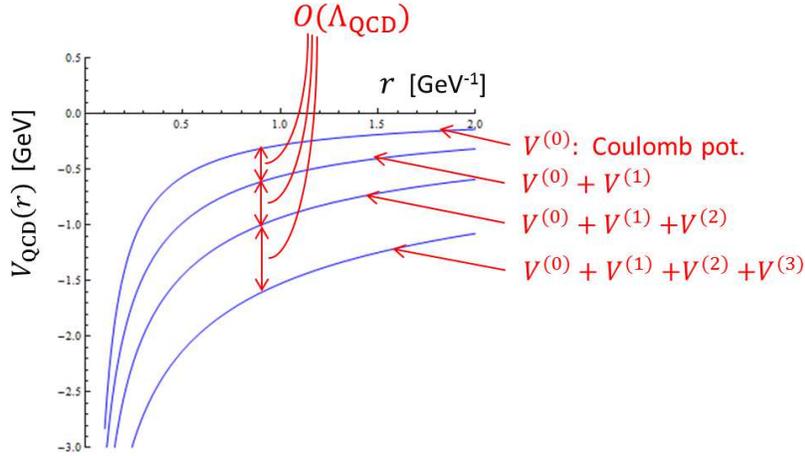}
\end{center}
\vspace*{-5mm}
\caption{\footnotesize
Truncated perturbative series of $V_{\rm QCD}(r)$
up to the first four terms.
At each order the potential receives a large 
correction,
which is an almost negative $r$-independent shift of order
$\Lambda_{\rm QCD}$.
\label{FixedOrderPotentials}}
\end{figure}

It is conjectured that in QCD $\alpha_s=0$ is also an essential
singularity.
Therefore, at best, perturbative expansions 
in $\alpha_s$ would be 
asymptotic series.
Explicit computations have shown that the 
perturbative expansion of $V_{\rm QCD}(r)$
receives large radiative 
corrections even at low orders of expansion,
which results in a theoretical uncertainty, 
$\delta V_{\rm QCD}(r)\sim\Lambda_{\rm QCD}\sim$200--300~MeV.
See Fig.~\ref{FixedOrderPotentials}.
This feature can be understood qualitatively in terms
of the renormalons.

\subsection[Renormalons in $V_{\rm QCD}(r)$: some details]{\boldmath Renormalons in $V_{\rm QCD}(r)$: some details}
\label{sec:somedetails}

According to the argument in Sec.~\ref{sec:RG},
the perturbative expansion of $V_{\rm QCD}(r)$ takes a
form:
\bea
&&
V_{\rm QCD}(r)=-\frac{C_F\alpha_s}{r}
\Biggl[ 1+ \frac{\alpha_s}{4\pi}\Bigl\{
\beta_0\log(\mu^2 r^2)+a_1 \Bigr\} 
+\biggl(\!\frac{\alpha_s}{4\pi}\!\biggr)^{\!\! 2}
\Bigl\{
\beta_0^2\log^2(\mu^2 r^2)+\cdots \Bigr\} + \cdots \Biggr]
,
\nonumber\\&&
\eea
where the Casimir operator of the fundamental representation is
defined as $T_F^aT_F^a=C_F\,\mathbf{1}$, and $C_F=4/3$ in QCD;
$\beta_0$ is the
coefficient of the one-loop beta function defined by
\bea
\mu \frac{d \alpha_s}{d\mu} = - \, \frac{\beta_0}{2\pi} \,\alpha_s^2 
- \cdots .
\eea

Let us
define a potential with LL resummation formally as
\bea
V_{\beta_0}(r) = - \int
\frac{d^3 \vec{q}}{(2\pi)^3} \, e^{i \vec{q} \cdot \vec{r}}
\, C_F \, \frac{ 4 \pi \alpha_{\rm 1L}(q)}{q^2}
~~~~~~
;
~~~~~~
q \equiv |\vec{q}| ,
\label{Vbeta0}
\eea
where
\bea
\alpha_{\rm 1L}(q) =
\frac{2\pi}{\beta_0 \log \bigl( {q}/{\Lambda_{\rm QCD}} \bigr)}
=
\frac{\alpha_s(\mu)}{ 1 + \frac{\beta_0\alpha_s(\mu)}{2\pi} \,
\log \bigl( \frac{q}{\mu} \bigr)}
\label{1LRunningCoupling}
\eea
denotes the one-loop running coupling constant.
In the second equality we substituted the relation
(\ref{1LoopLambda}).
Use of $\alpha_{\rm 1L}(q)$
corresponds to resumming LLs in momentum space, since the expansion of
$\alpha_{\rm 1L}(q)$ in $\alpha_s(\mu)$ gives a
geometric series 
$\sum_n
\Bigl[\frac{\beta_0\alpha_s(\mu)}{2\pi}\log \bigl( \frac{\mu}{q} \bigr)
\Bigr]^n$.

The integral in eq.~(\ref{Vbeta0}) is ill defined,
since $\alpha_{\rm 1L}(q)$ has a pole at $q=\Lambda_{\rm QCD}$.
Nevertheless, if we expand the integrand in $\alpha_s(\mu)$,
the integral of each term is well defined.
Thus, we obtain a perturbative series as
\bea
\rule[-8mm]{0mm}{6mm}
V_{\beta_0}(r) &=& - C_F \, 4 \pi \alpha_s(\mu) \sum_{n=0}^\infty
\int
\frac{d^3 \vec{q}}{(2\pi)^3} \, \,
\frac{e^{i \vec{q} \cdot \vec{r}}}{q^2}
\, \, \biggl[ \, \frac{\beta_0 \alpha_s(\mu)}{4\pi} \, \log
\biggl( \frac{\mu^2}{q^2} \biggr) \biggr]^n 
\nonumber \\ 
&=&
- C_F \, 4 \pi \alpha_s(\mu) \sum_{n=0}^\infty \,
\biggl[ \frac{\beta_0 \alpha_s(\mu)}{4\pi} \biggr]^n \,
f_n(r,\mu) \,\, n! \, .
\label{DefVbeta0}
\eea
Although each $f_n(r;\mu)$ is well defined, the perturbative
expansion is an asymptotic series.
This leads to an ${\cal O}(\Lambda_{\rm QCD})$ uncertainty.

To clarify this statement, we define a generating function:\footnote{
To derive the last expression, we rewrite
\bea
\left(\frac{1}{q^2}\right)^{u+1}=\frac{1}{\Gamma(1+u)}
\int_0^\infty d\alpha \, \alpha^u\, e^{-\alpha q^2} ,
\eea
then integrals over $\vec{q}$ reduce to Gaussian integrals,
which can be evaluated easily.
The remaining integral over $\alpha$ can be expressed
in terms of the $\Gamma$ function.
}
\bea
\rule[-6mm]{0mm}{6mm}
F(r,\mu;u) & \equiv &
\int \frac{d^3 \vec{q}}{(2\pi)^3} \, \,
\frac{e^{i \vec{q} \cdot \vec{r}}}{q^2} \, 
\biggl( \frac{\mu^2}{q^2} \biggr)^{u}
= \sum_n f_n(r,\mu) \, u^n 
\label{expF}
\\ \rule[-6mm]{0mm}{6mm}
&=& \frac{1}{4\pi^{3/2}\, r} \, \left(
\frac{\mu r}{2}\right)^{\!2u}
\frac{\Gamma(\frac{1}{2}-u)}{\Gamma(1+u)}
.
\label{analF}
\eea
This is called the Borel transform of $V_{\beta_0}(r)$.
The same $f_n(r;\mu)$ as in eq.~(\ref{DefVbeta0})
appears, since
$(\mu^2/q^2)^u=e^{u\log(\mu^2/q^2)}=\sum_n\frac{u^n}{n!}
\log^n \bigl( \frac{\mu}{q} \bigr)$.
The convergence of the series is accelerated by $1/n!$ compared
to $V_{\beta_0}(r)$.
As a result, it has a finite radius of convergence
around $u=0$.

\begin{figure}[t]
  \hspace*{\fill}
    \includegraphics[width=5cm]{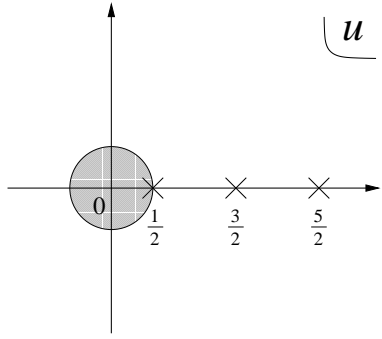}
  \hspace*{\fill}
  \\
  \hspace*{\fill}
\caption{\footnotesize
Analyticity of the generating function $F(r,\mu;u)$ 
shown in the complex $u$-plane.
Poles are located at $u=\frac{1}{2}, \frac{3}{2}, \frac{5}{2}, \cdots$.
Also the domain of convergence of the series expansion at $u=0$ is
shown.
      \label{borel}
}
  \hspace*{\fill}
\end{figure}

The behavior of $f_n(r,\mu)$ at $n\gg 1$ can be
determined from the analyticity of $F$ in the complex
$u$-plane (shown in Fig.~\ref{borel}).
In fact, the large-$n$ behavior of $f_n(r,\mu)$ determines the
domain of convergence of the series expansion (\ref{expF})
at $u=0$,
while the pole of (\ref{analF}) 
closest to $u=0$ determines the convergence radius.
The closest pole is located
at $u=1/2$, included in $\Gamma(\frac{1}{2}-u)$.
It follows that
\bea
f_n(r,\mu) ~\sim ~
\frac{\mu}{2\pi^2}  \times 2^n 
~~~\mbox{for}~~~
n \gg 1
.
\label{LeadingAsympt-fn}
\eea
This is extracted from the contribution of the pole
at $u=1/2$:\footnote{
If we subtract the pole at $u=1/2$ and
consider $\sum_n \Bigl[
f_n(r;\mu)-\frac{\mu}{2\pi^2}  \,2^n \Bigr]\,u^n
=F-(\mbox{pole at }u=1/2)$,
the convergence radius enlarges to $3/2$.
Namely, $|f_n-\frac{\mu}{2\pi^2}  \,2^n |$ at $n\gg 1$
decreases more rapidly than $|f_n|$.
This means that the leading behavior of $f_n$ at $n\gg 1$ is given by 
$\frac{\mu}{2\pi^2}  \,2^n $ [eq.~(\ref{LeadingAsympt-fn})].
} 
\bea
&&
F \sim 
\frac{1}{4\pi^{3/2}\, r} \, \left(
\frac{\mu r}{2}\right)^{\!1}
\frac{1}{\frac{1}{2}-u}\,\frac{1}{\sqrt{\pi}/2}
=\frac{\mu}{2\pi^2} \, \frac{1}{1-2u}
= \sum_{n=0}^\infty
\frac{\mu}{2\pi^2}  \times 2^n u^n
.
\label{LeadingContr-F}
\eea

Hence, asymptotically, the $n$-th term of $V_{\beta_0}(r)$ is
given by
\bea
&&
V^{(n)}_{\beta_0} \sim 
- C_F \, 4 \pi \alpha_s \times\frac{\mu}{2\pi^2} 
\biggl( \frac{\beta_0 \alpha_s}{2\pi} \biggr)^n \,
\times n! 
\nonumber \\&&
~~~~~
= {\rm const}.\times a^n \, n!
~~~~\mbox{with}~~~ a=\frac{\beta_0 \alpha_s}{2\pi}.
\label{LLest-LOrenormalon}
\eea
Note that it is independent of $r$.
The $r$-dependence is canceled when we evaluate the
residue of $F$ at $u=1/2$ in eq.~(\ref{LeadingContr-F}).
This means that, although each term
$V^{(n)}_{\beta_0}(r)$  of 
the potential is
a function of $r$, its dominant part for $n \gg 1$ is
only a constant potential;
see Fig.~\ref{FixedOrderPotentials}.
\pagebreak

The behavior of $-V^{(n)}_{\beta_0}$ for a small $\alpha_s$
is shown schematically
in the figure below.
\begin{wrapfigure}{l}{70mm}
\vspace*{-7mm}
\begin{center}
\includegraphics[width=6.5cm]{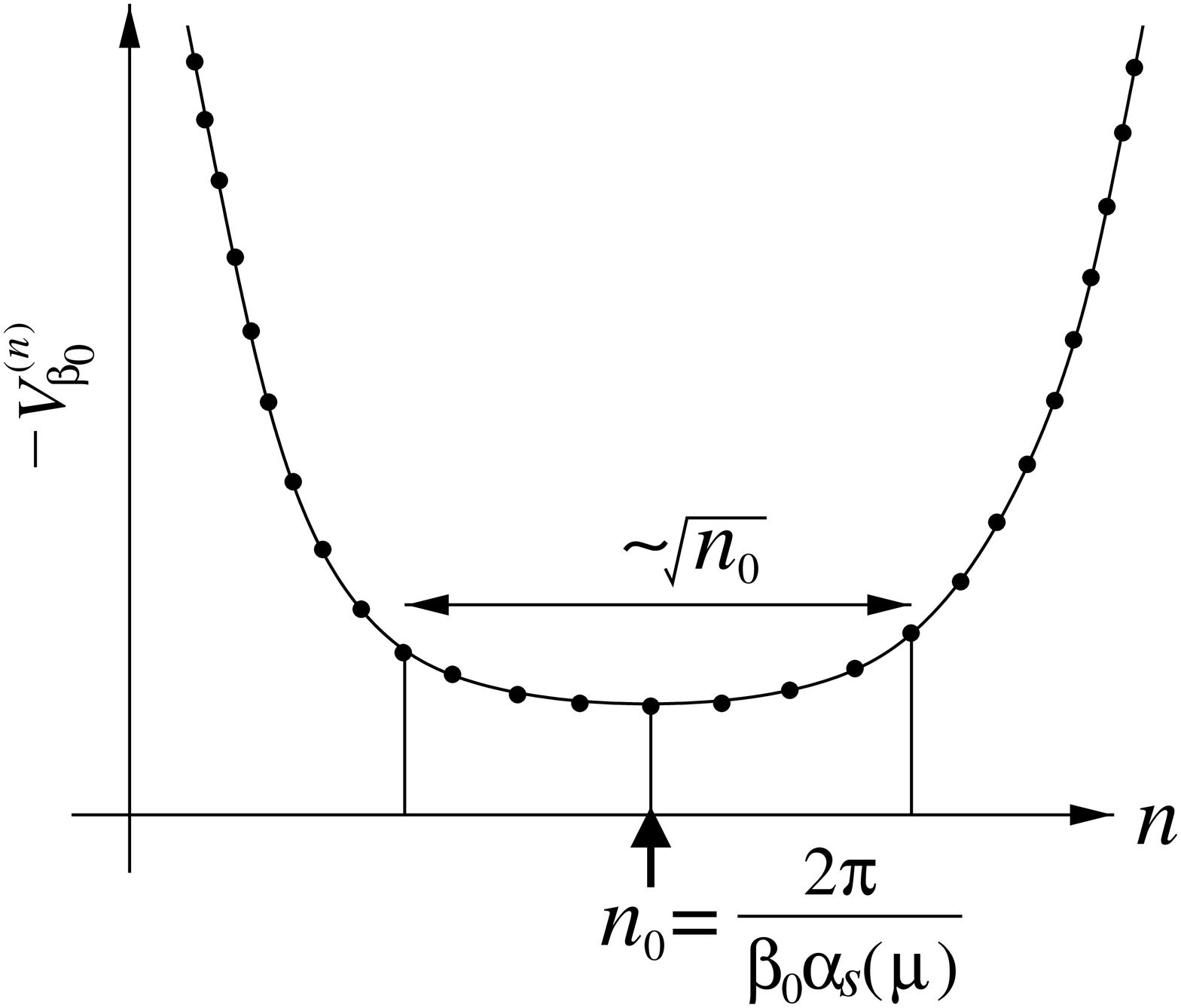}
\end{center}
\vspace*{-7mm}
\end{wrapfigure}
\noindent
As we raise $n$, first the term decreases due to powers of 
$\alpha_s$;
for large $n$ the term increases due to the factorial $n!$.
Around $n\approx n_0$,
the term becomes smallest.
The size of the term scarcely changes within the range 
$n \in ( n_0 - \sqrt{n_0}, n_0 + \sqrt{n_0} )$.

We may find $n_0$, which minimizes $-V^{(n)}_{\beta_0}$, as follows.
The terms are almost identical in the neighborhood of $n_0$, hence,
$a^{n_0-1}(n_0-1)! \approx a^{n_0} n_0!$, which means
\bea
n_0 \approx \frac{1}{a}=\frac{2\pi}{\beta_0\alpha_s(\mu)}
.
\eea
It follows that
\bea
a^{n_0} n_0!\, \approx \, a^{1/a}\times \sqrt{\frac{2\pi}{a}}\,
\biggl(\frac{1}{a}\biggr)^{\! \! 1/a}e^{-1/a}
=\sqrt{\frac{2\pi}{a}}\,e^{-1/a}
\eea
We may consider an uncertainty of this asymptotic series
as the sum of the terms within the range 
$ n_0 - \sqrt{n_0}<n< n_0 + \sqrt{n_0} $, since we
are not certain where to truncate the series within this range:
\bea
\delta V_{\beta_0}(r) \sim 
\sum_{ n = n_0 - \sqrt{n_0} }^{n_0 + \sqrt{n_0}} \, 
\left| V^{(n)}_{\beta_0} \right|
\sim \Lambda_{\rm QCD} .
\eea
The $\mu$-dependence vanishes in this sum, and this leads to the
claimed uncertainty.

The word ``renormalon'' denotes a pole (or more
generally a singularity)
in the complex $u$-plane (Borel plane), which dictates
the large-$n$ behavior with factorial growth
of a perturbative expansion.
The above estimate based on a renormalon reproduces well the
qualitative feature of the known first four terms of
the perturbative expansion of $V_{\rm QCD}(r)$ shown
in Fig.~\ref{FixedOrderPotentials} (which does not
use the LL approximation); see also Fig.~\ref{Comp-LLPot-ExactPot} 
in Sec.~\ref{Sec:PertPredicEtot} below.
In comparison, the level spacings in the measured
spectra of the bottomonium ($b\bar{b}$)
and charmonium ($c\bar{c}$) states are not larger
than order $\Lambda_{\rm QCD}$.
As a result, if the perturbative prediction for
$V_{\rm QCD}(r)$ is naively used to predict
these spectra, predictability turns out to be very poor.
This corresponds to the status before around 1998.

Before ending this section, we comment on the indication of
renormalons.
In the above argument, we start from an ill-defined integral and obtain
renormalons and asymptotic series.
One may wonder what is meant by a prediction by an asymptotic series
when the original integral is ill-defined.
We note that it is not the ill-defined nature of
the integral that directly leads to the renormalons.
In fact, one can construct examples, in which
(1) the integral is well-defined and renormalons exist,
and (2) the integral is ill-defined and there is no renormalon.
The former case is realized, for instance, when the two-loop coefficient 
$\beta_1$ of
the beta function is included
and has an opposite sign to the one-loop coefficient
$\beta_0$, so that the running coupling constant becomes large
in the region $q\simlt \Lambda_{\rm QCD}$ but is still
well defined down to $q=0$.\footnote{
In this case the renormalons are not poles but 
branch points in the Borel plane.
}
The latter example is realized, when we constrain the integral region
to $q>q_0$, where $0<q_0<\Lambda_{\rm QCD}$.
Considering these examples, it would be adequate to
interpret in the following way:
Renormalons signal
that an effective expansion parameter becomes large at
scale $q\simlt \Lambda_{\rm QCD}$ and validity of
the perturbative
expansion is breaking down in this part of the
integral region.
Later in this lecture, we will remedy this
problem by eliminating contributions from
the region $q\simlt \Lambda_{\rm QCD}$.
It is important to 
investigate and identify which part of the perturbative prediction
is free from
ambiguities in the IR region,
and in this context analysis of renormalons is useful.

\section{Cancellation of Renormalons in Total Energy}
\label{sec:CancellationRenormalon}
\clfn

We show that the uncertainty $\delta V_{\rm QCD}(r)\sim \Lambda_{\rm QCD}$
is canceled by the similar uncertainty included in the quark
pole mass, for a system of a color-singlet heavy quark-antiquark pair.
This leads to a higher predictive power of perturbative QCD
for the energy of this system.

\subsection[Quark pole mass and total energy of $Q\bar{Q}$ system]{\boldmath Quark pole mass and total energy of $Q\bar{Q}$ system}
\label{sec:m_pole+E_tot}

\begin{figure}[h]
\begin{center}
    \includegraphics[width=12cm]{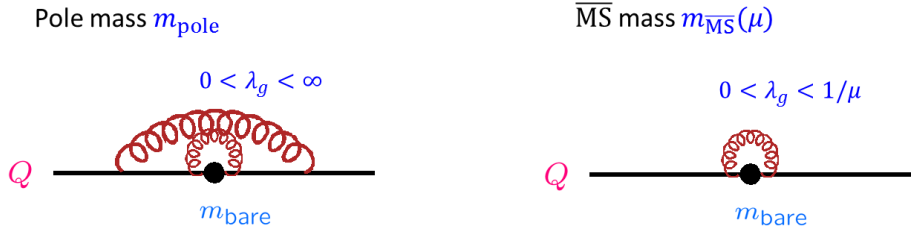}
\caption{\footnotesize
Since a quark has a color charge, gluons with arbitrarily
wavelength $\lambda_g$ can couple to the quark
and contribute to the quark self-energy.
$m_{\rm pole}$, being
the energy of a quark at rest, includes all the contributions from
$0<\lambda_g<\infty$.
On the other hand, $m_{\overline{\rm MS}}(\mu)$
is defined such that it includes only contributions from
short wavelength $0<\lambda_g\simlt \mu^{-1}$.
      \label{DepictMasses}
}
\end{center}
\end{figure}

Define the total energy of a heavy $Q\bar{Q}$ system
as
\bea
E_{\rm tot}(r) \equiv 2 m_{\rm pole} + V_{\rm QCD}(r) ,
\eea
where $m_{\rm pole}$ denotes the pole mass of the heavy quark.
This constitutes the major part of the energy of the 
$Q\bar{Q}$ system, in the limit where the masses of the
quarks are heavy.
If $m_{\rm pole}$ is expressed in terms of the $\overline{\rm MS}$
mass $m_{\overline{\rm MS}}(\mu)$, the 
renormalon at $u=1/2$ contained in
$V_{\rm QCD}(r)$ is canceled by the one contained in $m_{\rm pole}$.
Here, the two representative quark masses used
in perturbative QCD have the following meanings:
\begin{description}
\item[Pole mass:] 
the energy of a quark at rest, which
is equivalent to the pole position of the quark propagator.
All the contributions to the quark self-energy are included.
\item[$\overline{\text{MS}}$ mass:] 
a parameter in the Lagrangian
$\sim -m_q \bar{\psi}_q\psi_q$,
renormalized in the $\overline{\rm MS}$ scheme (i.e., only
UV divergence is subtracted).
Only UV part of the quark self-energy is included.
\end{description}
These are depicted schematically in Fig.~\ref{DepictMasses}.

The notion of the quark pole mass contradicts the quark
confinement picture.
In fact, if the energy of a single quark, which has a color charge,
is computed non-perturbatively, it is expected to be infinite.
It is almost equivalent to the statement that, to separate
a quark and an antiquark, one needs an energy proportional to
the distance between the two particles (provided that the string
breaking does not occur), so that to separate them infinitely apart,
one needs an infinite energy.

In perturbative QCD, the pole mass can be computed to be finite,
order by order in perturbative expansion.
Nevertheless, it is expected to contain at least ${\cal O}(\Lambda_{\rm QCD})$
uncertainty.
To illustrate this, let us compare measurements of the
masses of the $W$ boson and top quark, respectively.
\begin{figure}[h]
\begin{center}
\begin{tabular}{ccc}
\includegraphics[width=4cm]{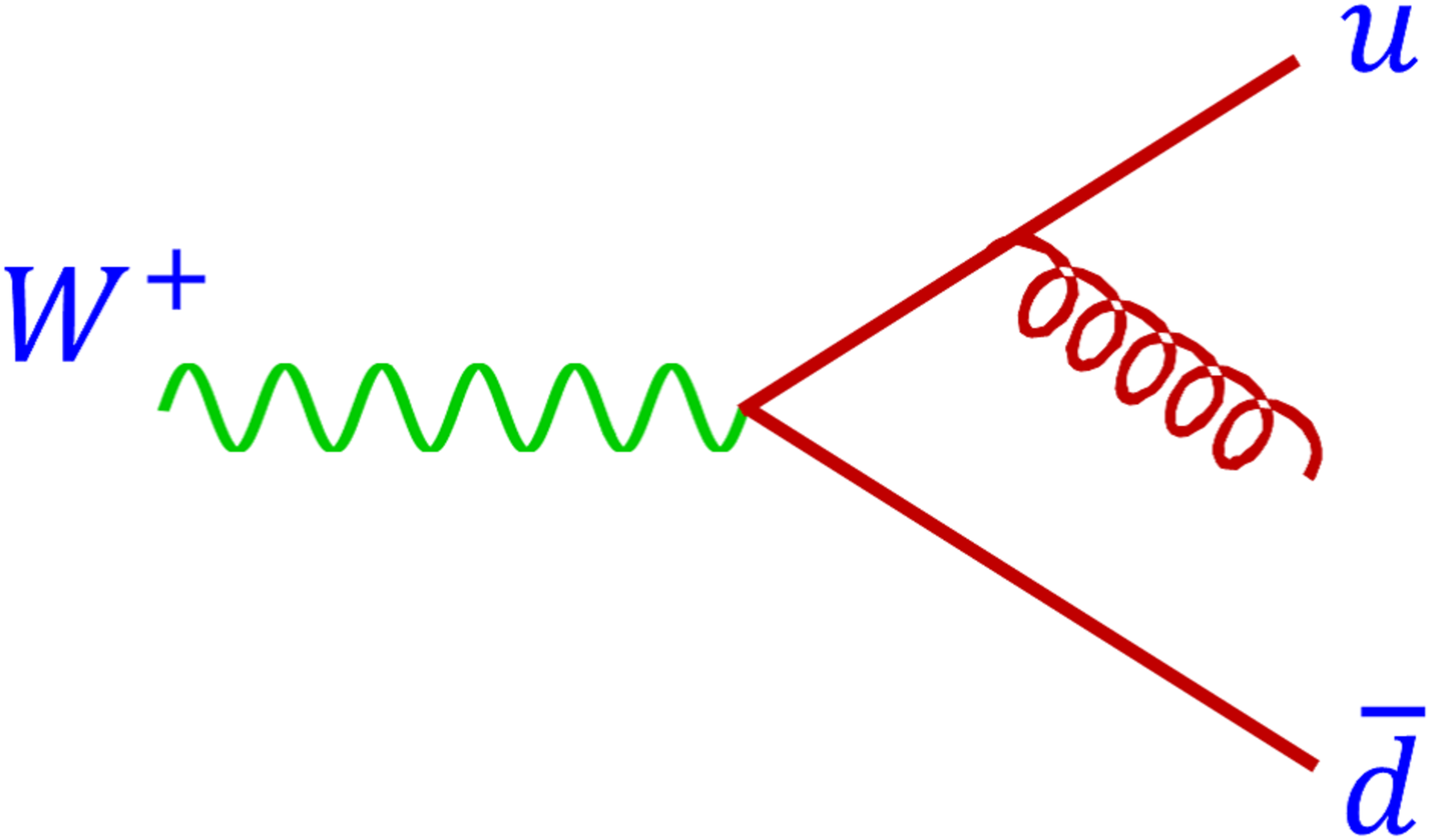}
&
\includegraphics[width=4cm]{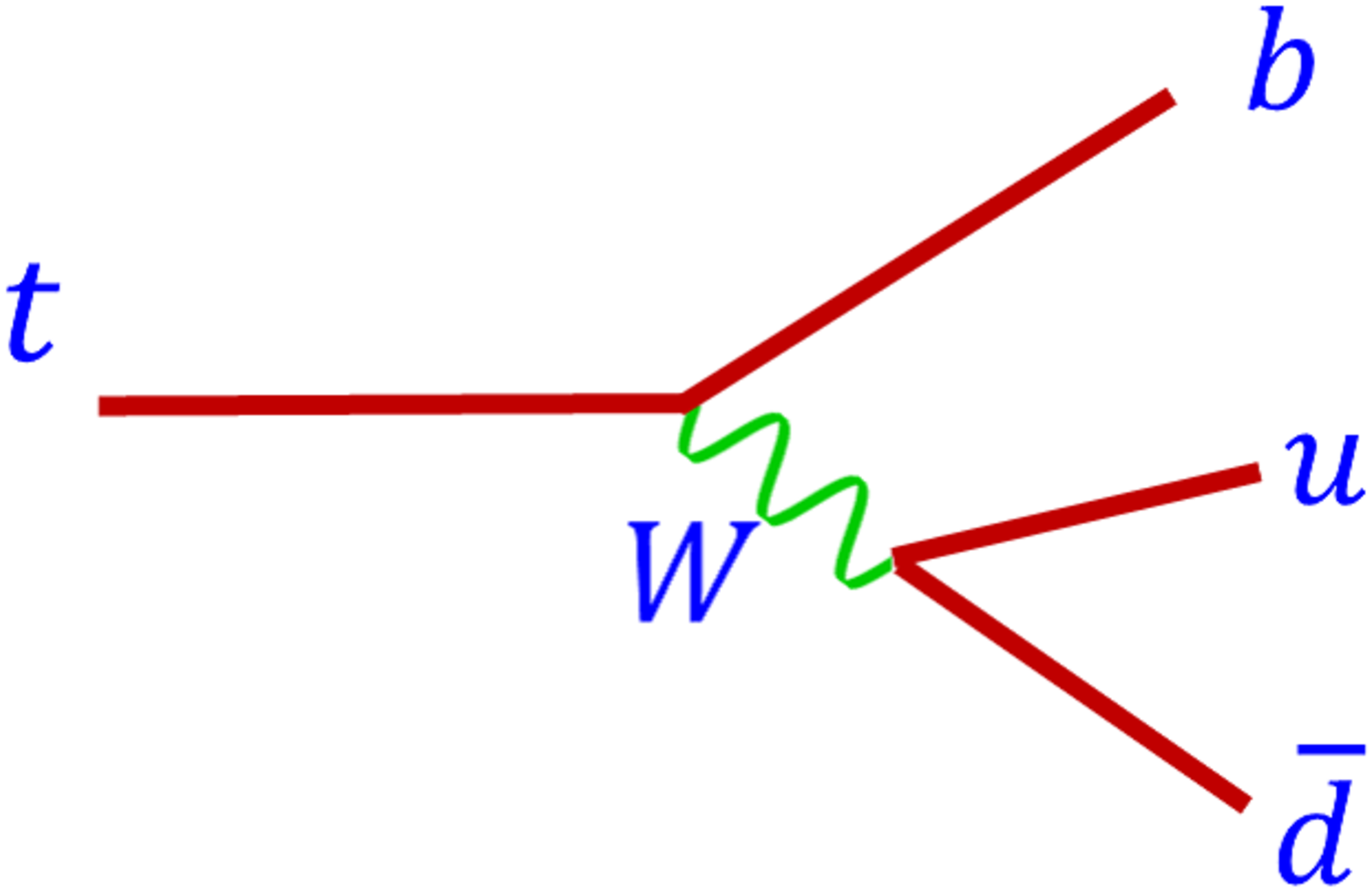}
~
&
\includegraphics[width=5cm]{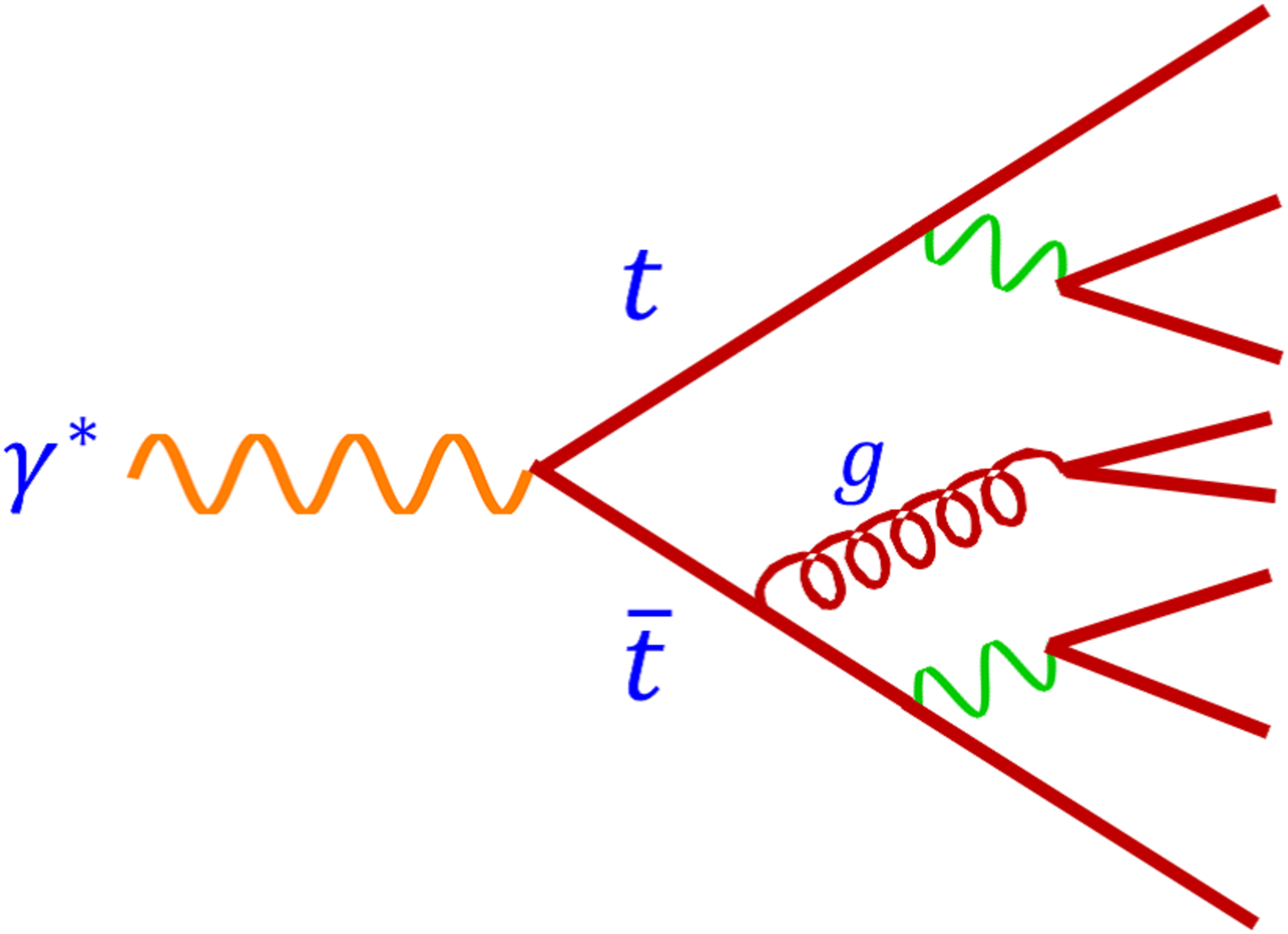}
\\
(a)&(b)&(c)
\end{tabular}
\end{center}
\vspace*{-.5cm}
\caption{\footnotesize
Relevant processes for measurements of the masses of the $W$ boson
and top quark from their hadronic decays.
In (a), (b) and (c), the initial states are color-singlet,
color-triplet and color-singlet, respectively.
Unlike $W$ boson, the momentum of top quark cannot be reconstructed
unambiguously from the final state hadrons in experiments.
\label{tWdecays}}
\end{figure}
As shown in Fig.~\ref{tWdecays}(a), the $W$ boson mass
can be measured from the invariant mass of all the hadrons
in its hadronic decay.
The invariant mass coincides with the $W$ mass, event by
event (if we ignore the decay width of $W$ and measurement
errors).
On the other hand,
if we measure the invariant
mass of all the hadrons, which come from the top quark
in its hadronic decay, it does not coincide with the top quark
mass.
The top quark has a color, while hadrons in the final state
are color singlet.
Hence, top quark momentum and the sum of the hadron momenta
do not coincide.
See Fig.~\ref{tWdecays}(b).
If we consider pair creation of $t\bar{t}$ from a color
singlet state, the total momenta of all the final state
hadrons coincide with the sum of the $t$ and $\bar{t}$
momenta.
However, it is non-trivial to separate the hadron momenta
into $t$ and $\bar{t}$ sides.
There should be uncertainty of at least order $m_\pi$
in this assignment, event by event.
See Fig.~\ref{tWdecays}(c).
Thus, the notion of the quark pole
mass would be ambiguous beyond ${\cal O}(\Lambda_{\rm QCD})$
accuracy,
even if we consider a 
realistic situation to extract a quark pole mass from a
physical process by comparison to perturbative QCD prediction,
due to the presence of string breaking and
color neutralization. 

The expression of the quark pole mass in terms of the $\overline{\rm MS}$
mass is given in perturbative series as
\bea
&&
m_{\rm pole}=m_{\overline{\rm MS}}(\mu)
\biggl[ 1 + \alpha_s(\mu) \, \biggl\{d_{11}\,\log\Bigl(\frac{\mu}{m_{\overline{\rm MS}}}
\Bigr)+d_{10}\biggr\} 
\nonumber\\&&
~~~~~~~~~~~~~~~~~~~~~~~
+ \alpha_s(\mu)^2
\biggl\{d_{22}\,\log^2\Bigl(\frac{\mu}{m_{\overline{\rm MS}}}
\Bigr)+d_{21}\,\log\Bigl(\frac{\mu}{m_{\overline{\rm MS}}}
\Bigr)+d_{20} \biggr\}+\cdots 
\biggr] .
\eea
We present the result for an estimate of the higher-order
terms of $m_{\rm pole}$.
In the LL approximation ($\beta_0>0$, $\beta_1=\beta_2=\cdots=0$),
similarly to the analysis of $V_{\rm QCD}(r)$,
the pole mass is given by
\bea
&&
m_{{\rm pole},\beta_0}=m_{\overline{\rm MS}}(\mu)\, (1+\delta_m)
,
\\ &&
\delta_m=\frac{C_F\alpha_s(\mu)}{2\pi}\sum_{n=0}^\infty
\biggl[\frac{\beta_0\alpha_s(\mu)}{4\pi}\biggr]^n
\,
\biggl[ n! \, G_{n+1} + \frac{(-1)^n}{n+1}\, g_{n+1} \biggr] 
.
\label{deltam}
\eea
The expansion coefficients $G_n$ and $g_n$ are given in terms of
generating functions as
\bea
&&
\sum_{n=0}^\infty G_n u^n = \biggl( \frac{\mu^2}{m_{\overline{\rm MS}}^2}
\biggr)^u \times
3(1-u) \, \frac{\Gamma(1+u)\Gamma(1-2u)}{\Gamma(3-u)},
\\ &&
\sum_{n=0}^\infty g_n y^n = \frac{3-2y}{6}\,
\frac{\Gamma(4-2y)}{\Gamma(1+y)\Gamma(2-y)^2\Gamma(3-y)} .
\eea
The series with $n!\,G_{n+1}$ in eq.~(\ref{deltam}) includes 
renormalons.
The contribution of the pole at $u=1/2$ in $m_{\overline{\rm MS}}
\times \sum_n G_nu^n$ is proportional to
$a^n n!$ with $a=\frac{\beta_0\alpha_s}{2\pi}$.
The proportionality coefficient is 
independent of 
$m_{\overline{\rm MS}}$, by evaluating 
$m_{\overline{\rm MS}}\!\times\!
\Bigl(\frac{\mu^2}{m_{\overline{\rm MS}}^2}
\Bigr)^{\! u}$ at $u=1/2$.

From the above result, we can estimate the uncertainty of the
pole mass to be $\delta m_{\rm pole}\sim
\Lambda_{\rm QCD} \sim 200$--300~MeV.
In particular
even if we take $m_{\overline{\rm MS}}$ to zero,
$\delta m_{\rm pole}$ remains to be order
$\Lambda_{\rm QCD} $, since $\delta m_{\rm pole}$ is independent of
$m_{\overline{\rm MS}}$.
Then, 
one may wonder if it is a constituent quark mass of Sec.~\ref{sec:setup}.
We remark that this is not the case, in that
$\delta m_{\rm pole}$ is an uncertainty and not a prediction.

Evaluating the residue at $u=1/2$, one verifies that
the uncertainties in $V_{\rm QCD}(r)$ and $m_{\rm pole}$
cancel in the perturbative series of $E_{\rm tot}(r)$.
The IR renormalon pole of $E_{\rm tot}(r)$ closest to the origin
is at $u=3/2$.\footnote{
For simplicity of the argument, let us ignore the contribution of the pole
at $u=-1$ included in $m_{\rm pole}$, which is  a
UV renormalon and
gives a much less harmful sign-alternating series.
}
Namely, the convergence radius is enlarged.
Consequently, convergence of the perturbative series of
$E_{\rm tot}(r)$ improves drastically compared to
those of $V_{\rm QCD}(r)$ and $m_{\rm pole}$
individually.
For this cancellation to happen, it is mandatory that 
(1) the uncertainty of $V_{\rm QCD}(r)$ is independent of $r$, and
that (2) the uncertainty of $m_{\rm pole}$ is independent of 
$m_{\overline{\rm MS}}$.
Both conditions are satisfied.

Let us examine the higher-order behavior of the
perturbative series of
$E_{\rm tot}(r)$.
Figure below shows schematically the
renormalon estimates of higher-order terms of
$2m_{\rm pole}$ or $V_{\rm QCD}(r)$ and $E_{\rm tot}(r)$.
The series converges more quickly and up to larger $n$
for $E_{\rm tot}(r)$.
\begin{center}
    \includegraphics[width=9cm]{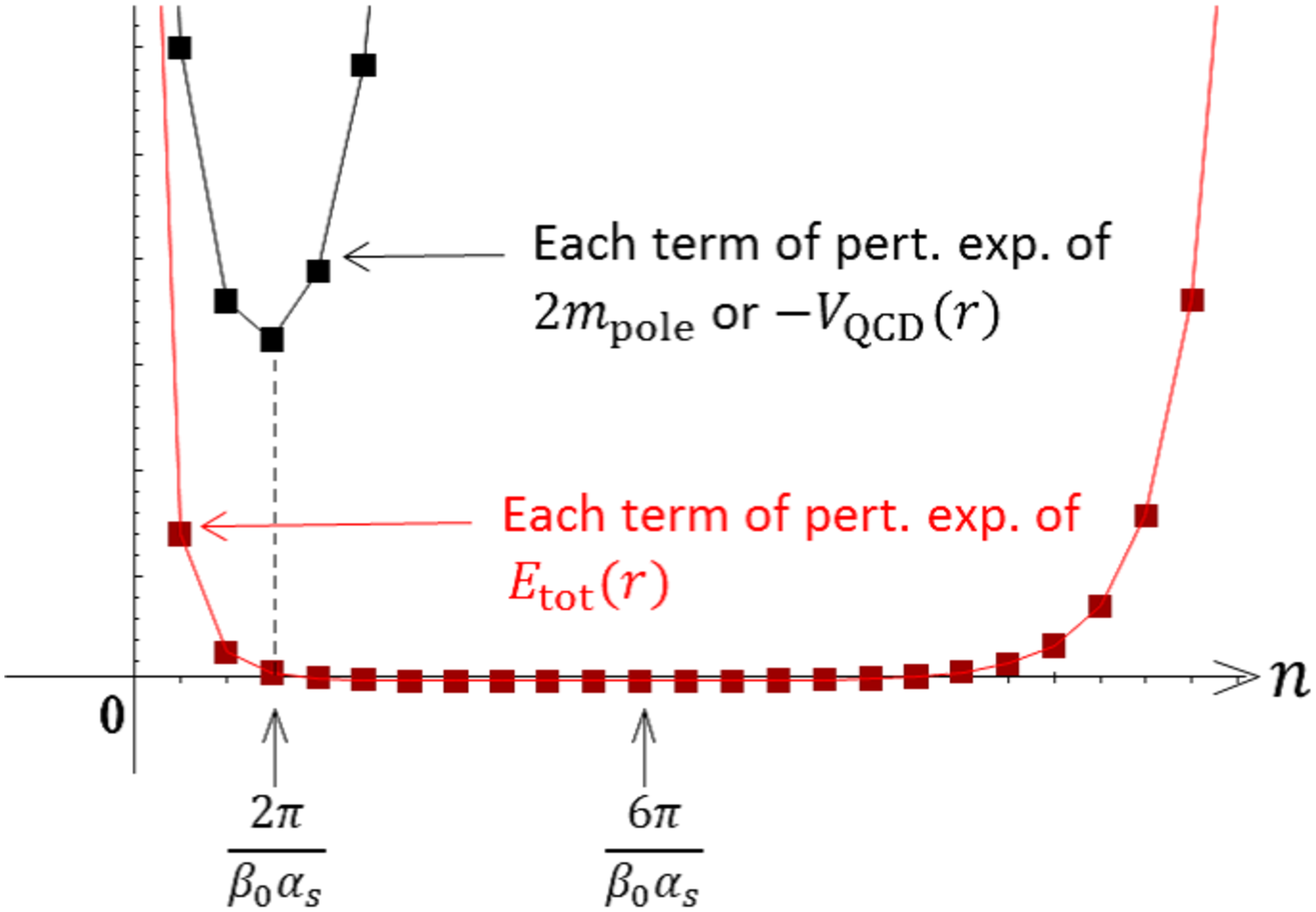}
\end{center}

\subsection{\boldmath Mechanism of cancellation of IR contributions}
\label{sec:IRcancellation}

We may understand qualitatively the mechanism of cancellation of 
the renormalons 
as follows.
The potential and the pole mass in the
LL approximation can be written as
\hspace*{35mm}
\begin{wrapfigure}{l}{50mm}
\vspace*{-7mm}
\begin{center}
\includegraphics[width=6cm]{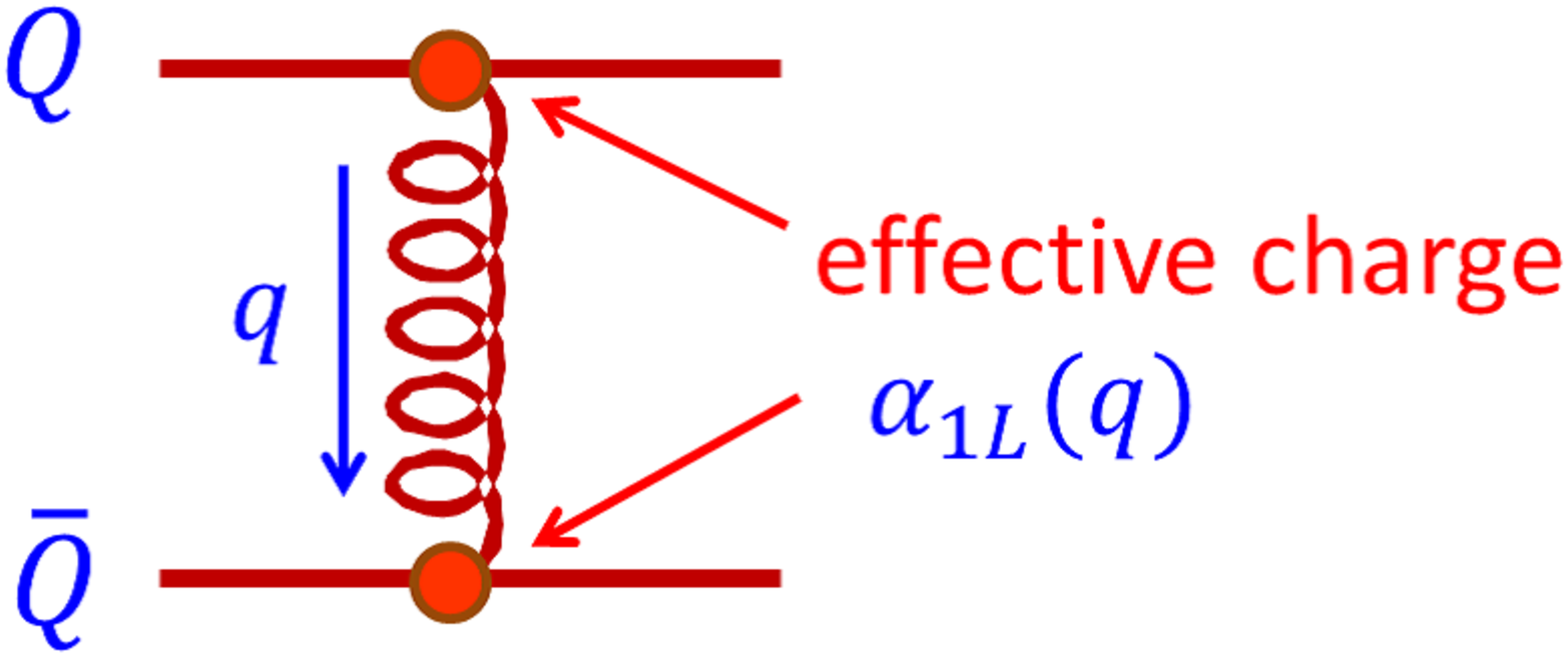}
\vspace*{-2mm}
\\
\hspace*{-6mm}
\includegraphics[width=4.5cm]{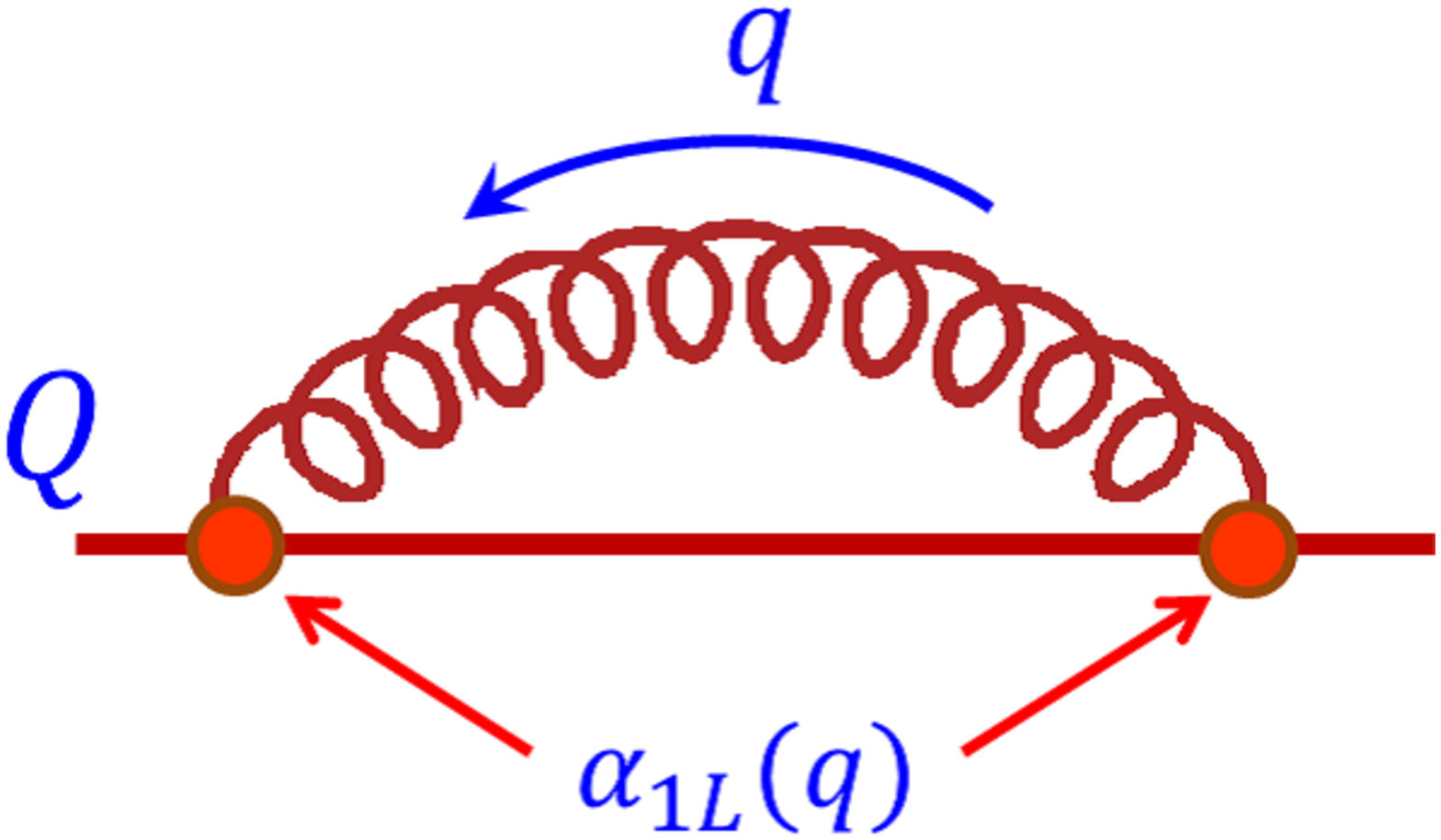}
\end{center}
\vspace*{-24mm}
\end{wrapfigure}
\begin{eqnarray}
&&
V_{\beta_0}(r) = - \int 
\frac{d^3\vec{q}}{(2\pi)^3} 
\, e^{i \vec{q} \cdot \vec{r}} \, 
C_F \frac{4\pi\alpha_{\rm 1L}(q)}{q^2} ,
\label{ApproxV}
\\
&&
m_{{\rm pole},\beta_0} \simeq m_{\overline {\rm MS}}(\mu ) +
\frac{1}{2} {\hbox to 18pt{
\hbox to -5pt{$\displaystyle \int$} 
\raise-15pt\hbox{$\scriptstyle {q}< \mu$} 
}}
\frac{d^3\vec{q}}{(2\pi)^3} \, 
C_F \frac{4\pi\alpha_{\rm 1L}(q)}{q^2} .
\label{Approx2M}
\end{eqnarray}
$V_{\beta_0}(r)$ is essentially Fourier transform of the
gluon propagator exchanged between quark and antiquark;
the difference of $m_{{\rm pole},\beta_0}$ and $m_{\overline {\rm MS}}$ is 
essentially the infrared part of the quark 
self-energy.
In both integrals, the charges are replaced by the one-loop
running coupling constant $\alpha_{\rm 1L}(q)$.
These are depicted schematically in the left figures.

As stated, 
the renormalon uncertainties stem from contributions from
the region $q\simlt \Lambda_{\rm QCD}$
in the above integrals, where $\alpha_{\rm 1L}(q)$ is large.
The signs of the renormalon contributions are opposite
between $V_{\beta_0}(r)$ and $m_{{\rm pole},\beta_0}$,
since the color charges are
opposite between quark and antiquark while the self-enregy is
proportional to the square of the same charge.
Their magnitudes differ by a factor of two because
both the quark and antiquark propagator poles contribute in the
calculation of the potential, whereas only one of the two contributes
in the calculation of the self-energy.
Since we are concerned with a small ${q}$ region,
we may expand the Fourier factor in $V_{\beta_0}(r)$
in  $\vec{q}$, 
\bea
e^{i \vec{q} \cdot \vec{r}} = 1 + {i \vec{q} \cdot \vec{r}}
+ \frac{1}{2}({i \vec{q} \cdot \vec{r}})^2 + \cdots .
\eea
The leading term (=1) 
is canceled against $2m_{{\rm pole},\beta_0}$, which
corresponds to the renormalon at $u=1/2$.
In fact, integral of this term 
over the region $q\simlt \Lambda_{\rm QCD}$
($\sim \int d^3\vec{q}~ 1/q^2$) 
evaluates to order $\Lambda_{\rm QCD}$.
The next term ($= i \vec{q} \cdot \vec{r}$)
vanishes upon integration over $\vec{q}$, due to rotational invariance.
The third term [$=\frac{1}{2}({i \vec{q} \cdot \vec{r}})^2$]
corresponds to the renormalon pole at
$u=3/2$ and is evaluated as
\bea
\delta V_{\beta_0}(r)\Bigr|_\text{contr.\ of $u=3/2$}
\sim 
{\hbox to 18pt{
\hbox to -1pt{$\displaystyle \int$} 
\raise-15pt\hbox{$\scriptstyle q\lesssim \Lambda_{\rm QCD}$} 
}}
d^3\vec{q}~\, (\vec{q} \cdot \vec{r})^2\,
\frac{1}{q^2}
~\sim ~
\Lambda_{\rm QCD} \,(\Lambda_{\rm QCD}\cdot r)^2 .
\eea
We obtain the same estimate if we perform an analysis
similar to Sec.~\ref{sec:somedetails}.
This uncertainty from the $u=3/2$ pole remains uncanceled.

Since the quark and antiquark are heavy, their typical
distance $r$ is small compared to the hadronization scale
$\Lambda_{\rm QCD}^{-1}$.
Thus, the residual uncertainty 
$\Lambda_{\rm QCD} \,(\Lambda_{\rm QCD}\cdot r)^2 $
is small compared to the original uncertainty 
$\Lambda_{\rm QCD}$, since $r\Lambda_{\rm QCD} \ll 1$.

Generally in perturbative QCD, 
convergence of perturbative series
become worse at small energy scale.  
(e.g., the prediction of $V_{\rm QCD}(r)$ at large $r$.)
This is due to increase of contributions from IR gluons.
Oppositely, if contributions from IR gluons are eliminated,
perturbative series show better convergence behaviors.

The cancellation of IR contributions in $E_{\rm tot}(r)$
is a general property of gauge theory, which holds
beyond the LL approximation.
This can be seen as follows.
A static current 
\begin{wrapfigure}{l}{70mm}
\vspace*{-3mm}
\includegraphics[width=7cm]{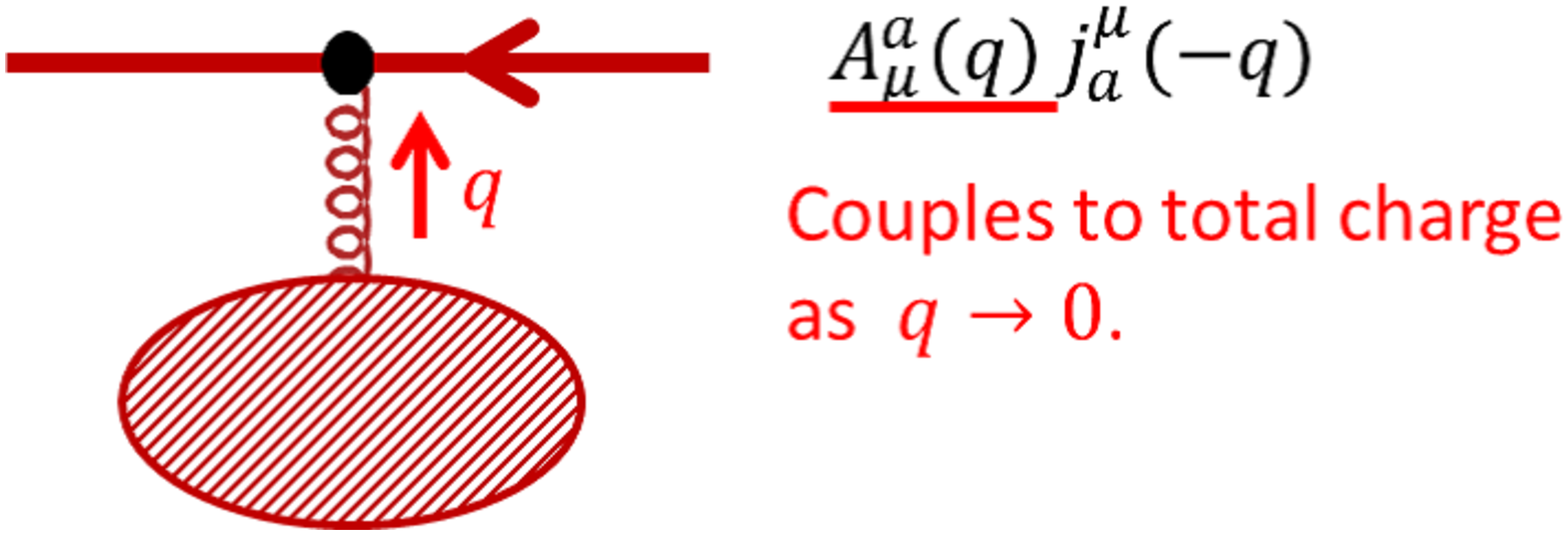}
\vspace*{-3mm}
\end{wrapfigure}
has only the time 
component,
\bea
j^\mu_a(x)=T_a\delta^{\mu 0}\delta^3(\vec{x}-\vec{r}/2) ,
\eea
since a static color charge has no spatial motion.
Hence,  an IR gluon, which couples to static 
currents 
via minimal coupling $A_\mu^a(q)\, j^\mu_a(-q)=A_0^a(q)\, j^0_a(-q)$,
 couples to
the total charge of the system in the IR limit $q\to 0$:
\bea
Q_a^{\rm tot}=\sum_{i=Q,\bar{Q}} j^0_{a,i}(q=0) .
\eea
Therefore, an 
IR gluon decouples
from a static color-singlet system.
\pagebreak
Diagrammatically, however, an IR gluon can
detect the total
charge of the system only when both self-energy 

\begin{wrapfigure}{l}{90mm}
\vspace*{-3mm}
\includegraphics[width=9cm]{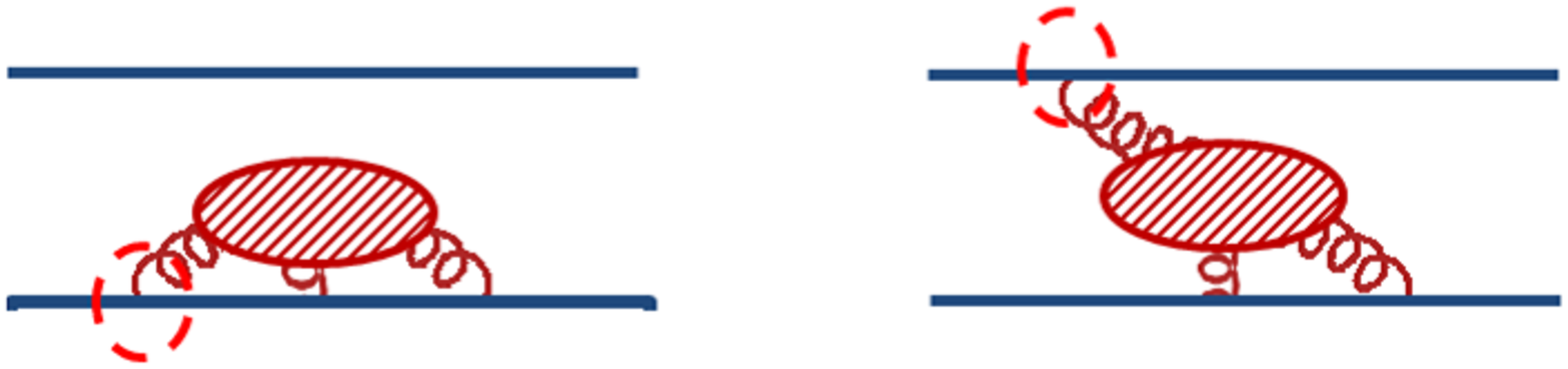}
\vspace*{-13mm}
\end{wrapfigure}
\noindent
diagrams\footnote{
In the large mass limit contributions from IR region
to the pole mass approximate 
IR contributions to the self-energy 
of a static charge.
} 
and 
potential-energy diagrams
are taken into account, as can be
seen from the left figures.
This means that 
a cancellation takes place between these two types of diagrams,
since the gluon couples to individual diagrams
but decouples from the sum of them.

Intuitively,
IR gluons with wavelengths of order $\Lambda_{\rm QCD}^{-1}(\gg r)$
cannot resolve the color charge of each particle, hence
they only see the total charge of the system.
More precisely,
coupling of IR gluons to the system can be expressed by
an expansion in $\vec{r}$ (multipole expansion)
for small $r$,
in which the zeroth multipole (=total charge) of
the color-singlet quark-antiquark pair is zero.

The modern approach (after around 1998)
to use the $\overline{\rm MS}$ mass
for the computation of $E_{\rm tot}(r)$
can be viewed as follows.
The total energy of the system is computed as the sum of 
(i) the
$\overline{\rm MS}$ masses of $Q$ and $\bar{Q}$, (ii) 
contributions to the self-energies
of $Q$ and $\bar{Q}$ which are not included in the $\overline{\rm MS}$ masses,
and (iii) the potential energy between $Q$ and $\bar{Q}$.
Contributions of IR gluons with wavelengths larger than $r$
automatically cancel between (ii) and (iii) in this computation.
In this way we can eliminate IR contributions from the
computation of $E_{\rm tot}(r)$.

\subsection[Perturbative prediction for $E_{\rm tot}(r)$]{\boldmath Perturbative prediction for $E_{\rm tot}(r)$}
\label{Sec:PertPredicEtot}

Let us demonstrate the improvement of accuracy of the perturbative
prediction for the total energy
$E_{\rm tot}(r) = 2 m_{\rm pole} + V_{\rm QCD}(r)$ 
up to ${\cal O}(\alpha_s^3)$ (without using LL approximation).
[Presently the perturbative series of $V_{\rm QCD}(r)$ is
known up to ${\cal O}(\alpha_s^4)$ and
$m_{\rm pole}$ 
up to ${\cal O}(\alpha_s^3)$.]

\begin{figure}[h]\centering
\begin{tabular}{ccc}
\includegraphics[width=7cm]{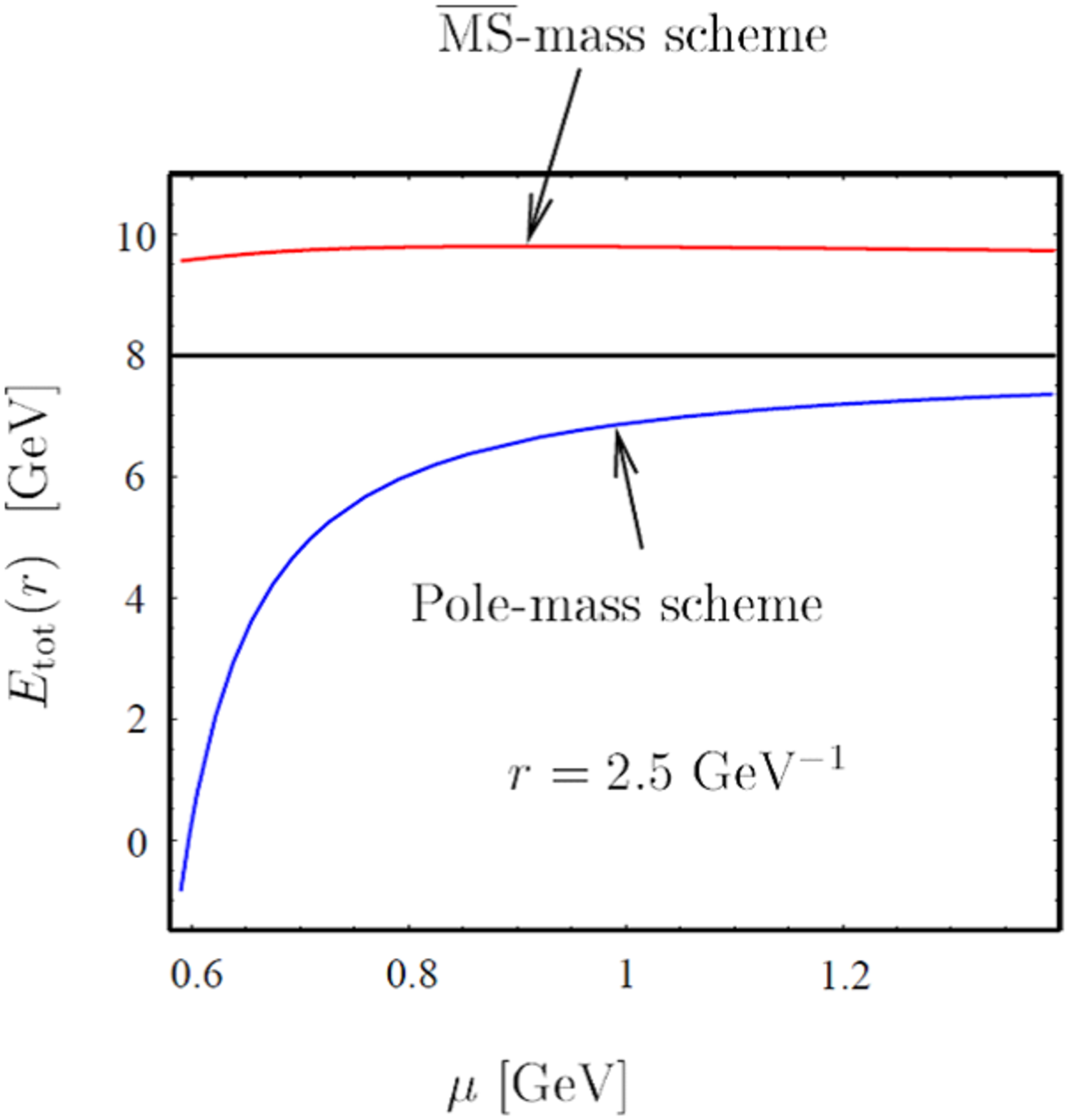}
&& \vspace*{-55mm}
\\
&~~~~&
\includegraphics[width=6.5cm]{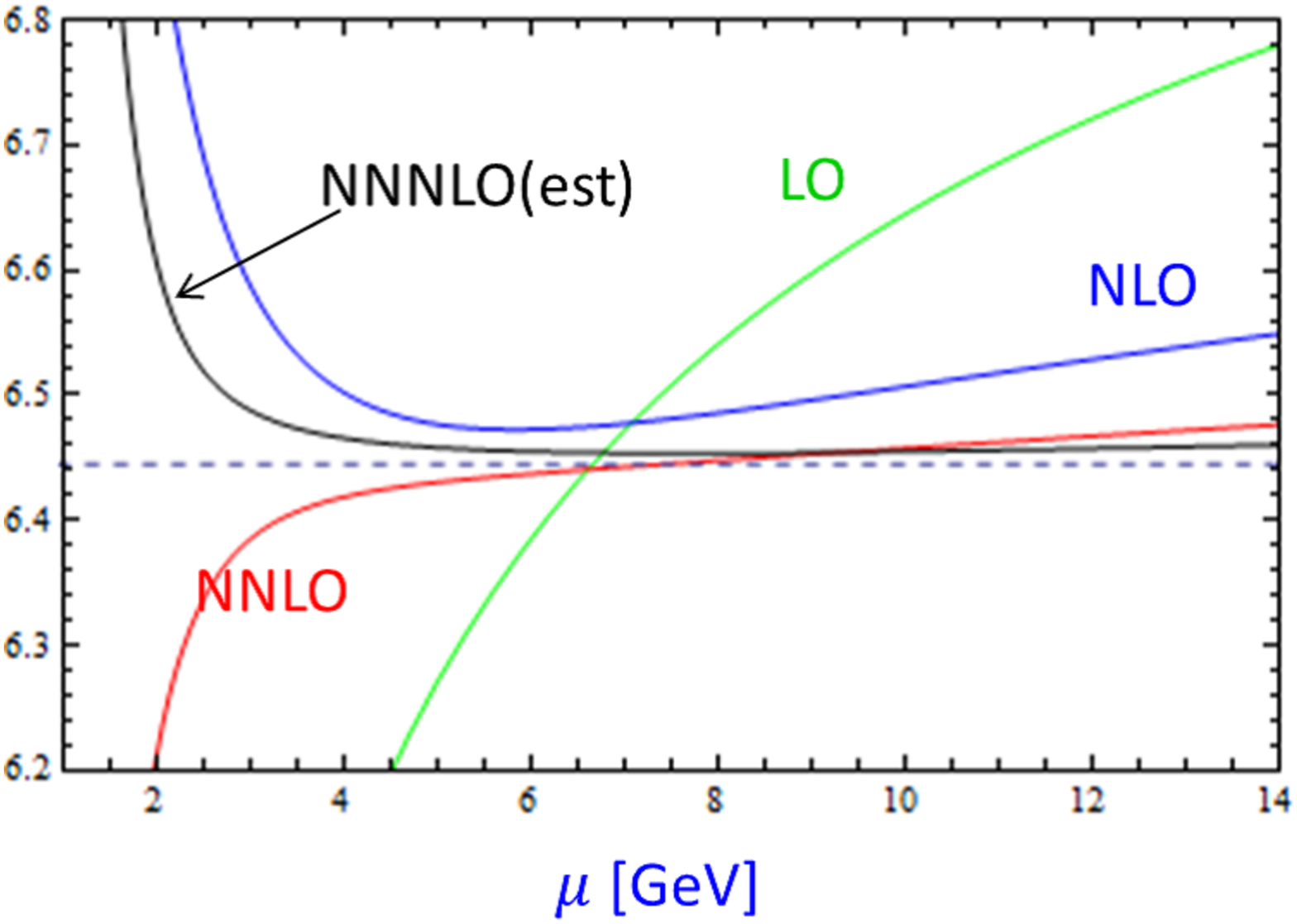}
\vspace*{4mm}
\\
~~~(a)&&(b)\\
\end{tabular}
\caption{\footnotesize 
(a)
Scale dependences of $E_{\rm tot}(r)$ 
up to ${\cal O}(\alpha_s^3)$ at 
$r =2.5~{\rm GeV}^{-1} \approx 0.5$~fm, in the pole-mass and 
$\overline{\rm MS}$-mass schemes.
(b)Scale dependences of $E_{\rm tot}(r)$ at $r=0.1$~GeV$^{-1}$ in the
$\overline{\rm MS}$-mass scheme 
at LO, NLO, NNLO and NNNLO.
The NNNLO prediction uses the estimate by the large-$\beta_0$
approximation for the ${\cal O}(\alpha_s^4)$ term of the
relation between the pole and $\overline{\rm MS}$ masses.
In both figures, horizontal lines are shown for guides.
\label{scale-dep}}
\end{figure}

As an example, we take the bottomonium case:
We choose the $\overline{\rm MS}$
mass of the $b$-quark, renormalized at the  $b$-quark $\overline{\rm MS}$
mass, as
$\overline{m}_b \equiv 
m_b^{\overline{\rm MS}}(m_b^{\overline{\rm MS}})
= 4.190$~GeV;
in internal loops, four flavors of light quarks are included
with $\overline{m}_u=\overline{m}_d=\overline{m}_s=0$
and $\overline{m}_c=1.243$~GeV.
In Fig.~\ref{scale-dep}(a), we fix $r=2.5~{\rm GeV}^{-1} \approx 0.5$~fm
and examine the $\mu$-dependence of $E_{{\rm tot}}(r)$,
before rewriting $m_{b,{\rm pole}}$
{\bf [}Pole-mass scheme{\bf ]}, and after
rewriting $m_{b,{\rm pole}}$ by $\overline{m}_b$
{\bf [}$\overline{\rm MS}$-mass scheme{\bf ]}.
We see that $E_{{\rm tot}}(r)$ is much less scale dependent when we
use the $\overline{\rm MS}$ mass (after cancellation
of renormalons) than when we use the pole mass 
(before cancellation of renormalons).
This shows clearly that the perturbative prediction of $E_{{\rm tot}}(r)$
is much more
stable if we
use the $\overline{\rm MS}$ mass.

We also compare the convergence behaviors
of the perturbative series of $E_{\rm tot}(r)$ for the same $r$
and when $\mu$ is fixed to the scale where $\mu$ dependence
vanishes
(the minimal-sensitivity scale).
At $r =2.5~{\rm GeV}^{-1}$, this scale is
$\mu = 0.90$~GeV for the $\overline{\rm MS}$-mass scheme.
Convergence of the series turns out to be close to optimal
for this scale choice:\footnote{
In the pole-mass scheme, there exists no minimal-sensitivity scale within a
wide range of $\mu$, and the convergence behavior of the series is
qualitatively similar to eq.~(\ref{conv-polemass-scheme}) within this range.
}
\bea
E_{{\rm tot}}^{b\bar{b}}(r) &=& 
10.408 - 0.275 - 0.362 - 0.784 ~~{\rm GeV}
~~~~~
\mbox{(Pole-mass scheme)}
\label{conv-polemass-scheme}
\\
&=&
~\,8.380 + 1.560 -0.116 - 0.022~~{\rm GeV}
~~~~~
\mbox{($\overline{\rm MS}$-mass scheme)} .
\eea
The four numbers represent the ${\cal O}(\alpha_s^0)$,
${\cal O}(\alpha_s^1)$, ${\cal O}(\alpha_s^2)$ and 
${\cal O}(\alpha_s^3)$ terms of the series expansion in each
scheme.
The ${\cal O}(\alpha_s^0)$ terms represent the twice of
the pole mass and of the $\overline{\rm MS}$ mass, respectively.
The ${\cal O}(\alpha_s^1)$, ${\cal O}(\alpha_s^2)$, 
${\cal O}(\alpha_s^3)$ terms in eq.~(\ref{conv-polemass-scheme}) 
come solely from $V_{\rm QCD}(r)$.

As can be seen, if we use the pole mass, the series is not
converging beyond ${\cal O}(\alpha_s^1)$, whereas
if we use the $\overline{\rm MS}$ mass, the series is converging.
One may further verify that, 
when the series is converging ($\overline{\rm MS}$-mass scheme),
$\mu$-dependence of $E_{\rm tot}(r)$ decreases
as we include more terms of the perturbative series,
whereas when the series is diverging
(pole-mass scheme), $\mu$-dependence does not decrease with
increasing order.

We observe qualitatively the same features at
different $r$ and for different number of light quark
flavors $n_f$, or if we change the values 
of the masses $\overline{m}_b$, $\overline{m}_c$.
Generally, at smaller $r$,
$E_{\rm tot}(r)$ becomes less
$\mu$-dependent and more convergent, 
due to the asymptotic freedom of QCD.
See Fig.~\ref{scale-dep}(b).

The stability against scale
variation and convergence of the 
perturbative series are closely connected with each other.
Formally, scale dependence vanishes 
at all order of perturbation series.
This means that,
for a truncated perturbative series up to ${\cal O}(\alpha_s^N)$,
scale dependence is of ${\cal O}(\alpha_s^{N+1})$.
Hence, the scale dependence decreases for larger $N$
as long as the series is converging.
Thus, the truncated perturbative series is expected to become 
less $\mu$-dependent with increasing order
when the series is converging, and {\it vice versa}.

\begin{figure}[h]
\begin{center}
\includegraphics[width=14cm]{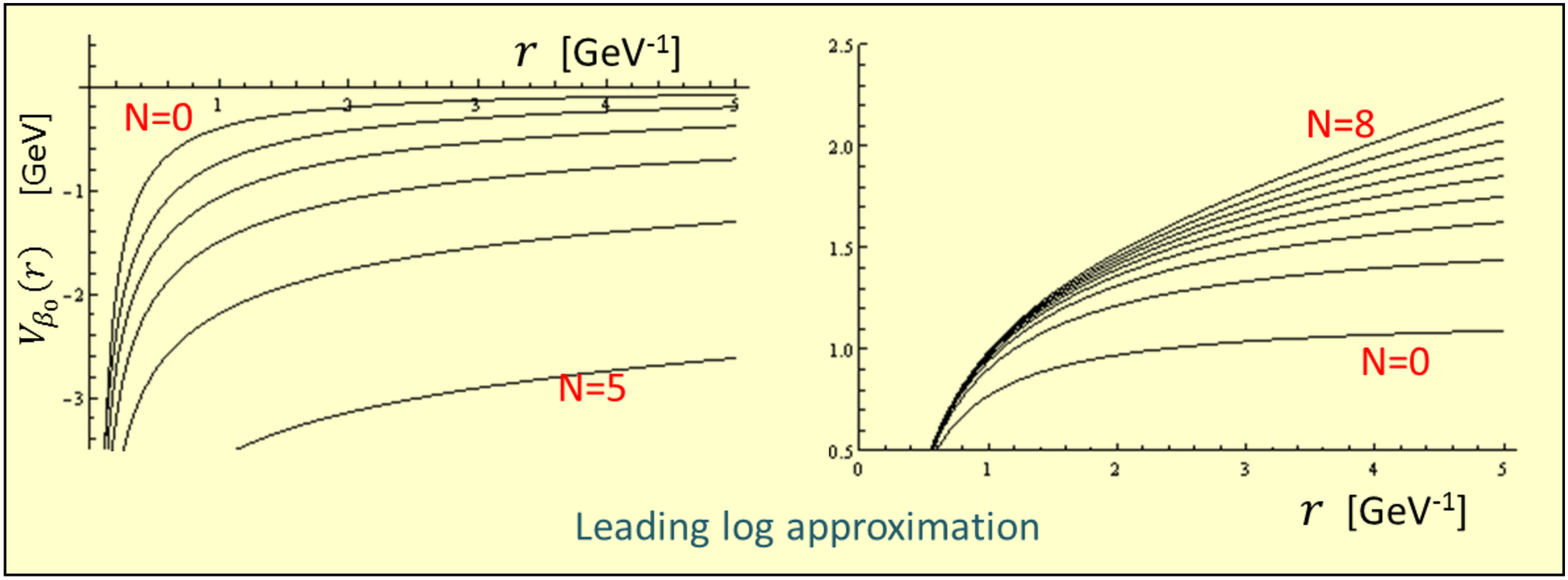}
\vspace*{5mm}
\\
\includegraphics[width=14cm]{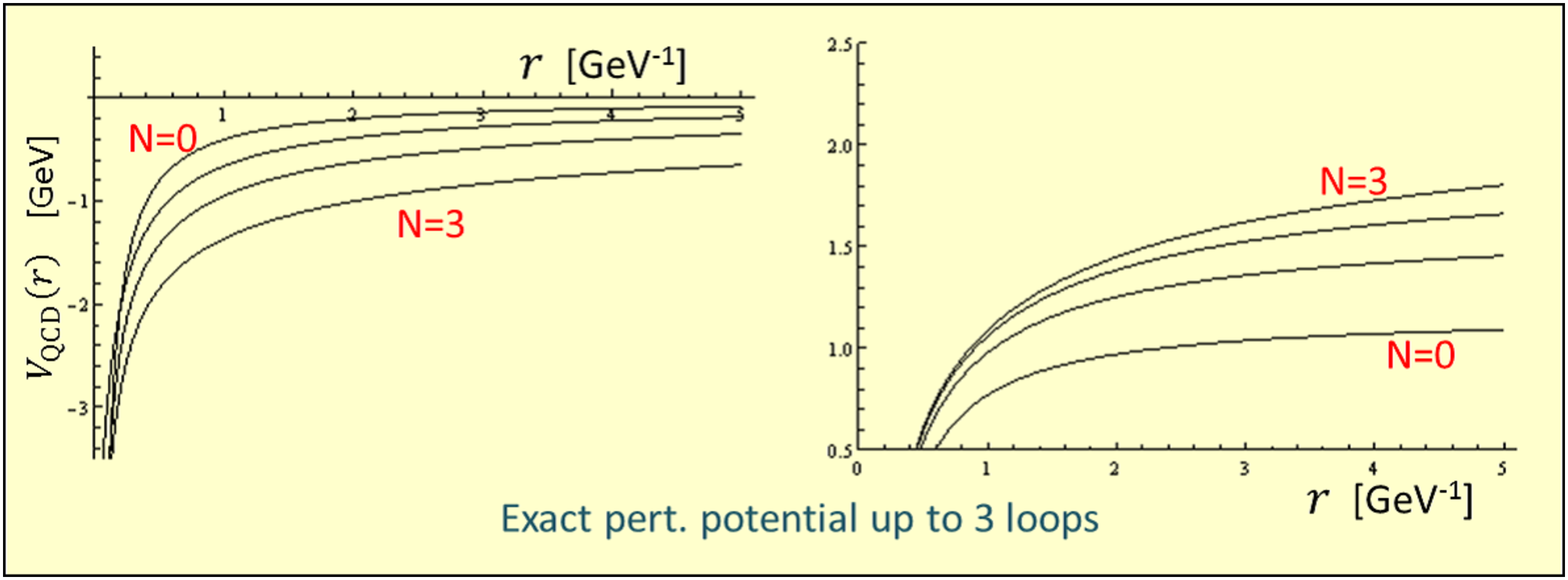}
\end{center}
\caption{\footnotesize
The potentials in the LL approximation and the
exact perturbative
potentials.
(The former incorporates an improvement by the ``large-$\beta_0$ 
approximation,'' where $\mu$ is replaced by $\mu\, e^{5/6}$.)
The potentials are given as the sum of the perturbative
series up to ${\cal O}(\alpha_s^{N+1})$.
We set $\mu=2$~GeV, $\alpha_s=0.3$ and $n_f=4$
(all the internal quarks are taken to be massless).
In the right figures, the dominant $r$-independent 
part [eq.~(\ref{LOrenorm-Largebeta0})] is subtracted.
\label{Comp-LLPot-ExactPot}}
\end{figure}
Finally we compare predictions of $V_{\rm QCD}(r)$
in the LL approximation and exact perturbative series
in Fig.~\ref{Comp-LLPot-ExactPot}.
The left figures show the QCD potential truncated at 
${\cal O}(\alpha_s^{N+1})$, while
the right figures show the same potentials but 
after subtracting an $r$-independent ${\cal O}(\LQ)$ renormalon contribution
\bea
V_{0}^{(n)}=- C_F \, 4 \pi \alpha_s \times\frac{\mu \,e^{5/6}}{2\pi^2} 
\biggl( \frac{\beta_0 \alpha_s}{2\pi} \biggr)^n \,
\times n! \, 
\label{LOrenorm-Largebeta0}
\eea
from each term of the perturbative series.
Compare eq.~(\ref{LLest-LOrenormalon}).\footnote{
The replacement $\mu\to\mu\, e^{5/6}$ is incorporated in order to improve 
the estimate
of the normalization of the renormalon using the so-called
``large-$\beta_0$ approximation.''
The potentials in the LL approximation in
Fig.~\ref{Comp-LLPot-ExactPot} are also improved
by the same method.
(This does not change the position of the renormalon poles
in the Borel plane.)
}

We see that the LL approximation gives good estimates of the
known perturbative corrections.
Convergence improves in the right figures.
 (Note that the
vertical scale is magnified by factor two compared to
the left figures, and also look in particular at
the region $r\simlt 2.5$~GeV$^{-1}\approx 0.5$~fm.)
Furthermore, after subtracting the dominant $r$-independent
part,
the potential becomes steeper at larger $r$
as we include more terms.
This is a desirable feature, in the sense that it
approaches the correct shape at larger $r$
as we include higher-order terms,
c.f.~Sec.~\ref{sec:IntroQCDpot}.
(We collect some formulas for $E_{{\rm tot}}(r)$
in Appendix~\ref{AppA}, such that one can reproduce
the features demonstrated in this subsection.)

\section{\!\!``Coulomb+Linear'' Potential by Log Resummation}
\clfn

We have seen that, 
by including higher-order terms,
$E_{\rm tot}(r)$ becomes steeper at larger
$r$ (within the range $r<\Lambda_{\rm QCD}^{-1}$).
Numerically it approximates a ``Coulomb+linear''
shape.
We show that a resummation of LLs, after subtracting renormalons,
indeed leads to a ``Coulomb+linear'' potential.\footnote{
This can be extended to the cases NLL, NNLL, etc.
}
This part of the potential is determined solely by UV contributions
and is a genuine prediction of perturbative QCD.

There are two ways to derive the 
``Coulomb+linear'' potential:
one is based on resummation of LLs purely in perturbative prediction,
the other uses the framework of a Wilsonian effective
field theory (EFT).
Both methods lead to the same formula.
For pedagogical reasons, we explain the latter method in this
lecture.

\subsection[Analysis of $V_{\rm QCD}(r)$ in an EFT framework
(Outline)]{\boldmath Analysis of $V_{\rm QCD}(r)$ in an EFT framework
(Outline)}
\label{sec:Outline-pNRQCD}

We explain an outline of the analysis of $V_{\rm QCD}(r)$
within a Wilsonian low energy EFT, called ``potential-NRQCD (pNRQCD).''
Its formulation will be explained extensively in Sec.~\ref{Sec:pNRQCD}.
Here, we sketch some aspects in advance, required for the analysis
in this section.

Let us first explain the general concept of a Wilsonian
low energy EFT, written in terms of light quarks
and gluons.
Formally such an EFT can be constructed from full QCD by
integrating out high-energy modes above a factorization scale
$\mu_f\, (\gg \! \Lambda_{\rm QCD})$, in a path-integral formulation of the theory.
The Lagrangian of the EFT can be written in a form
\bea
{\cal L}_{\rm EFT}(\mu_f)=\sum_i g_i(\mu_f) \, {\cal O}_i
(\psi_q,\bar{\psi}_q,A_\mu),
\label{EFT-Lagrangian}
\eea
which is a sum of operators ${\cal O}_i$ composed of
light quarks and gluons, whose energies and momenta are restricted
to be below $\mu_f$.
The effective coupling constant $g_i(\mu_f)$ multiplying
each operator is called a Wilson coefficient, which is
determined such that physics at $E<\mu_f$ is unchanged
from full QCD.
Since the Wilson coefficients
$g_i(\mu_f)$'s include only effects of UV degrees of
freedom ($E>\mu_f$), they can be computed reliably using perturbative
QCD.
In practice
there are two ways to compute the Wilson coefficients.
One way is to compute various $S$ matrix elements with
external momenta of order $E$, where
$\mu_f \simgt E \gg \LQ$,
in both EFT and full QCD in expansions in $\alfs$,
and to require that both computations give the same results.
This is known as a matching procedure.
The other method is to apply the technique called
integration by regions, 
which determines the operators and Wilson coefficients
of EFT in an efficient way by expanding Feynman diagrams
in terms of small parameters.
This method is explained briefly in Appendix~\ref{AppC}.

The Wilson coefficients computed using perturbative QCD
should be free of 
uncertainties by IR renormalons,
since the region of integration (above $\mu_f$)
does not include the domain
where the strong coupling constant is large.
Thus, the EFT Lagrangian eq.~(\ref{EFT-Lagrangian})
consists of Wilson coefficients, which effectively
contain information on UV degrees of freedom
($E>\mu_f$), and
operators composed of dynamical variables representing
IR degrees of freedom ($E<\mu_f$).

Generally a physical observable computed in the EFT
is expressed by a sum of
products of Wilson coefficients and matrix elements
of IR operators.
In this way we can factorize UV contributions (Wilson
coefficients) and IR contributions 
(matrix elements which include non-perturbative contributions).
If the observable includes a high mass scale $M~(\gg\mu_f)$, 
the expression can be organized systematically
in an expansion in powers of $1/M$.
Since matrix elements of operators include scales
only below $\mu_f$, this expansion generates
powers of small parameters ($\simlt \mu_f/M$).
As compared to the purely perturbative computation of the
same observable,
in the EFT formulation, only UV part of the purely perturbative prediction
is encoded in the Wilson coefficients, whereas 
uncertainties originating from IR renormalons in perturbative QCD
are replaced
by non-perturbative matrix elements.
Hence, intrinsic uncertainties are
eliminated from perturbative computations of the Wilson coefficients,
and we obtain more accurate predictions as we compute
higher-orders corrections (provided that non-perturbative matrix elements
can be determined in some way).

In the case of the static QCD potential,
since the perturbative prediction is more accurate at short distances,
$r\ll\LQ^{-1}$,
we consider a short-distance expansion of
$V_{\rm QCD}(r)$:
\bea
V_{\rm QCD}(r)\sim\frac{c_{-1}}{r} + c_0 + c_1r + c_2 r^2 +\cdots .
\label{naive-expansion}
\eea
This
expansion in $r$ is (at best) qualitative, 
in the sense that there must be a logarithmic
correction at least to the Coulomb term
$\sim 1/[r \log (\LQ r)]$, as designated by the
renormalization-group equation.
We have seen in previous sections that,
in purely perturbative evaluation of $V_{\rm QCD}(r)$,
the leading uncertainty at $r\ll\LQ^{-1}$
is included in the $r$-independent constant, while the next-to-leading
uncertainty is included in the $r^2$ term,  induced by the renormalons:
\bea
&&
c_0 ~~~\sim {\cal O}(\LQ) ,
\\ &&
c_2\, r^2 \sim {\cal O}(\LQ^3r^2) .
\eea
The renormalon in $c_0$ is canceled against the renormalon
in $2m_{\rm pole}$ in the total energy $E_{\rm tot}(r)$.
The renormalon in $c_2$ is replaced by 
a non-perturbative matrix element if we compute $E_{\rm tot}(r)$
in the pNRQCD EFT framework.

pNRQCD is a low energy EFT which describes interaction between
a heavy $Q\bar{Q}$ system and IR gluons.
In this EFT, IR gluons, whose wavelengths are
larger than $r$, 
\begin{wrapfigure}{l}{50mm}
\vspace*{-4mm}
~~\includegraphics[width=7.5cm]{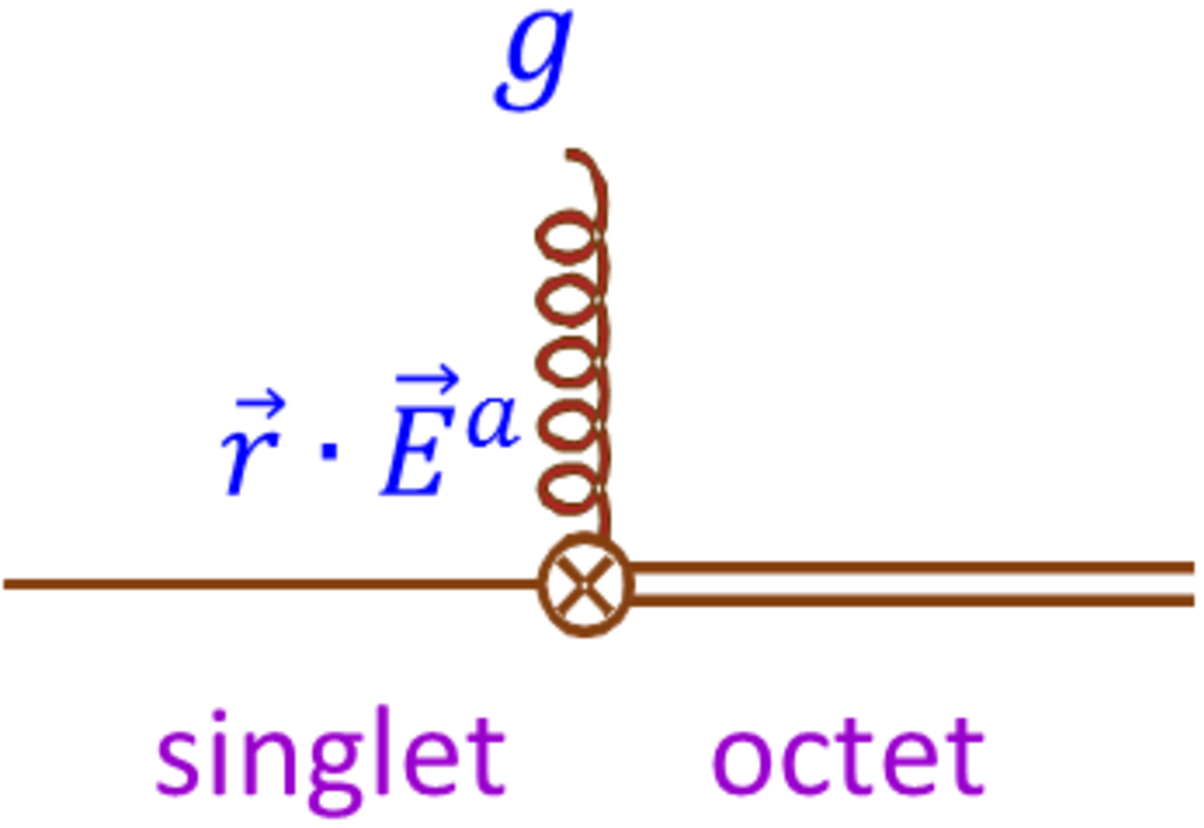}
\vspace*{-13mm}
\end{wrapfigure}
interact with
color multipoles of the 
$Q\bar{Q}$ system.
Since the color charge of a $Q\bar{Q}$ bound state is zero (color singlet),
the leading interaction of the singlet bound state with
IR gluons
is a dipole-type, $\vec{r}\cdot\!\vec{E}^a$,
where $\vec{E}^a$ denotes the color electric field.
This is depicted as a vertex of pNRQCD in the left figure.

The leading contribution to the $Q\bar{Q}$ bound state energy

\vspace*{-4.5mm}$~$
\begin{wrapfigure}{l}{50mm}
~\vspace*{4mm}
\\
\hspace*{-8mm}
\includegraphics[width=6cm]{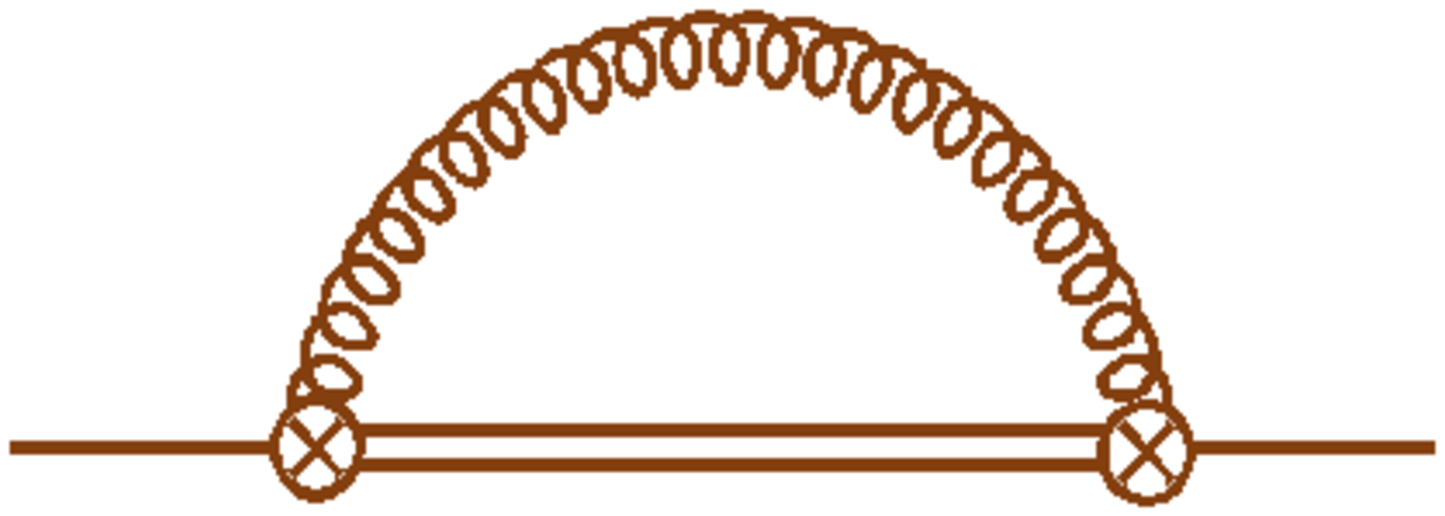}
\vspace*{-13mm}
\end{wrapfigure}
from IR degrees of freedom 
is given by the self-energy diagram with two
insertions of the dipole interactions
(see the left figure), which
is expressed in terms of the matrix element
\bea
\langle\, \vec{r}\cdot\!\vec{E}^a(t) \,\vec{r}\cdot\!\vec{E}^a(0)
\,\rangle .
\label{ErEr}
\eea
We will see in Sec.~\ref{Sec:pNRQCD} that
it has a correct form to replace
the ${\cal O}(r^2)$ renormalon in the
purely perturbative prediction.

\subsection[UV contributions to $V_{\rm QCD}(r)$ and OPE]{\boldmath UV contributions to $V_{\rm QCD}(r)$ and OPE}
\label{sec:UVcontrV_QCD}

Let us use the framework of  pNRQCD with a
factorization scale $\mu_f$ to compute $V_{\rm QCD}(r)$.
We assume
\bea
r^{-1} \gg \mu_f \gg \LQ .
\eea
In this framework the Wilson coefficient, which corresponds to the UV
contribution to
$V_{\rm QCD}(r)$,
can be computed using perturbative QCD.
We intoduce 
a cut-off in the momentum-space integral and resum LLs:
\hspace*{75mm}
\begin{wrapfigure}{l}{28mm}
\vspace*{-2mm}
\hspace*{0mm}
\includegraphics[width=2.3cm]{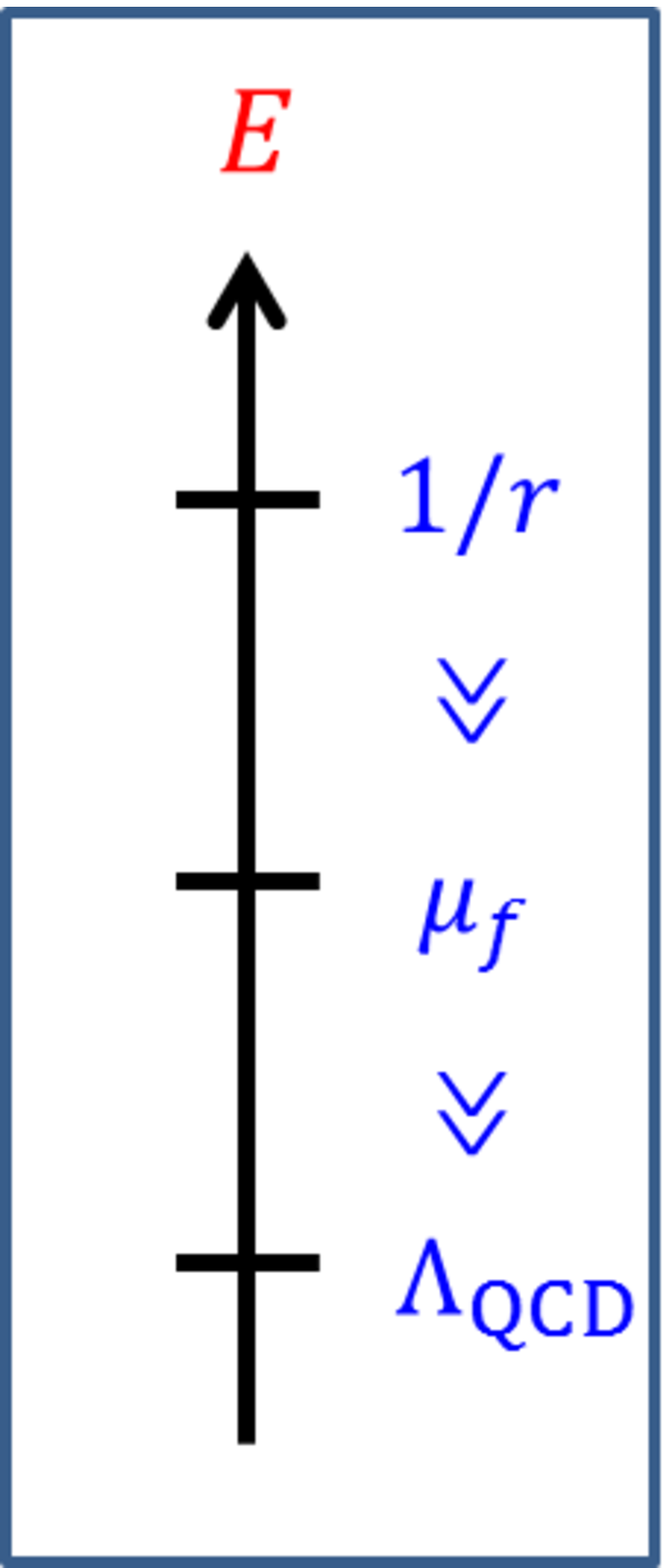}
\vspace*{-10mm}
\end{wrapfigure}
\bea
V_{\rm UV}(r;\mu_f)\equiv -
{\hbox to 18pt{
\hbox to -1pt{$\displaystyle \int$} 
\raise-15pt\hbox{$\scriptstyle q>\mu_f$} 
}}~
\frac{d^3\vec{q}}{(2\pi)^3} 
\, e^{i \vec{q} \cdot \vec{r}} \, 
C_F \frac{4\pi\alpha_{\rm 1L}(q)}{q^2} ,
\label{V_UV}
\eea
where
\bea
\alpha_{\rm 1L}(q) =
\frac{2\pi}{\beta_0 \log \bigl( {q}/{\Lambda_{\rm QCD}} \bigr)} .
\label{alpha1L}
\eea
The integral is well defined, since the pole of
$\alpha_{\rm 1L}(q)$ is not included in the integral region.

\begin{figure}[t]
\begin{center}
\includegraphics[width=7cm]{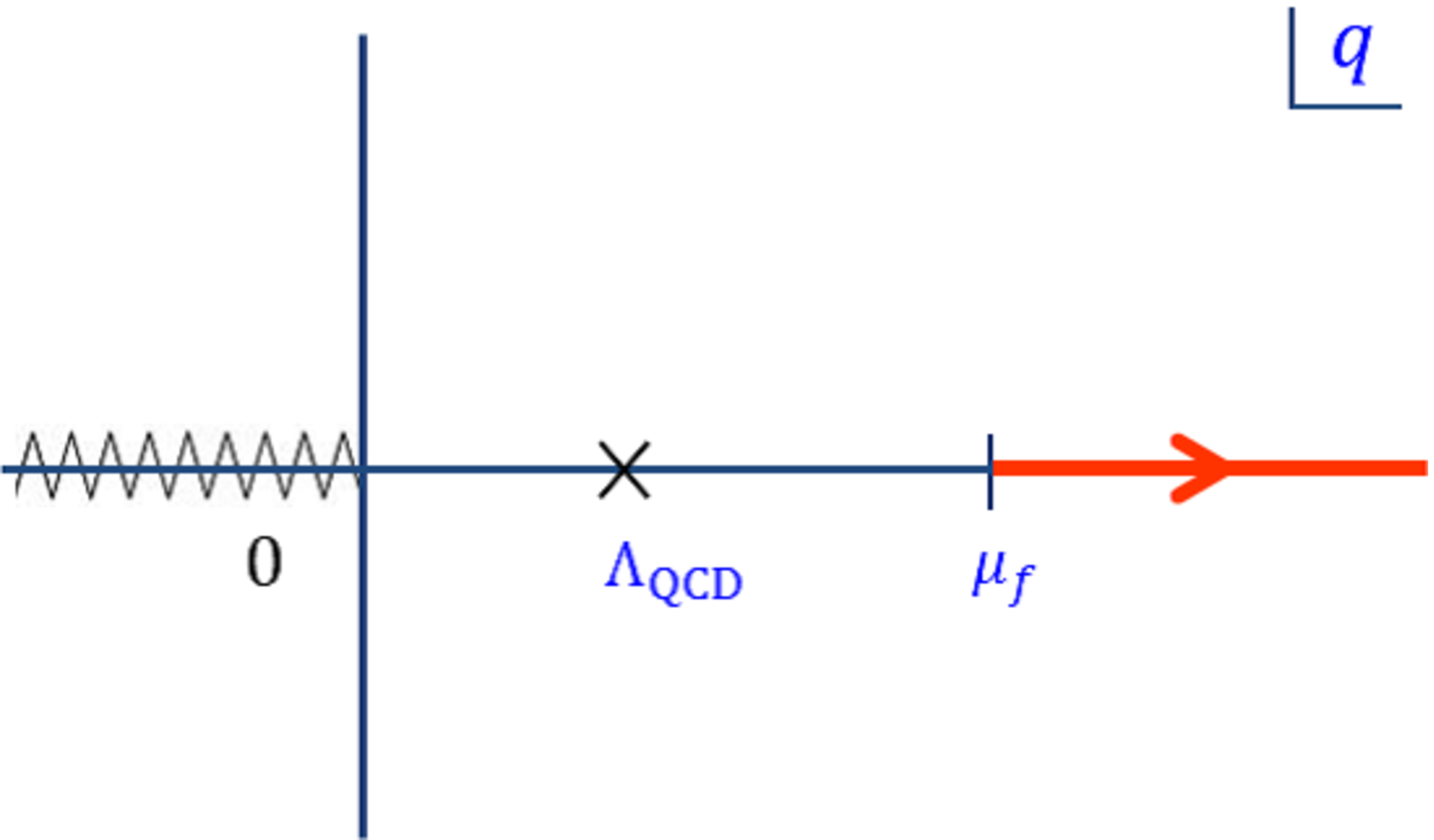}
~~~
\includegraphics[width=7cm]{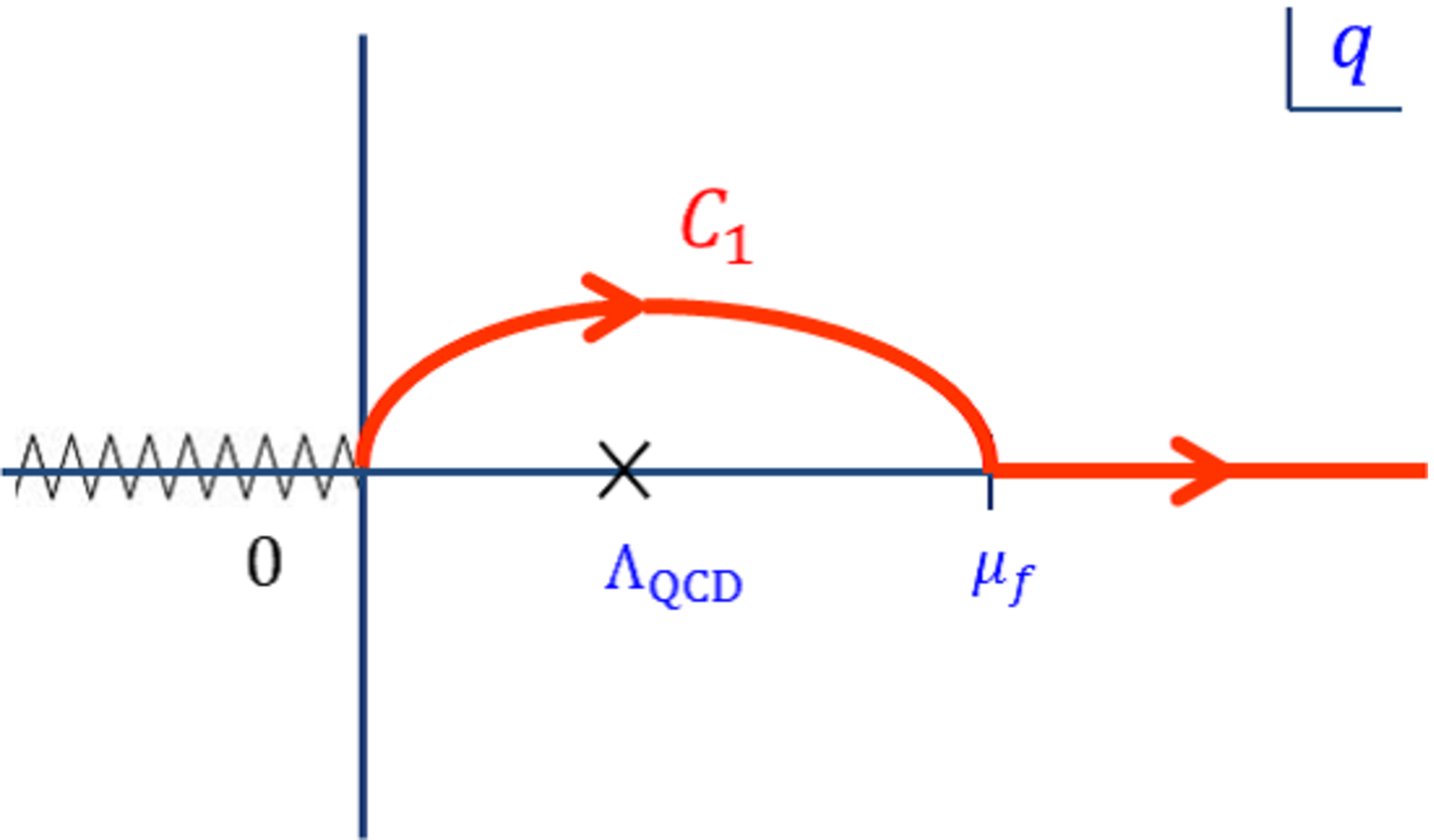}
\vspace*{5mm}
\\
\includegraphics[width=7cm]{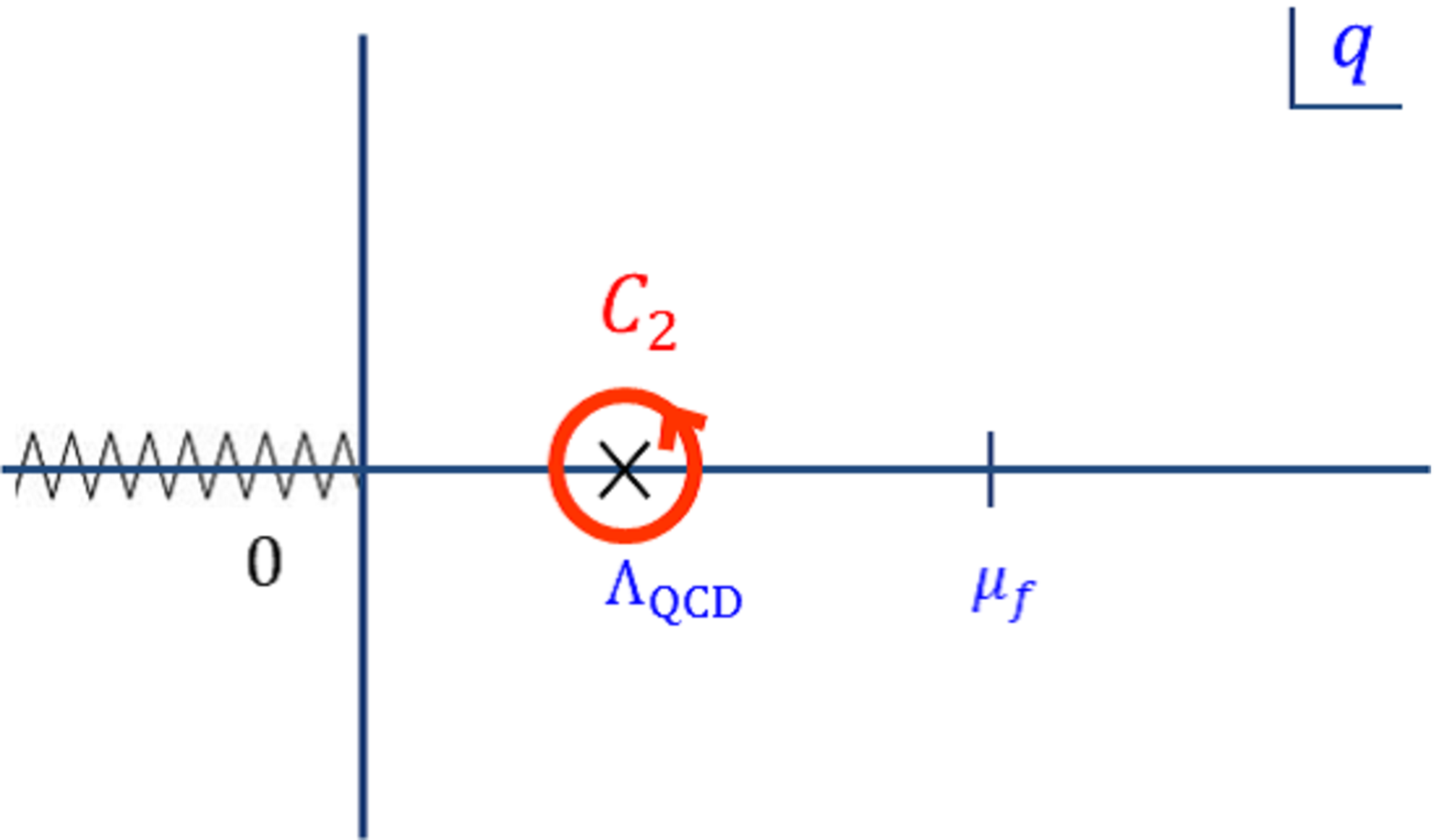}
~~~
\includegraphics[width=7cm]{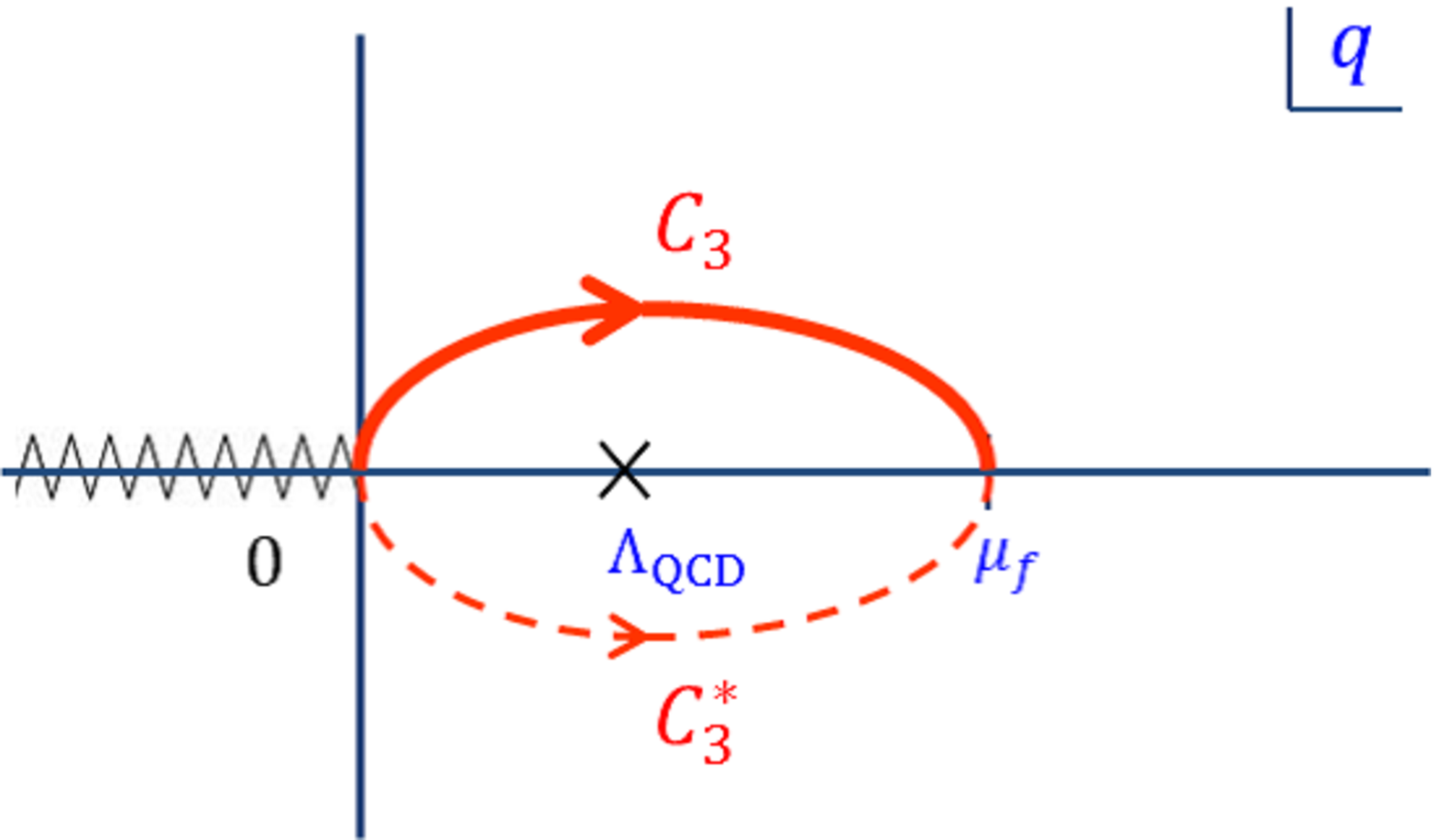}
\end{center}
\caption{\footnotesize
Integral contours in the complex $q$-plane shown by
red lines.
$C_3^*$ (dashed line) represents the complex conjugate
of $C_3$.
A pole and branch cut of the integrand are also shown.
\label{Contours}}
\end{figure}

We can obtain a short-distance expansion of $V_{\rm UV}$
as follows.
After integrating over the angular variables, we obtain
a one-parameter integral
\bea
&&
V_{\rm UV}(r;\mu_f)=-\frac{2C_F}{\pi}\int_{\mu_f}^\infty
\!dq\, \frac{\sin(qr)}{qr}\, \alpha_{\rm 1L}(q)
=-\frac{2C_F}{\pi}{\rm Im}\,\int_{\mu_f}^\infty
\!dq\, \frac{e^{iqr}}{qr}\, \alpha_{\rm 1L}(q)
.
\eea
This is an integral along the real 
axis in the complex $q$-plane, ranging from $q=\mu_f$ to $q=\infty$.
We separate the integral contour
into the difference of two contours $C_1-C_3$
in the complex $q$-plane;
see Fig.~\ref{Contours}.
Since $\mu_f r\ll 1$, along the contour $C_3$ it
is justified to expand the Fourier factor as
\bea
e^{iqr}=1+iqr+\frac{1}{2}(iqr)^2 +\cdots .
\eea
Then the integral along $C_3$ can be written as
\bea
\frac{2C_F}{\pi} {\rm Im}\int_{C_3}\!\!dq\,
\frac{1+iqr+\frac{1}{2}(iqr)^2 +\cdots}{qr}\,\alpha_{\rm 1L}(q)
=\frac{A}{r}+B+\sigma r + D r^2 + {\cal O}(r^3) .
\eea
The coefficient $A$ can be computed analytically as\footnote{
Note that $\alpha_{\rm 1L}(q)^*=\alpha_{\rm 1L}(q^*)$
in the domain ${\rm Re}\,q>0$ according to eq.~(\ref{alpha1L}).
}
\bea
&&
A=\frac{2C_F}{\pi}
 {\rm Im}\int_{C_3}\!\!dq\,
\frac{\alpha_{\rm 1L}(q)}{qr}
=\frac{C_F}{\pi i}
 \int_{C_3-C_3^*}\!\!dq\,
\frac{\alpha_{\rm 1L}(q)}{qr}
\nonumber \\ &&
~
=-\frac{C_F}{\pi i}
 \int_{C_2}\!\!dq\,
\frac{\alpha_{\rm 1L}(q)}{qr}
=-\frac{4\pi C_F}{\beta_0} .
\label{ExpCoeffA}
\eea
The integral contour can be deformed to $C_2$, which surrounds
the pole of $\alpha_{\rm 1L}(q)$. 
Then we used the
Cauchy theorem in the last equality.

We can evaluate the coefficient $\sigma$ in a similar manner:
\bea
&&
\sigma=\frac{2C_F}{\pi}
 {\rm Im}\int_{C_3}\!\!dq\,
\Bigl(-\frac{1}{2}q\Bigr)\alpha_{\rm 1L}(q)
=\frac{C_F}{2\pi i}
 \int_{C_2}\!\!dq\,
q\,{\alpha_{\rm 1L}(q)}
=\frac{2\pi C_F}{\beta_0} \LQ^2 .
\label{sigma}
\eea
Although originally the expressions for $A$
and  $\sigma$ appear
to be dependent on $\mu_f$, in fact they reveal to be
independent of $\mu_f$, since they can be expressed by
integrals along the closed contour $C_2$.

In contrast, $B$ and $D$ are dependent on $\mu_f$, since 
they cannot be expressed
as integrals along a closed contour:\footnote{
The difference is that the integrals can be written only
as ones along $C_3+C_3^*$ rather than
$C_3-C_3^*$.
}
\bea
&&
B=\frac{2C_F}{\pi}\,{\rm Re}\int_{C_3}dq\,\alpha_{\rm 1L}(q)
,
\\ &&
D=-\frac{C_F}{3\pi}\,{\rm Re}\int_{C_3}dq\,q^2\,\alpha_{\rm 1L}(q)
.
\eea

Combining these, we obtain
\bea
&&
V_{\rm UV}(r;\mu_f)=
V_C(r)+B+\sigma r+Dr^2+{\cal O}(r^3) ,
\label{expVUV}
\\ &&
V_C(r)=\frac{A}{r}-
\frac{2C_F}{\pi}\,{\rm Im}\,\int_{C_1}
\!dq\, \frac{e^{iqr}}{qr}\, \alpha_{\rm 1L}(q) .
\label{Vc(r)}
\eea
Note that $V_C(r)$ is
independent of $\mu_f$,
since the integral along $C_1$ is also independent 
\begin{wrapfigure}{l}{65mm}
\vspace*{-2mm}
\hspace*{0mm}
\includegraphics[width=6cm]{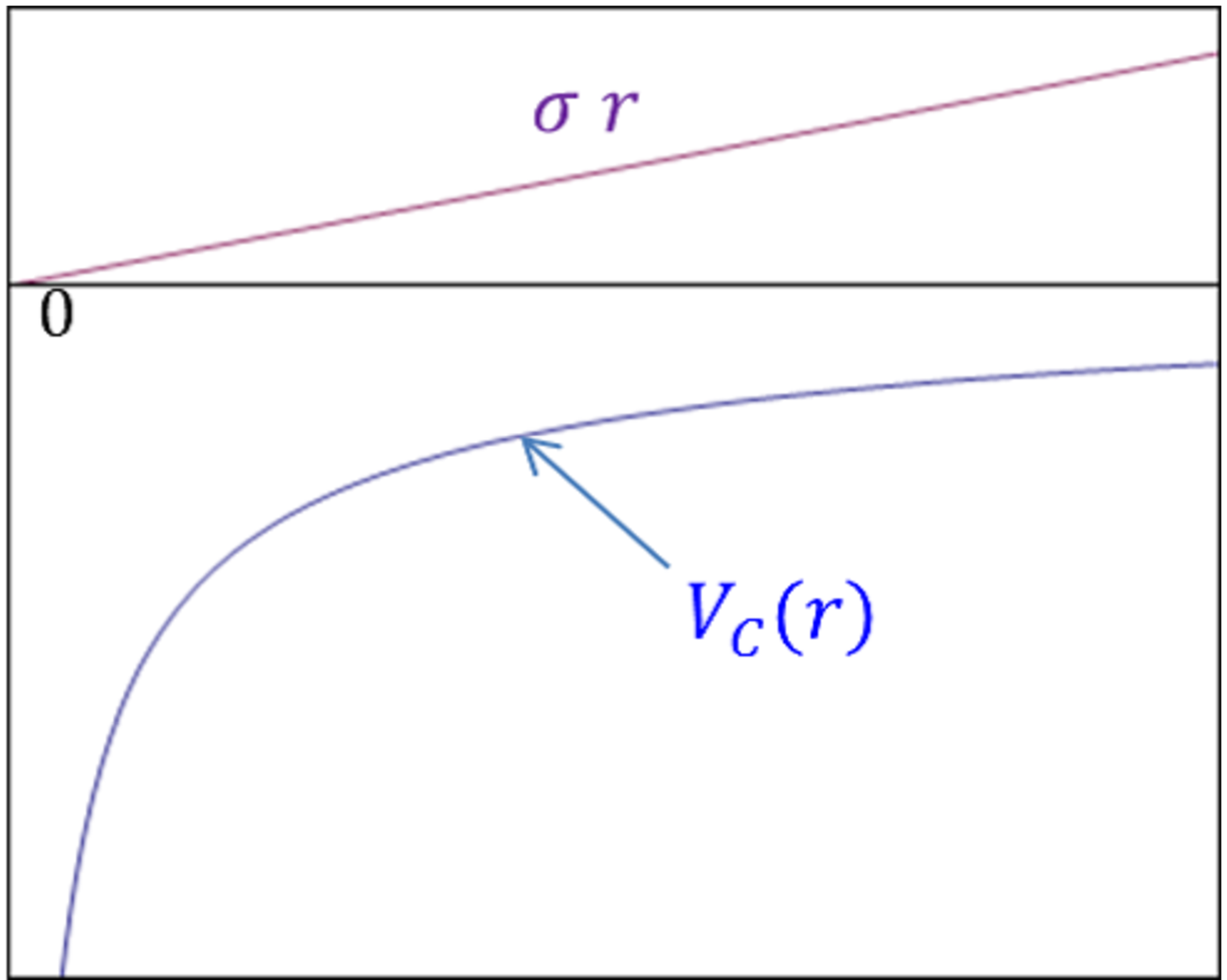}
\vspace*{-27mm}
\end{wrapfigure}
of the intermediate
point $\mu_f$.
$V_C(r)$ and $\sigma r$ are shown in the left figure.
The asymptotic behaviors of $V_C(r)$ as $r\to 0$ and $r \to\infty$
can be computed analytically as
\bea
V_C(r)
\to
\left\{
\begin{array}{l}\displaystyle
-\frac{2\pi C_F}{\beta_0 }\,
\frac{1}{r \,\bigl|\log (
\Lambda_{\rm QCD}
\,r)\bigr|}
,
~~~\,
r \to 0 ,
\\ \displaystyle
\rule{0mm}{10mm}
 - \frac{4\pi C_F}{\beta_0 r} ,
~~~
~~~~~~~~~~
~~~~~~~~~
r \to \infty .
\end{array}
\right.
\nonumber\\ 
\eea
[See Appendix~\ref{AppB} for computation of $V_C(r)$.]
At $r\to 0$, $V_C(r)$ tends to a Coulomb potential with
the correct logarithmic correction as determined by the RG equation;
at $r\to\infty$, it approaches a pure Coulomb potential;
in the intermediate distance region these asymptotic behaviors are
smoothly interpolated.
Thus, eq.~(\ref{expVUV}) can be regarded as a (qualitative)
expansion of $V_{\rm UV}(r)$ in $r$, at short-distances.

As will be shown in Sec.~\ref{Sec:pNRQCD}, the 
operator-product expansion (OPE) of 
$V_{\rm QCD}(r)$ in $r$ within pNRQCD framework
takes a form
\bea
V_{\rm QCD}(r) =
V_{\rm UV}(r;\mu_f) + V_{\rm IR}(r;\mu_f),
\eea
where the leading term of the IR contribution
$V_{\rm IR}$ is expressed in terms of the non-perturbative matrix element
of eq.~(\ref{ErEr}).
Substituting the expansion (\ref{expVUV}), we obtain
\bea
V_{\rm QCD}(r) = 
{\rm const.}+
\underbrace{V_C(r)+\sigma r}_{\mu_f\text{-indep.}} + 
\underbrace{D r^2
+V_{\rm IR}^{\rm (LO)}(r;\mu_f)}_{\mu_f\text{-indep.}}  
+ {\cal O}(r^3) .
\label{OPEofV_QCD}
\eea
We can show explicitly 
that the $\mu_f$-dependences cancel between $Dr^2$ and
the leading non-perturbative contribution
(see Sec.~\ref{sec:RenormalonInV_IR}).
It is consistent, since in total $V_{\rm QCD}(r)$
should be independent of the factorization scale $\mu_f$.
We also note that, if the ``Coulomb'' and linear
terms were dependent on $\mu_f$, that would have
led to inconsistencies, since there are no 
IR contributions which can cancel such $\mu_f$-dependences.

Thus, $V_C(r)+\sigma r$ is included in 
a short-distance expansion of $V_{\rm UV}(r;\mu_f)$
and insensitive to $\mu_f$.
Therefore, it is
a genuinely UV contribution, determined by
perturbative QCD.\footnote{
We are not concerned about
the constant part of
$V_{\rm QCD}(r)$, since it is also dominated by UV contributions
when combined with $2m_{\rm pole}$
in $E_{\rm tot}(r)$.
}
We also emphasize that, since $\sigma r$ arises in the
short-distance expansion, a priori
this linear potential has nothing to do with
the linear potential as determined from the
large-distance behavior of $V_{\rm QCD}(r)$,
which is closely connected with quark confinement.

We show (without derivation)
the result of analysis including
subleading logarithms via RG equation.
\begin{figure}[h]
\begin{center}
\includegraphics[width=10cm]{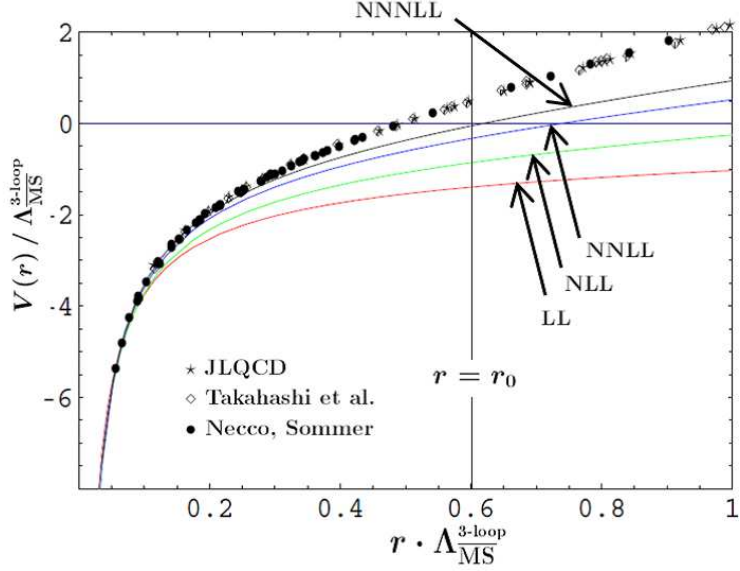}
\vspace*{-5mm}
\end{center}
\caption{\footnotesize
Comparison of lattice computations
of $V_{\rm QCD}(r)$ (data points) and 
perturbative QCD predictions for ${V_C(r)} + {\sigma\,r}$
(solid lines)
in the case $n_f=0$ (quenched approximation).
$r_0$ denotes the Sommer scale, which is 
considered to be roughly 0.5~fm\,$\approx (400$~MeV$)^{-1}$.
\label{LogResumPots}}
\end{figure}
Fig.~\ref{LogResumPots} shows a
comparison of lattice data
and ${V_C(r)} + {\sigma\,r}$ in different orders
of log resummations.
As can be seen, with increasing  order, the range where
${V_C(r)} + {\sigma\,r}$
(perturbative prediction) agrees with the lattice data extends
to larger $r$.
The difference between the lattice
computations and the purely perturbative computation
corresponds to the ``${\cal O}(r^2)$-term,''
$D r^2
+V_{\rm IR}^{\rm (LO)}(r;\mu_f)$, up to
higher-order terms in $r$.
At NNNLL,
there is no room in this difference to accommodate
a linear potential $Kr$ of
eq.~(\ref{Coulomb+LinearForm}) at $r\simlt 0.5$~fm.
Namely, one should not add a linear term to the perturbative
potential in this region.
To our current knowledge, $\sigma$ increases
as we include more subleading logarithms.
For example, at NNNLL and $n_f=0$, $\sigma/K$ is
between 1/3 and 1/2, where $K$ is determined from the
large-distance behavior of $V_{\rm QCD}(r)$ by lattice
computations.

As mentioned, it is not essential to introduce the
factorization scale $\mu_f$ to derive the 
``Coulomb+linear'' potential.
It can be shown that the truncated
series expansion of $E_{\rm tot}(r)$,
\bea
\sum_{n=0}^N \Bigl[ 2m^{(n)}_{{\rm pole},\beta_0}+V_{\beta_0}^{(n)}(r)
\Bigr]
~~~~~~~
\mbox{for}~~~ N< n_0=\frac{6\pi}{\beta_0\alpha_s},
\eea
approaches toward $V_C(r)+\sigma r$,
up to an uncertainty $\sim \text{const.}+{\cal O}(\LQ^3r^2)$ 
(c.f., Sec.~\ref{sec:m_pole+E_tot}).

Let us present an interpretation of the formula for
$V_C(r)+\sigma r$, eqs.~(\ref{ExpCoeffA}), (\ref{sigma})
and (\ref{Vc(r)}).
The integral contours in these formulas originate from
a relation
depicted schematically as below:
\begin{center}
\includegraphics[width=15cm]{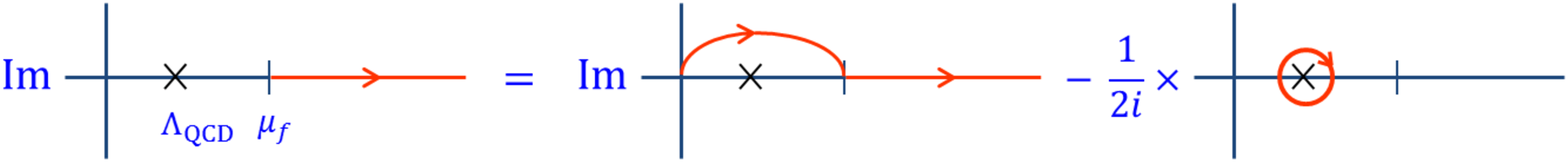}
\end{center}
This has a form where contributions from IR region are
subtracted as contour integrals surrounding the pole
at $q=\LQ$.
Hence, it can be regarded as a generalization of
the method of ``integration by regions,''
in which contributions from different scales are
factorized as contour integrals surrounding
the corresponding poles in each Feynman diagram.
(See Appendix~\ref{AppC}.)
In computing $V_{\rm UV}(r;\mu_f)$ we need to
separate UV and IR contributions in $V_{\rm QCD}(r)$ 
and subtract the
latter.
It is, however, not possible to identify IR contributions
by the standard integration-by-regions technique, since
the scale $\LQ$ never  appears as a singularity
at each order of perturbative expansion.
In our formula, the scale $\LQ$ appears
as a singularity by log resummation, and
its contributions are subtracted as contour integrals.

\section{Implication and Interpretation}
\clfn

Using the result of the previous section,
we can compute the total energy of a static
$Q\bar{Q}$ pair as
\bea
E_{\rm tot}(r)=2\overline{m}+{\rm const.}+
{V_C(r)}+{\sigma\,r}+{\cal O}(\LQ^3r^2) 
\eea
within perturbative QCD,
where the $r$-independent part is also
UV dominant and accurately predictable.
(The ${\cal O}(\LQ^3r^2)$ uncertainty can be replaced by
a non-perturbative matrix element within the EFT approach.)
The spectrum of a heavy quarkonium system,
such as bottomonium,
can be computed roughly as the energy eigenvalues
of the quantum mechanical Hamiltonian\footnote{
More accurate prediction is possible using the
potential-NRQCD framework.
Currently
the spectrum is known up to NNNLO.
}
\bea
H=\frac{\vec{p}^{\,2}}{m_{\rm pole}}+E_{\rm tot}(r) .
\label{LOHamiltonian}
\eea

\begin{figure}[t]\centering
\vspace*{5mm}
\includegraphics[width=6cm]{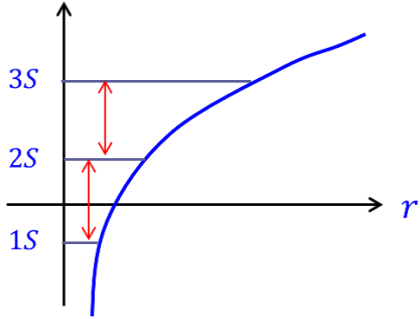}
\caption{\footnotesize
A schematic diagram showing the energy levels of
the Hamiltonian Eq.~(\ref{LOHamiltonian}).
The level spacings are order $\LQ(\LQ/m)^{1/3}$ if
they are predominantly determined by the linear part
$\sigma r$ of the potential.
\label{Levels-in-C+L_Pot}
}
\end{figure}
A linear potential of order $\LQ^2\,r$ generates
level spacings between different $S$-states
of order $\LQ(\LQ/m)^{1/3}$.\footnote{
This estimate can be derived using a semi-classical
approximation (WKB approximation) in quantum mechanics.
}
On the other hand, a Coulomb potential $\sim -\alpha_s/r$ generates
level spacings of
order $\alpha_s^2 m$.
For the bottomonium states, the linear potential
is estimated to be comparable to or
more important than the Coulomb potential
in generating these level spacings.
In contrast, if we consider
(would-be) toponium states, the 
Coulomb potential by far dominates over
the linear potential.
Thus, a major part of the perturbative QCD predictions for
the level spacings between different bottomonium $S$-states
is order $\LQ(\LQ/m)^{1/3}$.
See Fig.~\ref{Levels-in-C+L_Pot}.

We may develop a microscopic understanding on 
the composition of the energy
of a bottomonium state,
using eqs.~(\ref{ApproxV}) and (\ref{Approx2M}).
After integration over angular variables, we obtain
\bea
&&
2m_{\rm pole}\approx 2m_{\overline{\rm MS}}(\mu)+
\frac{2C_F}{\pi}\int_0^\mu \!\! dq\, \alpha_{1L}(q) ,
\\ &&
V_{\rm QCD}(r) \approx 
-\frac{2C_F}{\pi}\int_0^\infty \!\!\!\! dq\, \frac{\sin(qr)}{qr}\,\alpha_{1L}(q) 
.
\eea
For a bottomonium state $X$,
\bea
&&
E_X^{b\bar{b}}=\bra{X}\hat{H}\ket{X}
\nonumber\\ &&
~~~~
\approx
\bra{X}
\frac{\vec{p}^{\,2}}{m_{\rm pole}}+
2m_{\overline{\rm MS}}(\mu)+
\frac{2C_F}{\pi}\int_0^\mu\!\! dq\, \alpha_{1L}(q)
-\frac{2C_F}{\pi}\int_0^\infty \!\!\!\! dq\, \frac{\sin(qr)}{qr}\,\alpha_{1L}(q) 
\ket{X}
\nonumber\\ &&
~~~~
= 2m_{\overline{\rm MS}}(\mu)+
\bra{X}
\frac{\vec{p}^{\,2}}{m_{\rm pole}}\ket{X}+
\frac{2C_F}{\pi}\int_0^\infty \!\!\!\! dq\, \alpha_{1L}(q) \, f_X(q) ,
\label{Ecomposition}
\eea
where
\bea
f_X(q)=\theta(\mu-q)-\int_0^\infty dr\, r^2|R_X(r)|^2
\, \frac{\sin(qr)}{qr} .
\eea
$R_X(r)$ represents the radial part of the wave function of $X$.
We find that the kinetic energy $\bra{X}
{\vec{p}^{\,2}}/{m_{\rm pole}}\ket{X}$ turns out to be numerically
small, which we ignore in the following
discussion.\footnote{
This does not contradict the virial theorem for the
Coulomb system, since our
potential is significantly deviated from the Coulomb potential.
}

Let us set $\mu=m_{\overline{\rm MS}}(\mu)(\equiv \overline{m}_b)$
for the bottomonium case.
According to the discussion in Sec.~\ref{sec:IRcancellation},  
infrared gluons decouple
in the computation of the energy of a  bottomonium state $X$.
\begin{figure}[t]
\vspace*{0.5cm}
\begin{center}
    \includegraphics[width=12cm]{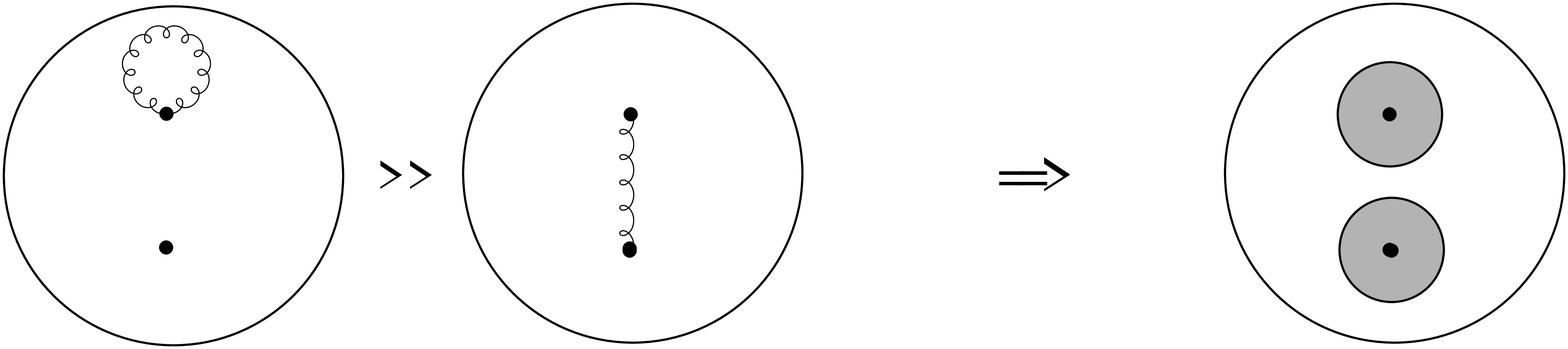}
\end{center}
\caption{\footnotesize
The total energy of a heavy quarkonium state is carried by the
$\overline{\rm MS}$ masses of $Q$ and $\bar{Q}$ and by 
the gluons whose wavelengths are smaller than the bound-state
size.
In the latter contributions the self-energies of $Q$ and $\bar{Q}$
dominate over the potential energy between the two particles.
\label{physpic}}
\end{figure}
The energy consists of the self-energies
of $b$ and $\bar{b}$ and the potential energy between $b$ and $\bar{b}$,
where gluons whose wavelengths $\lambda$ are smaller than the bound state
size $a_X$ contribute.
At IR ($\lambda>a_X$) 
the sum of the self-energies and the potential energy cancel.
On the other hand, at UV ($\lambda<a_X$), 
the potential energy quickly dumps
due to the rapid oscillation factor $e^{i\vec{q}\cdot\vec{r}}$ for large $q$
in the potential energy.
See eqs.~(\ref{ApproxV}) and (\ref{Approx2M}).
It means that
the major contribution 
to the bottmonium energy comes from the region (in momentum space)
$1/a_X$ $ \simlt$ $ q$ $\simlt$ $\overline{m}_b$ of the self-energy corrections 
of $b$ and $\bar{b}$,
apart from the constant contribution $2 \overline{m}_b$.
See Fig.~\ref{physpic}.

\begin{figure}[h]
\begin{center}
\begin{tabular}{cc}
\includegraphics[width=8cm]{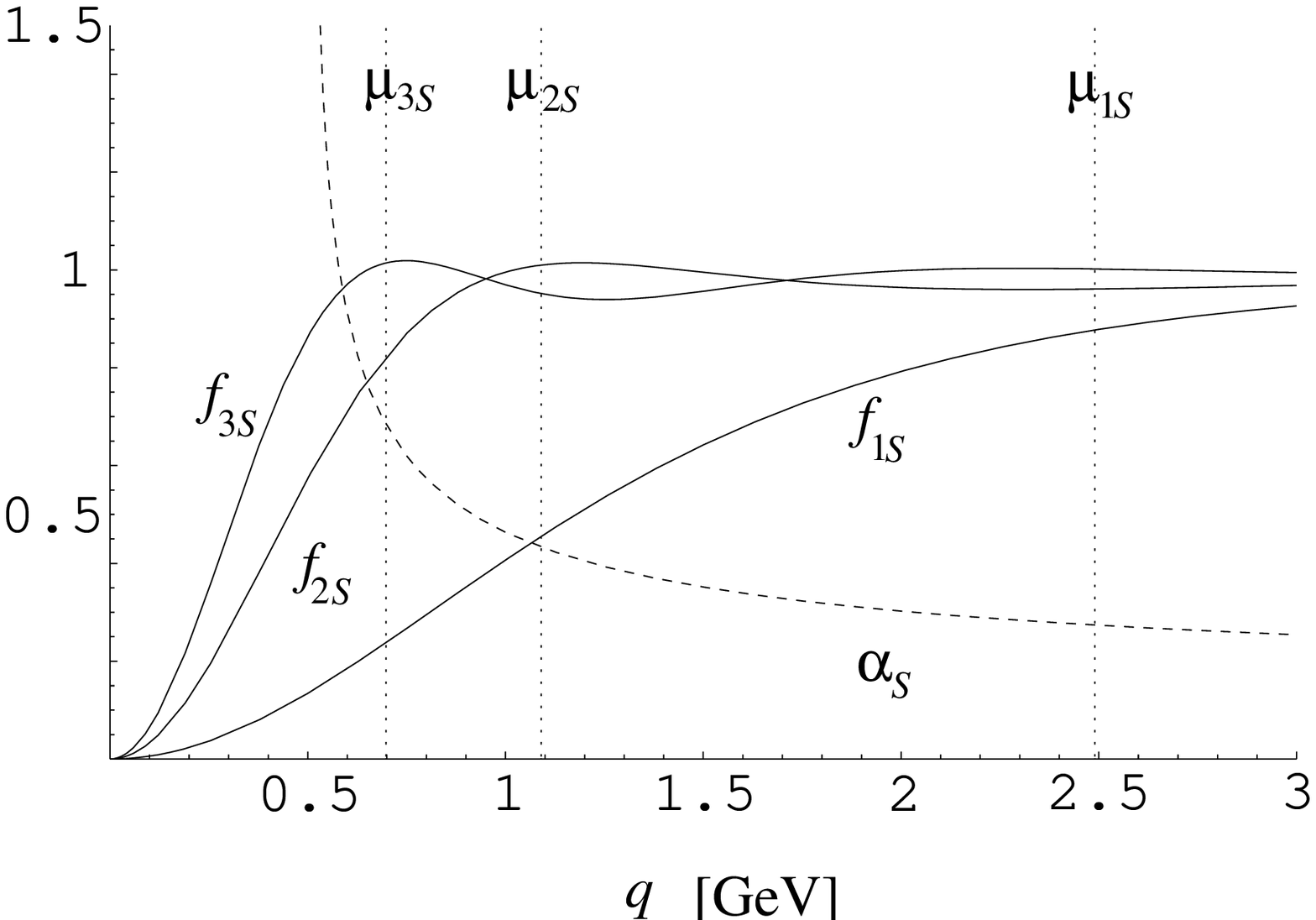}~~
&
\includegraphics[width=5.5cm]{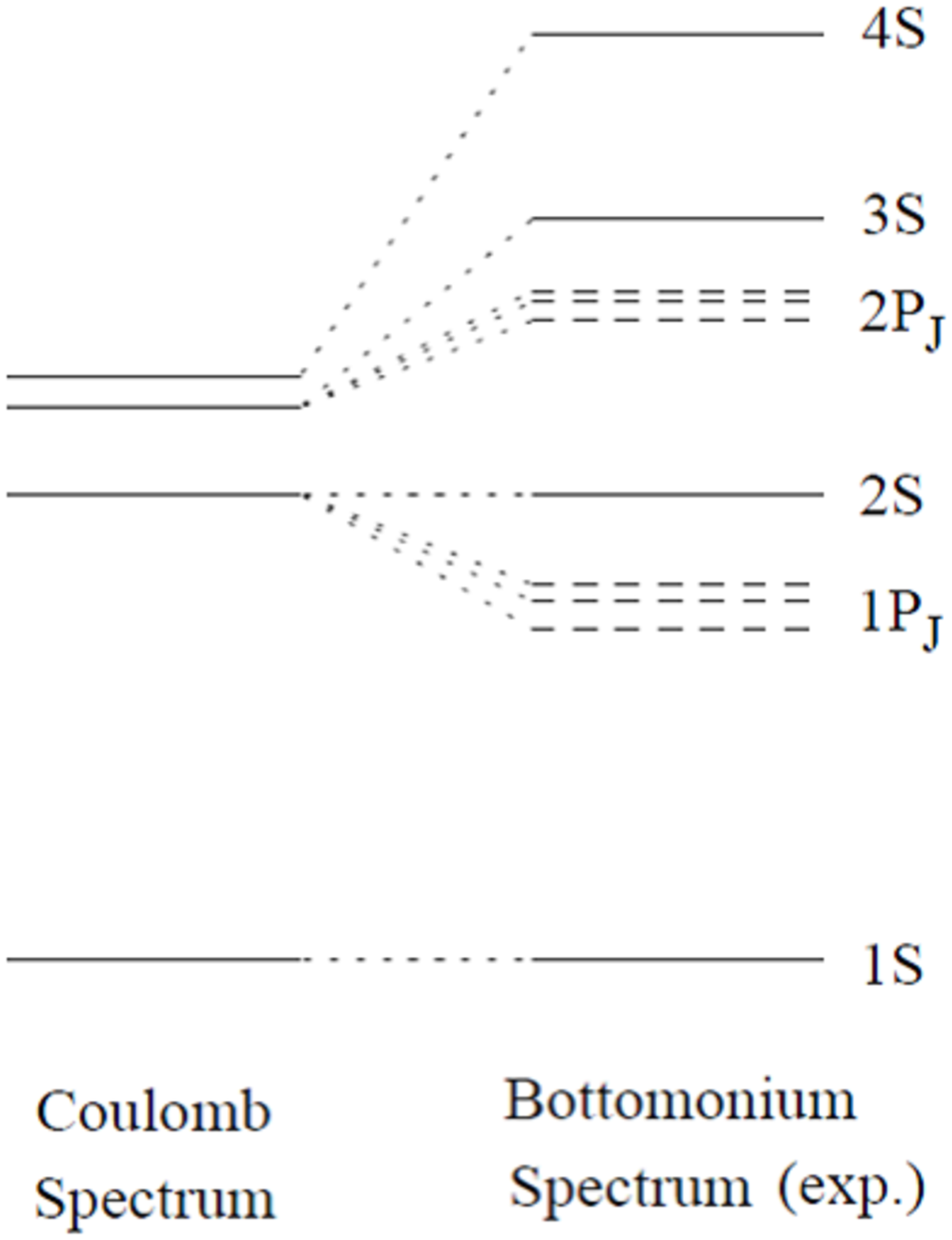}\\
(a)~~&~(b)
\end{tabular}
\end{center}
\caption{\footnotesize
(a) The support functions $f_X(q)$ used to express the
energy of the bottomonium state $X$ in Eq.~(\ref{Ecomposition})
for $X=1S$, $2S$ and $3S$.
The running coupling constant $\alpha_s(q)$ 
close to the dumping scale of $f_X(q)$ grows rapidly
as $X$ varies from the $1S$ to $3S$ states.
$\mu_X$ represents the typical momentum scale of $X$.
(b)~Comparison of the Coulomb spectrum and observed
bottomonium spectrum.
The Coulomb spectrum is scaled such that the $1S$--$2S$ level
spacing coincides with that of the bottomonium spectrum.
\label{BottomoniumEnergy}}
\vspace*{.5cm}
\end{figure}

In fact, the composition of the energy in momentum space, 
eq.~(\ref{Ecomposition}) when $\langle p^2/m \rangle$ is
neglected, has exactly this form,
since $f_X(q)$ is a support function constructed
from the wave function of the bound state $X$, which is roughly
unity in the region 
$1/a_X$ $ \simlt$ $ q$ $\simlt$ $\overline{m}_b$;
see Fig.~\ref{BottomoniumEnergy}(a).

A characteristic feature of the bottomonium spectrum
in comparison to the Coulomb spectrum is that
the level spacings among the bottomonium
excited states are much wider than those of the Coulomb spectrum.
The level spacings of the Coulomb spectrum 
shrink quickly for higher levels.
The difference from the
Coulomb spectrum results from the linear rise of the potential.
See Figs.~\ref{Levels-in-C+L_Pot} and \ref{BottomoniumEnergy}(b).

The size $a_X$ of the state $X$ becomes larger
for higher excited states.
Then gluons with longer wavelengths can contribute
to the energy of $X$.
Positive contributions to the self-energies increase
rapidly since interactions of IR gluons become
stronger by the running of the coupling constant.
In Fig.~\ref{BottomoniumEnergy}(a) also $\alpha_s(q)$ is shown. 
We see that as the state varies from
$X= 1S$ to $3S$, the coupling $\alpha_s(q)$, close to the 
dumping scale of $f_X(q)$, grows rapidly.
According to Eq.~(\ref{Ecomposition}), as the integral region 
extends down to smaller $q$, the self-energy contributions grow
rapidly in comparison to the non-running case.
(Note that the non-running case corresponds to the
Coulomb spectrum.)
The self-energies push up the energy levels of the excited states considerably
and widen the level spacings among the excited states
as compared to the Coulomb case.

Hence, we may draw the following 
qualitative pictures for the energies of the bottomonium states:
\begin{itemize}
\item[(I)]
The energy of a bottomonium state mainly  consists of
(i) the $\overline{\rm MS}$ masses of $b$ and $\bar{b}$ 
$(=2\overline{m}_b)$, and 
(ii) contributions to the self-energies of $b$ and $\bar{b}$
from gluons with wavelengths $1/\overline{m} \simlt \lambda_g \simlt a_X$.
The latter contributions may be regarded as the difference between
the (state-dependent) constituent quark masses and the current quark masses.
\item[(II)]
The energy levels between excited states are widely separated
as compared to the Coulomb spectrum. 
This is because the self-energy contributions (from
$1/\overline{m} \simlt \lambda_g$ $\simlt a_X$)
grow rapidly as the physical size $a_X$ of the bound-state increases.
\end{itemize}
We conjecture that 
the conventional picture, that
the mass of a light hadron consists of the constituent quark masses,
can be viewed as an extrapolation of picture (I),
although it lies outside the validity range of perturbative QCD.
An important point is that it is clear in
perturbative QCD to which extent the prediction can be made
quantitative.
Namely, the potential energy is predictable
up to the linear potential $\sim \LQ^2 r$
[corresponding to the bound state energy
$\sim\LQ(\LQ/m_Q)^{1/3}$], while
the quadratic potential $\sim \LQ^3r^2$ cannot
be predicted.

\section{Potential-NRQCD EFT in Static Limit}
\label{Sec:pNRQCD}
\clfn

As already mentioned in Sec.~\ref{sec:Outline-pNRQCD},
pNRQCD is an EFT appropriate for describing interactions between a
static quark-antiquark system and IR gluons.
In fact, Wilsonian low energy EFTs 
can be constructed from full QCD
particularly clearly for this system, and
one can trace a number of steps in perturbative
expansions in $\alpha_s$.

In this section we mainly explain the construction and
structure of the EFT when we integrate out the
scale $1/r$.
We clarify how the renormalon is replaced
by a non-perturbative matrix element.
We also discuss briefly the case where
the binding energy scale $\sim -\alpha_s/r$
is further integrated out.

\subsection{Historical background}
\label{sec:HistoricalBkg}

Soon after the birth of perturbative QCD,
it was pointed out that perturbative
series of $V_{\rm QCD}(r)$ contains IR divergences
starting from three-loop order
(Appelquist, Dine, 
\begin{wrapfigure}{l}{70mm}
\vspace*{-2mm}
\hspace*{0mm}
\includegraphics[width=65mm]{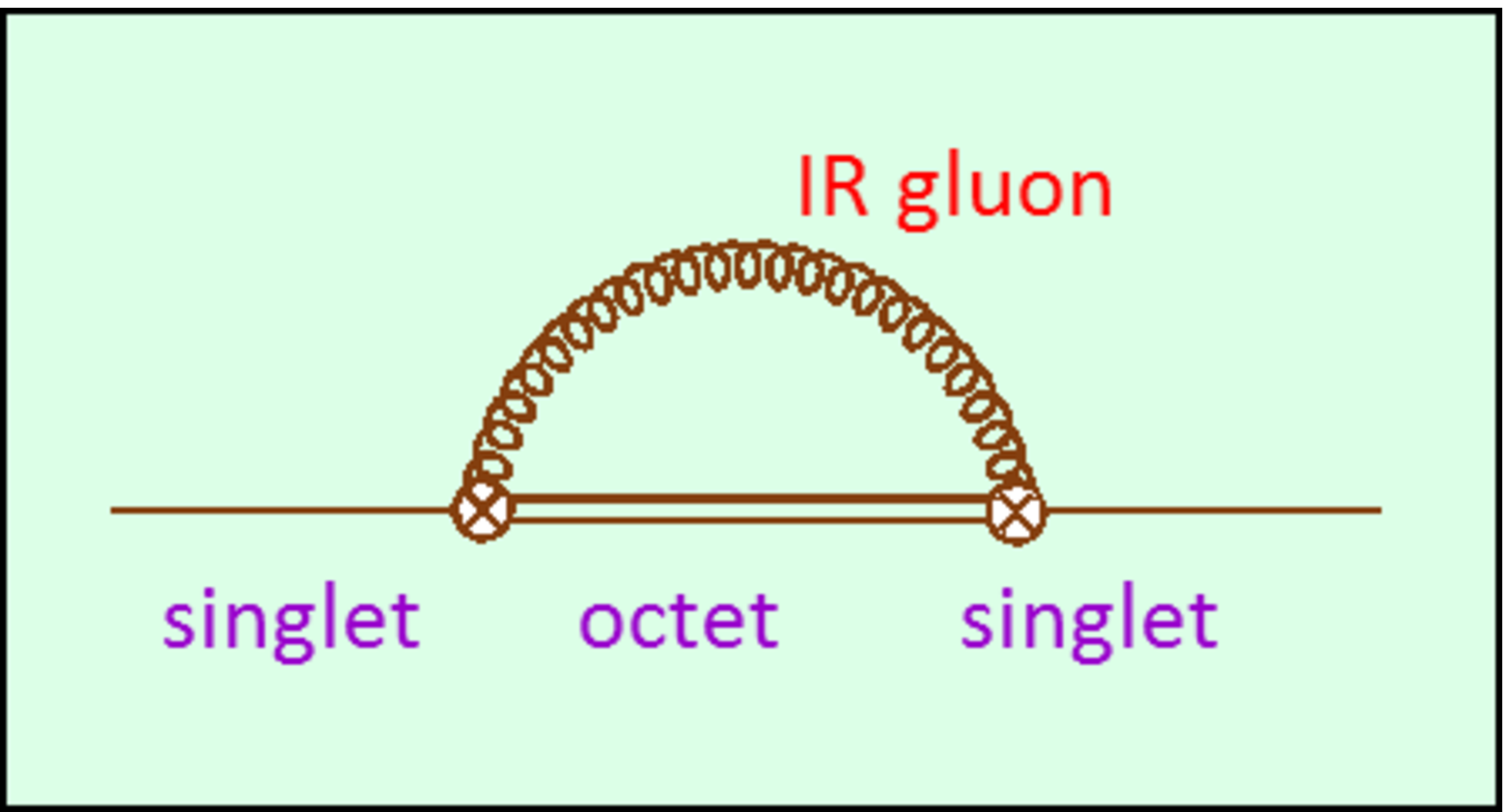}
\vspace*{-5mm}
\end{wrapfigure}
Muzunich).
The IR divergences originate from
degeneracy of the energy of the initial- or final-state
with those of intermediate states.
At the same time, the divergences are unique to
the  non-abelian gauge theory and is absent in QED.
Consider a process as depicted in the left figure:
the color-singlet static $Q\bar{Q}$ pair emits 
and re-absorbs an IR gluon; in the intermediate state,
the $Q\bar{Q}$ pair turns to a color-octet state due to color
conservation.
According to time-independent perturbation theory,
this process contributes to the energy of the system
as
\bea
\delta V_{\rm QCD}(r)\sim
\sum_{\vec{k}_g} \frac{|\bra{S}H_{\rm int}\ket{Og}|^2}
{(E_O + |\vec{k}_g|)-E_S}
=
\sum_{\vec{k}_g} \frac{|\bra{S}H_{\rm int}\ket{Og}|^2}
{|\vec{k}_g|+\Delta V} ,
\label{OriginIRdiv}
\eea
where
\bea
&&
E_S=V_S(r)=-C_F\frac{\alpha_s}{r} + {\cal O}(\alpha_s^2)
,
\label{V_S}
\\ &&
E_O=V_O(r)=\Bigl(\frac{C_A}{2}-C_F\Bigr)
\frac{\alpha_s}{r} + {\cal O}(\alpha_s^2)
,
\label{V_O}
\\ &&
\Delta V=E_O-E_S=\frac{C_A}{2}
\frac{\alpha_s}{r} + {\cal O}(\alpha_s^2)
\label{Pert-DeltaV}
\eea

\noindent
denote, respectively, the energy of the singlet state,
that of the octet state, and the difference of the two energies.
(The Casimir operator for the adjoint represesntation
\begin{wrapfigure}{l}{55mm}
\vspace*{-2mm}
\hspace*{0mm}
\includegraphics[width=5cm]{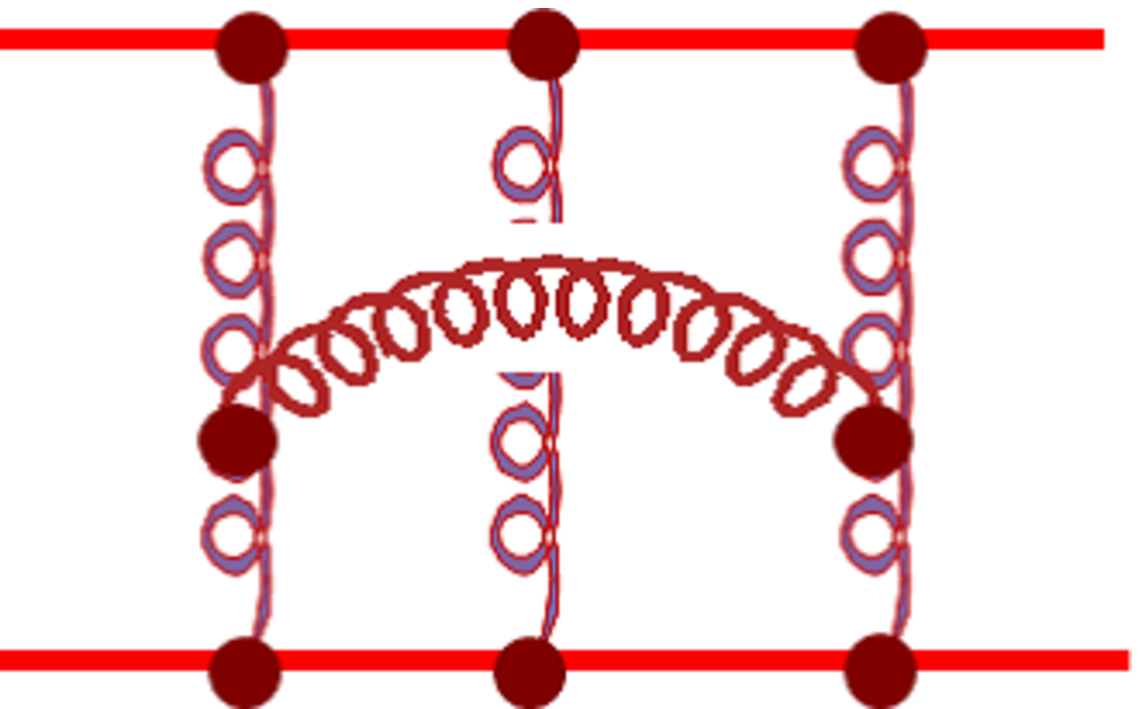}
\end{wrapfigure}
is defined by $T^a_{\rm adj}T^a_{\rm adj}=C_A {\bf 1}$, and
$C_A=3$ for QCD.)
In eq.~(\ref{OriginIRdiv}),
if we expand $(|\vec{k}_g|+\Delta V)^{-1}$ in $\Delta V\propto \alpha_s$, 
we find that
an IR divergence emerges corresponding to the three-loop diagram
shown in the left figure.
If we retain $\Delta V$ in the denominator, it becomes
IR finite.
Thus, the energy difference between the initial (final) and intermediate
states regularizes the divergence, if we do not expand in
$\alpha_s$.
We may compare this feature with the collinear divergence in the
process where an electron emits a photon:
if the electron propagator $(p_e^2-m_e^2)^{-1}$ is expanded in $m_e$,
the electron self-energy diagram
exhibits a collinear divergence, while it behaves as
$\log (E_e/m_e)$ if we do not expand in $m_e$.\footnote{
The IR divergences in $V_{\rm QCD}(r)$ are different
from the usual IR divergences which cancel between virtual
corrections and real emission processes according to
Kinoshita-Lee-Nauenberg theorem, since
there are no real emission processes contributing to $V_{\rm QCD}(r)$.
The mechanism for the absence of IR divergence is closer to that of the
collinear divergence for finite electron mass, as described here.
}

The above argument indicates existence of a non-trivial
IR dynamics in this system, and pNRQCD is suited
to clarify its nature.
It is also closely related to the Lamb shift in 
QED bound states.

\subsection{Basic concept of pNRQCD for static quarks}
\label{sec:Concept-pNRQCD}

We consider a
system composed of a static $Q\bar{Q}$ pair and
IR gluons.\footnote{We may also include
massless quarks, which can be treated similarly
to gluons.}
We assume that
$1/r \gg \Delta V(r)$\footnote{This is equivalent to 
$\frac{1}{2}C_A\alpha_s(1/r)\ll 1$, which holds at
sufficiently small $r$.}
and simultaneously energies 
of IR gluons satisfy $E_g \simlt \Delta V(r)$.
\begin{wrapfigure}{l}{45mm}
\vspace*{-5mm}
\hspace*{0mm}
\includegraphics[width=7cm]{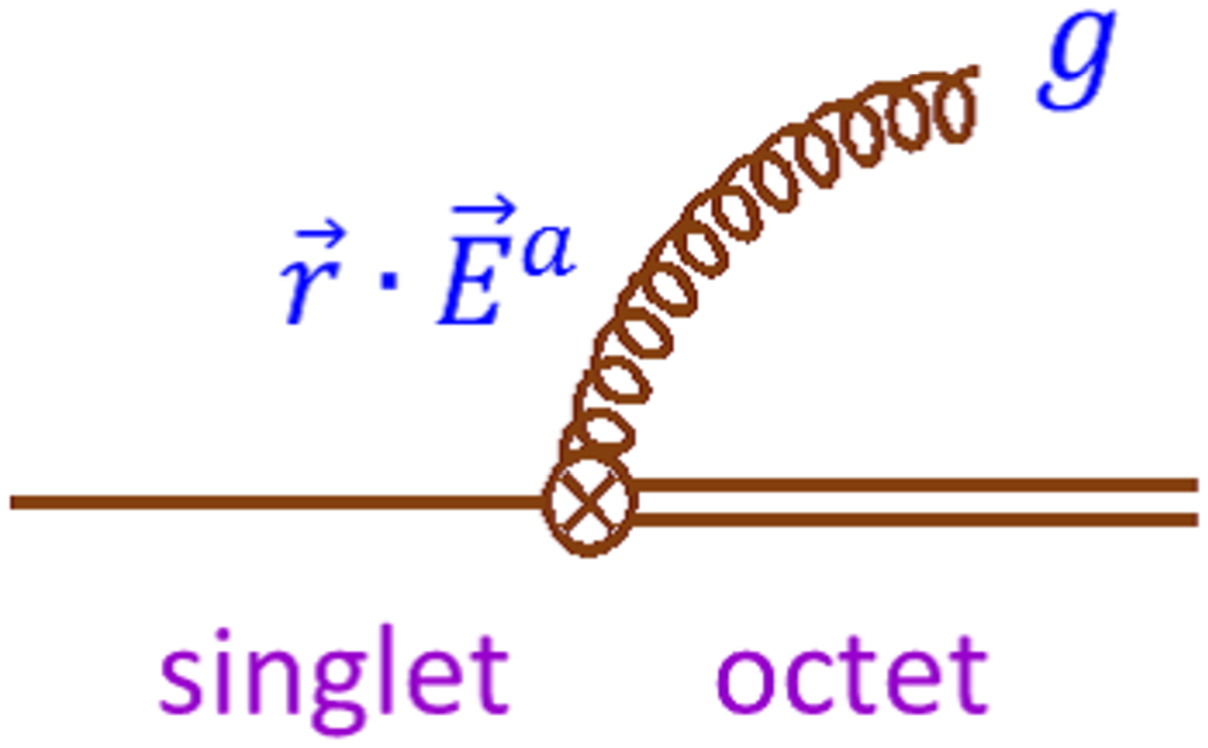}
\end{wrapfigure}
The EFT describes dynamics such as the one shown in the left figure,
namely the $Q\bar{Q}$ system emits or absorbs IR gluons
whose energies are comparable to or smaller than
energy differences of 
different $Q\bar{Q}$ states.
We choose the factorization scale $\mu_f$ as 
\bea
\Delta V(r) \ll \mu_f \ll \frac{1}{r}
\eea

\begin{wrapfigure}{l}{25mm}
\vspace*{-10mm}
\hspace*{0mm}
\includegraphics[width=22mm]{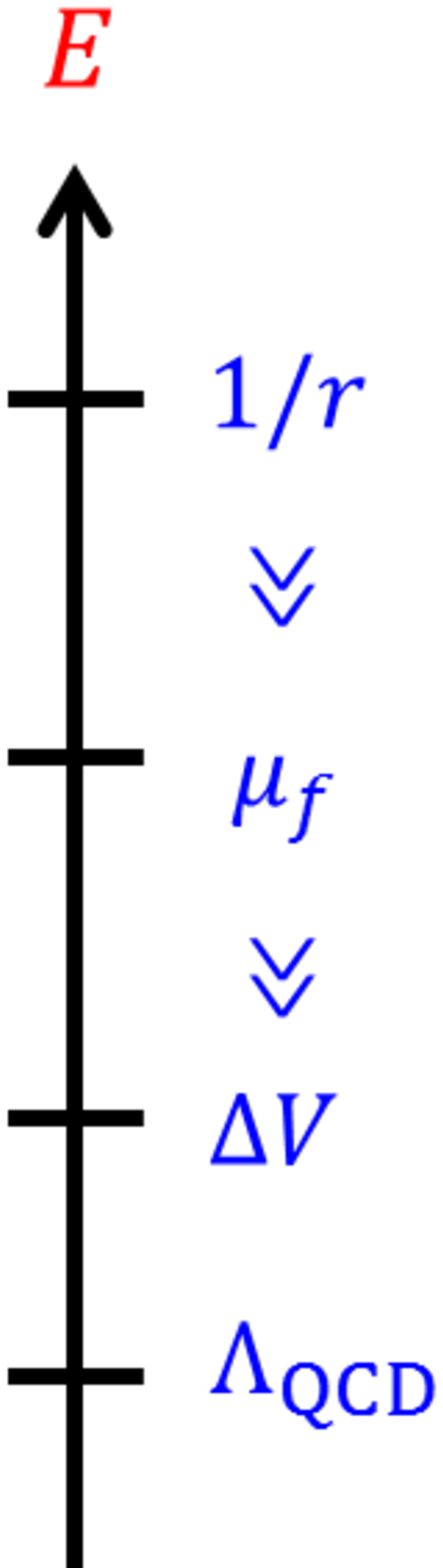}
\vspace*{-15mm}
\end{wrapfigure}
\noindent
and integrate out dynamical degrees of freedom above
the scale $\mu_f$.
The assumed scale 
hierarchy is shown in the left figure.

The dynamical degrees of freedom of the EFT, which
remain after integrating out the higher modes, are written in 
terms of the following quantum fields:
\bea
&&
\mbox{$Q\bar{Q}$ composite fields}
\left\{
\begin{array}{ll}
\text{color-singlet field}: &
~S(\vec{X},\vec{r};t),
\\
\text{color-octet field}: &
O(\vec{X},\vec{r};t),
\end{array}\right.
\\ &&
\mbox{IR gluon}:~ 
A_0(t,\vec{x}), \,\vec{A}(t,\vec{x}) .
\eea
We denote the positions of $Q$ and $\bar{Q}$ as
$\vec{X}\pm \vec{r}/2$, namely, $\vec{X}$ stands for the
c.m.\ coordinate of $Q$ and $\bar{Q}$, 
while $\vec{r}$ stands for the relative coordinate.
The $Q\bar{Q}$ composite fields are bilocal in space coordinates
but local in time coordinate.

The expansion parameters of this
EFT are as follows.
\begin{itemize}
\item[(i)]
Since $r E_g = rp_g \ll 1$, the time and spatial derivatives
acting on gauge fields, such as
$\vec{r}\cdot \partial_t \vec{A}$ and $\vec{r}\cdot \vec{\partial} A_0$,
are considered as ``small.''
These are induced by multipole expansion of 
$A_\mu(t,\vec{X}\pm \vec{r}/2)$ in $\vec{r}$.

Intuitively, since IR gluons have long wavelengths,
$\lambda_g (=E_g^{-1})\gg r$, 
they cannot resolve internal structures of the $Q\bar{Q}$ state
but rather couple to color multipoles.

\item[(ii)]
On the other hand, since $r p_r\sim {\cal O}(1)$ by uncertainty
principle,
we do not expand the $Q\bar{Q}$ fields $S$ and $O$ in $\vec{r}$.

\item[(iii)]
Since $r E_{S,O}\sim r \Delta V \ll 1$, expansions
such as $(r\partial_t)^n S$ are legitimate.
Nevertheless, we can eliminate more than one time derivative $\partial_t$
acting on $S$ or $O$ using the equation of motion.
\end{itemize}

\subsection{Derivation of Lagrangian and Feynman rules}

There are two methods for constructing the 
Lagrangian of pNRQCD EFT:
\begin{itemize}
\item[1.]
List up all the operators which are consistent with the symmetry
of the theory.
\item[2.]
Start from $\langle W[A_\mu]\rangle$ in full QCD;
redefine fields as appropriate for this system;
perform multipole expansion; 
supplement operators, which originate from scales
above $\mu_f$.
\end{itemize}
In this lecture we explain the latter method.\footnote{
This is a simplified version of a more solid method
based on the asymptotic expansion (integration by regions) 
of Feynman diagrams; 
see Appendix~\ref{AppC}.
}

To set up the necessary formulation, we express the
Wilson loop using a bilocal field.
(Note that we expressed the Wilson loop using static fields $\psi$ and
$\chi$ in Sec.~\ref{Sec:MeaningV_QCD}.)
We first introduce a bilocal field
$\psi (\vec{x},\vec{y};t)$, which is a complex scalar field 
given as an $N$ by $N$ matrix in the $SU(N)$ color space, and which
transforms in the same way as $\psi\chi^\dagger$, namely,
\bea
\psi (\vec{x},\vec{y};t)\sim \psi(t,\vec{x})\,
\chi^\dagger(t,\vec{y})
.
\eea
Using this bilocal field, we can express the Wilson loop
eq.~(\ref{WilsonLoop})
as
\bea
&&
W[A_\mu]\, \delta^3(\vec{x}_1-\vec{x}_1^{\,\prime})
\delta^3(\vec{x}_2-\vec{x}_2^{\,\prime})
\nonumber\\&&
=\int\!
{\cal D}\psi{\cal D}\psi^\dagger\,
\exp\!\left[ i\!\int \! dt\, d^3\vec{x}\, d^3\vec{y}\,
{\cal L}[\psi,\psi^\dagger,A_\mu]
 \right]
\nonumber\\&&
~~~~~~~~~
\times 
{\rm Tr}\!\left[ 
\psi^\dagger(\vec{x}_1,\vec{x}_2;0)\,
\phi(\vec{x}_1,\vec{x}_2;0)\right]
\, 
{\rm Tr}\!\left[ 
\psi(\vec{x}_1^{\,\prime},\vec{x}_2^{\,\prime};T)\,
\phi^\dagger(\vec{x}_1^{\,\prime},\vec{x}_2^{\,\prime};T)
\right]
,
\label{WilsonLoopInBilocalF}
\eea
with
\bea
&&
{\cal L}[\psi,\psi^\dagger,A_\mu]=
{\rm Tr}\!\left[ 
\psi^\dagger(\vec{x},\vec{y};t)\, iD_t
\psi(\vec{x},\vec{y};t)
\right]
\nonumber \\ &&
~~~~~~~~~~~~~~~
= {\rm Tr}[ 
\psi^\dagger(\vec{x},\vec{y};t)\, i\partial_t
\psi(\vec{x},\vec{y};t)
+g\, \psi^\dagger(\vec{x},\vec{y};t)\, A_0(t,\vec{x})\,
\psi(\vec{x},\vec{y};t)
\nonumber \\ &&
~~~~~~~~~~~~~~~~~~~~~
-g\, \psi^\dagger(\vec{x},\vec{y};t)\,
\psi(\vec{x},\vec{y};t)\, A_0(t,\vec{y})
].
\eea
Note that the gauge transformation of the bilocal field
is given by
\bea
\psi(\vec{x},\vec{y};t)
\to
U(\vec{x},t)\,
\psi(\vec{x},\vec{y};t)\,
U^\dagger(\vec{y},t) ,
\eea
so that the covariant derivative
in
the above Lagrangian is also a bilocal operator.
The expression of the Wilson loop in terms of 
$\psi(\vec{x},\vec{y};t)$ can be justified in a similar manner
as in Sec.~\ref{Sec:MeaningV_QCD}, by examining
the time evolution of the Wilson loop in terms of
a differential equation.

We decompose the bilocal field into the singlet
and octet components as
\bea
\psi(\vec{x},\vec{y};t) =
\phi(\vec{x},\vec{y};t)S(\vec{X},\vec{r};t)
+
\phi(\vec{x},\vec{X};t)O(\vec{X},\vec{r};t)
\phi(\vec{X},\vec{y};t) ,
\eea
where $S$ is proportional to the identity matrix and $O$ is
traceless.
Hence, the gauge transformations of these fields are given by
\bea
&&
S(\vec{X},\vec{r};t) \to S(\vec{X},\vec{r};t) ,
\\ &&
O(\vec{X},\vec{r};t)\to 
U(\vec{X},t)\,
O(\vec{X},\vec{r};t)\,
U^\dagger(\vec{X},t)
.
\eea
The singlet and octet fields are defined
such that their gauge transformations
depend only on $\vec{X}$ and
not on $\vec{r}$.
In this way, we can maintain gauge invariance
of the theory explicitly at each order of multipole expansion
in $\vec{r}$.
(Otherwise different orders of expansion in $\vec{r}$
mix under gauge transformation.)
This helps greatly to simplify the interactions of
the EFT.

Let us expand ${\cal L}[\psi,\psi^\dagger,A_\mu]$
in $\vec{r}$.
The expansion of the color string can be computed as
\bea
&&
\phi(\vec{x},\vec{y};t)=
{\rm P}\, \exp\left[ ig\int_{-1/2}^{1/2} ds\, \vec{r}\!\cdot\!
\vec{A}(t,\vec{X}+s\vec{r}) \right]
\nonumber\\ &&
~~~~~~~~~~~
= {\bf 1} + ig\int_{-1/2}^{1/2} ds\, \, \vec{r}\!\cdot\!
\vec{A}(t,\vec{X}+s\vec{r}) + {\cal O}(r^2)
\nonumber\\ &&
~~~~~~~~~~~
= {\bf 1} + ig\, \vec{r}\!\cdot\!
\vec{A}(t,\vec{X}) + {\cal O}(r^2) .
\eea
Similarly, we obtain
\bea
&&
\phi(\vec{x},\vec{X};t)
= {\bf 1} + \frac{1}{2}ig\, \vec{r}\!\cdot\!
\vec{A}(t,\vec{X}) + {\cal O}(r^2) ,
\\ &&
\phi(\vec{X},\vec{y};t)
= {\bf 1} + \frac{1}{2}ig\, \vec{r}\!\cdot\!
\vec{A}(t,\vec{X}) + {\cal O}(r^2) .
\eea
We may then express $\psi$ by $S$ and $O$ in expansion in $\vec{r}$:
\bea
&&
\psi(\vec{x},\vec{y};t) =
S(\vec{X},\vec{r};t)
+ ig\, \vec{r}\!\cdot\!
\vec{A}(t,\vec{X}) 
S(\vec{X},\vec{r};t)
\nonumber\\ &&
~~~~~~~~~~~~~~~
+O(\vec{X},\vec{r};t)
+ \frac{1}{2}ig\, \Bigl\{ \vec{r}\!\cdot\!
\vec{A}(t,\vec{X}),\,
O(\vec{X},\vec{r};t)\Bigr\}
+ {\cal O}(r^2) .
\eea
By substituting this to ${\cal L}$ and expanding
in $\vec{r}$, we obtain the following 
interaction terms up to ${\cal O}(r)$:
\bea
&&
{\cal L}_{O S}=
g\,{\rm Tr} \Bigl[ O^\dagger\, \vec{r}\!\cdot\! \vec{E}\, S\Bigr] 
+g\,{\rm Tr} \Bigl[ O \,\vec{r}\!\cdot\! \vec{E}\, S^\dagger\Bigr] ,
\\&&
{\cal L}_{S S}=
{\rm Tr} \Bigl[ S^\dagger\,i\partial_t\, S\Bigr] ,
\\&&
{\cal L}_{O O}=
{\rm Tr} \Bigl[ O^\dagger\,iD_t O\Bigr] 
+ \frac{1}{2}g\,
{\rm Tr} \Bigl[ O^\dagger O\, \vec{r}\!\cdot\! \vec{E}\Bigr] 
+ \frac{1}{2}g\,
{\rm Tr} \Bigl[ O O^\dagger\, \vec{r}\!\cdot\! \vec{E}\Bigr] 
,
\eea
where
$\vec{E}=-\partial_t \vec{A}-\vec{\partial}A_0-ig[A_0,\vec{A}]$
represents the color electric field.
All the gauge fields are evaluated at $(t,\vec{X})$.

We should add to the above interactions the
``potential terms'' 
\bea
{\cal L}_{\rm pot}= -
{\rm Tr} \Bigl[ S^\dagger\,V_S(r)\, S
+O^\dagger\,V_O(r)\, O
\Bigr] ,
\eea
\begin{wrapfigure}{l}{40mm}
\vspace*{2mm}
\hspace*{3mm}
\includegraphics[width=2.8cm]{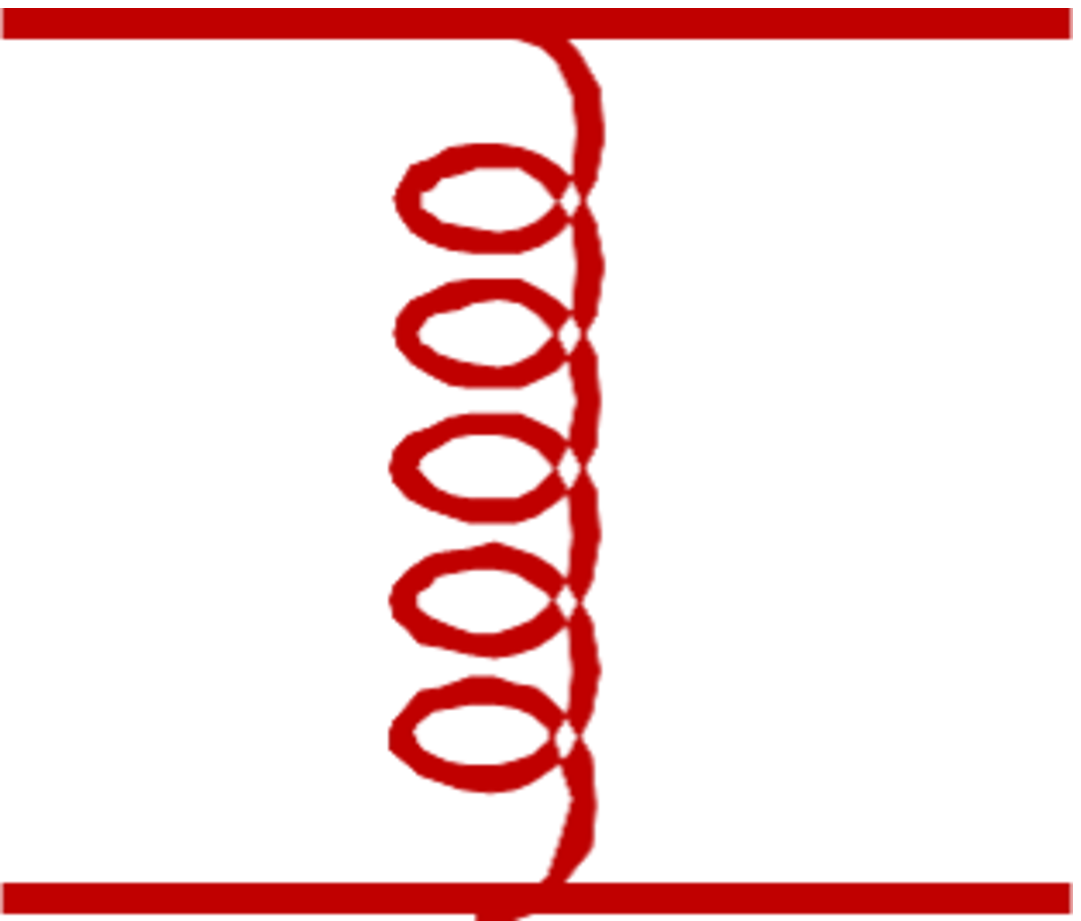}
\end{wrapfigure}
where $V_S(r)$ and $V_O(r)$ represent the singlet and octet
potentials, respectively.
If we evaluate them in expansions in $\alpha_s$,
the leading terms are given by
Coulomb potentials; see eqs.~(\ref{V_S}) and (\ref{V_O}).
These arise as contributions from scales above $\mu_f$,
from the one-gluon exchange diagram shown left with
$E_g>\mu_f$.
Intuitively, the singular behaviors as $r\to 0$
cannot arise from IR gluons whose wavelengths
are larger than $r$, but  arise from UV gluons whose
wavelengths are smaller than $r$.
In practice, we identify ${\cal L}_{\rm pot}$ by
matching to full QCD or by using the integration-by-regions
method.

Collecting all the terms, we find
\bea
&&
{\cal L}=
{\rm Tr} \Bigl[ S^\dagger\,\{ i\partial_t-V_S(r)\}\, S
+O^\dagger\,\{ iD_t-V_O(r)\}\, O
\nonumber\\&&
~~~~~~~~~~~~
+g\,S^\dagger\,O \,\vec{r}\!\cdot\! \vec{E} +
g\,O^\dagger\, \vec{r}\!\cdot\! \vec{E}\, S
+ \frac{1}{2}g\,O^\dagger O\, \vec{r}\!\cdot\! \vec{E}
+ \frac{1}{2}g\,OO^\dagger \, \vec{r}\!\cdot\! \vec{E}
\Bigr]
\nonumber\\ &&
~~~~~~
 + {\cal O}(r^2) 
,
\eea
Note that $d^3\vec{x}\, d^3\vec{y}=d^3\vec{X}\, d^3\vec{r}$
and $\delta^3(\vec{x})\,\delta^3(\vec{y})=
\delta^3(\vec{X})\,\delta^3(\vec{r})$.
As stated, the interactions are gauge invariant at
each order of $r$, and
the leading interaction of $S$ and
gauge field is of dipole type $\sim gS^\dagger
O^a \,\vec{r}\!\cdot\! \vec{E}^a$.

Feynman rules of pNRQCD can be obtained in a
straightforward manner from the above Lagrangian.
These are given as follows.
\begin{center}
\includegraphics[width=15cm]{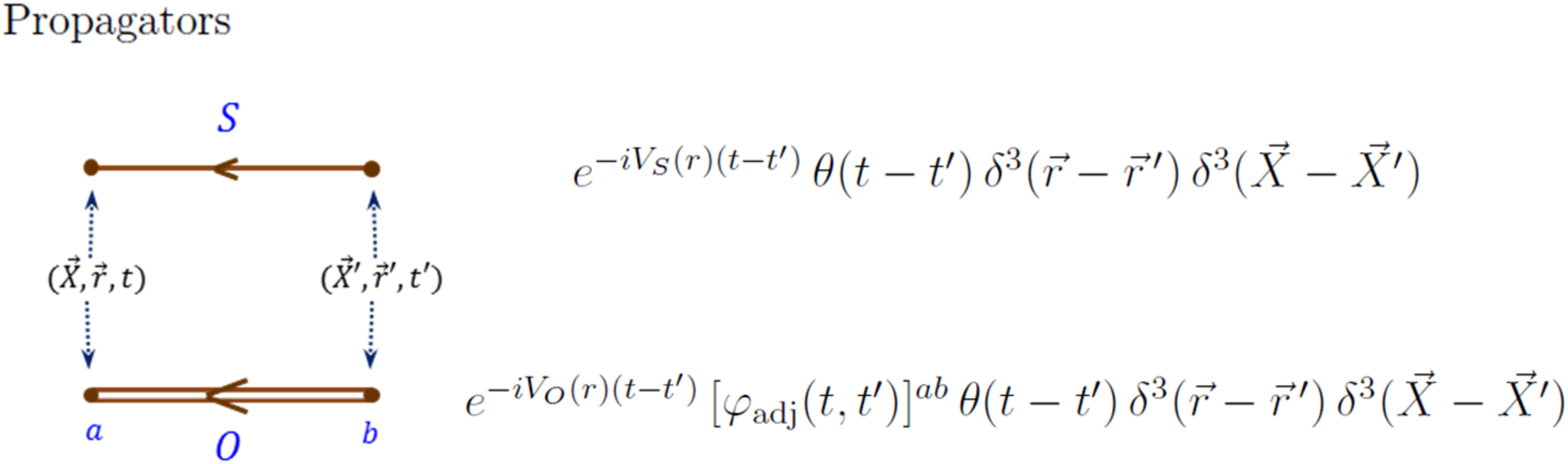}
\\
\includegraphics[width=15cm]{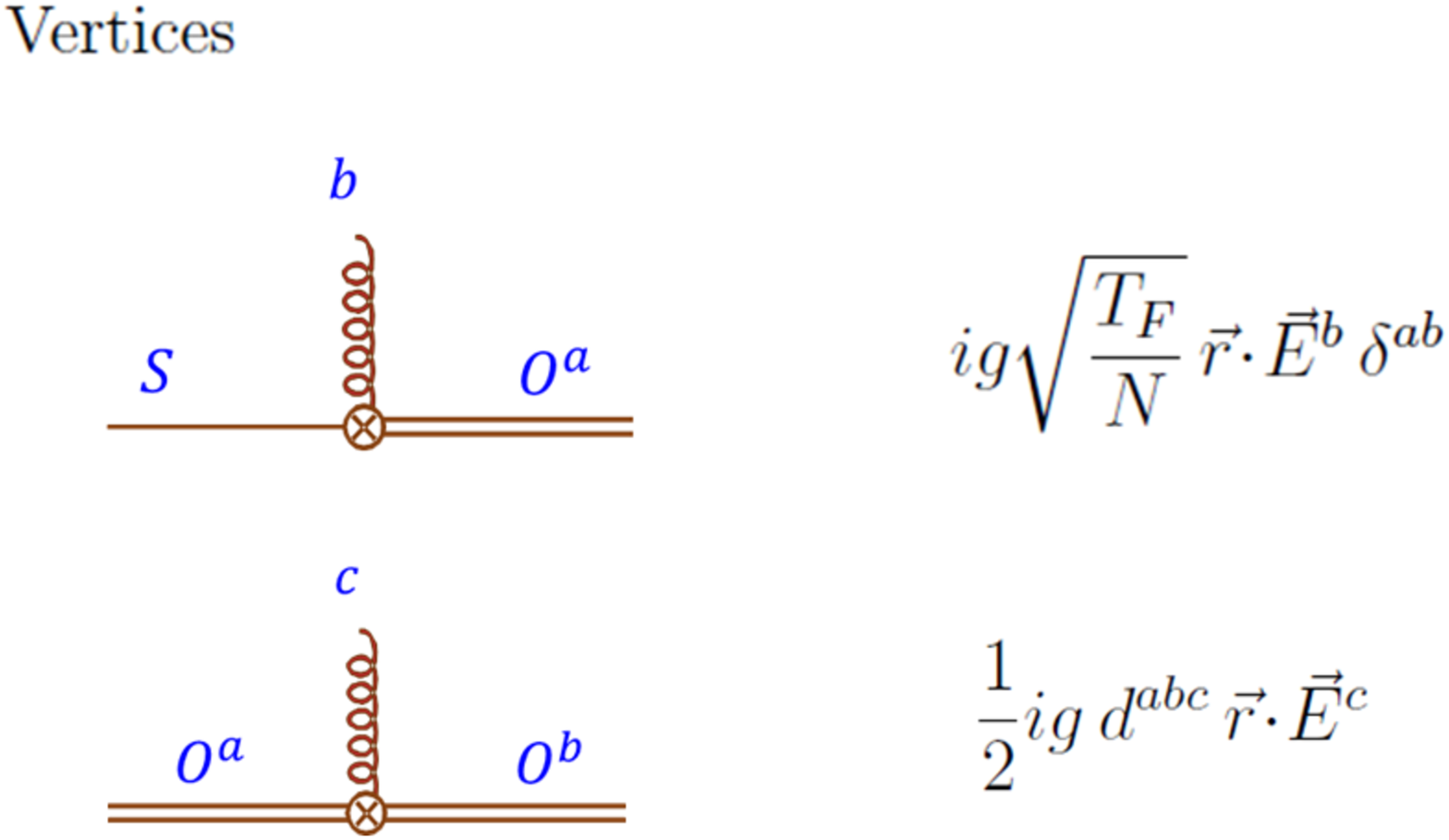}
\end{center}
Here, $T_F=\frac{1}{2}$ denotes the trace normalization 
in the fundamental representation;
$d^{abc}=2{\rm Tr}(\{T^a_F,T^b_F\}T^c_F)$ denotes the 
symmetric invariant tensor;
color string for the adjoint representation is given by
$
\varphi_{\rm adj}(t,t')=
{\rm T}\, \exp\Bigl[ig\int_{t'}^t d\tau\,
A_0^c(\tau,\vec{X})\,T^c_\text{adj}
\Bigr]
$.

\subsection[Computation of $V_{\rm QCD}(r)$ in pNRQCD]{\boldmath Computation of $V_{\rm QCD}(r)$ in pNRQCD}

Using the expression of the Wilson loop and Feynman rules derived
above,
we can compute 
the expectation value of the Wilson loop
in expansion in $\vec{r}$:
\bea
&&
\langle W[A_\mu]\rangle\,\delta^3(\vec{X}-\vec{X}')\,
\delta^3(\vec{r}-\vec{r}\,')
\nonumber\\ &&
=
{\rm Tr}\,
\bra{0}S(\vec{X},\vec{r};T)\,  S^\dagger(\vec{X}',\vec{r}\,';0)
\ket{0}+{\cal O}(r^3)
\nonumber\\ &&
=N \, e^{-iV_S(r)T}\,\delta^3(\vec{X}-\vec{X}')\,
\delta^3(\vec{r}-\vec{r}\,')
\nonumber\\ &&
~~~~~~
\times \biggl[ \,{\bf 1}-g^2\,\frac{T_F}{N}\int_0^T \!\!\! dt
\int_0^t \!\! dt'\, e^{-i(V_O-V_S)(t-t')}
\nonumber\\ &&
~~~~~~~~~~~~~~~~~~~~~~
\times
\bra{0}\vec{r}\!\cdot\! \vec{E}^a(t,\vec{X})\,
[\varphi_{\rm adj}(t,t')]^{ab}\,
\vec{r}\!\cdot\! \vec{E}^b(t',\vec{X})\ket{0} \biggr]
+{\cal O}(r^3)
,
\eea
where $N=3$ for QCD.
The order $r^0$ and $r^2$ terms in the last line correspond to the diagrams
below.
\begin{center}
\hspace*{-15mm}
\includegraphics[width=10cm]{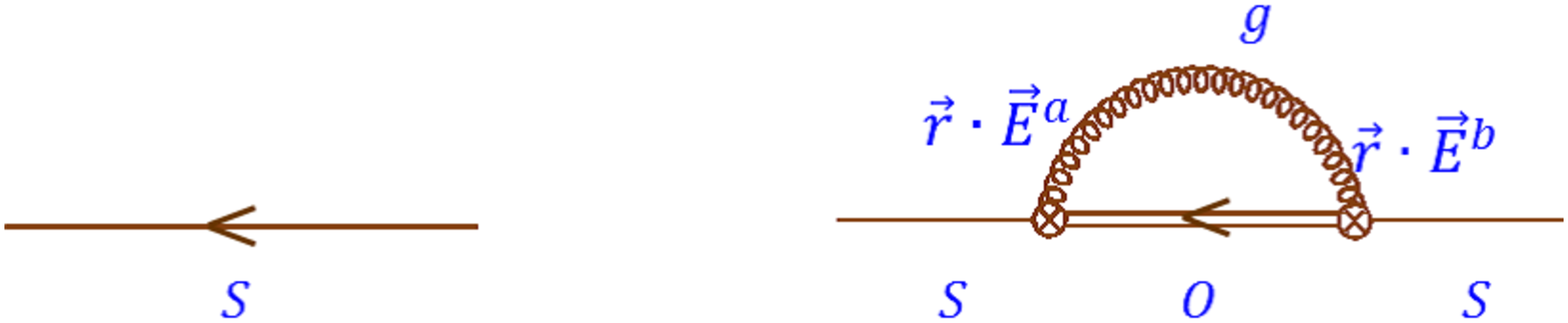}
\end{center}
On the other hand, in full QCD, we have
\bea
\langle W[A_\mu]\rangle={\rm const.}\times
e^{-iV_{\rm QCD}(r)T}
~~~~~
\text{as}~~~~T\to\infty .
\eea
Hence, by equating both quantities,
we obtain the matching relation of the
static potential as
\bea
&&
V_{\rm QCD}(r)=V_S(r)
-i g^2\,\frac{T_F}{N}\!\int_0^\infty \!\!\! dt
\, e^{-i(V_O-V_S)t}
\langle\,\vec{r}\!\cdot\! \vec{E}^a(t,\vec{0})\,
\varphi_{\rm adj}(t,0)^{ab}\,
\vec{r}\!\cdot\! \vec{E}^b(0,\vec{0})\,\rangle 
\nonumber\\ &&
~~~~~~~~~~~~~~
+{\cal O}(r^3)
.
\label{MatchingRel}
\eea
Note that, in deriving this relation, we only used
expansion in $\vec{r}$ but 
did not
use expansion in $\alpha_s$.

Let us call the second term of eq.~(\ref{MatchingRel})
as $V_{\rm IR}(r)$ and proceed with
computation of this quantity, using expansion in $\alpha_s$
and in dimensional regularization.
(Expansion in $\alpha_s$ makes sense only in the case
that the relevant scale of $V_{\rm IR}$ is much larger than
$\LQ$ scale, but here we are more interested in the
formal structure of $V_{\rm IR}$ in $\alpha_s$ expansion.)
Since $\varphi_{\rm adj}(t,0)^{ab}=\delta^{ab}+
{\cal O}(\alpha_s)$, we drop the ${\cal O}(\alpha_s)$
term and compute
\bea
V_{\rm IR}(r)\approx
-i g^2\,\frac{T_F}{N}\!\int_0^\infty \!\!\! dt
\, e^{-it\,\Delta\! V(r)}\,
\langle\,\vec{r}\!\cdot\! \vec{E}^a(t)\,
\vec{r}\!\cdot\! \vec{E}^a(0)\,\rangle 
~;~~~~\Delta V=V_O-V_S .
\label{V_IR}
\eea
The following formulas can be used
for the computation:
\bea
&&
\langle\,\vec{r}\!\cdot\! \vec{E}^a(t)\,
\vec{r}\!\cdot\! \vec{E}^a(0)\,\rangle 
=\frac{r^2}{d}
\langle\,\vec{E}^a(t)\cdot
\vec{E}^a(0)\,\rangle ,
\\ &&
\langle\,{E}^{ia}(x)\,
{E}^{ja}(y)\,\rangle 
=\delta^{aa}\,\langle\,
\bigl(\partial^0A^i(x)-\partial^iA^0(x)\bigr)
\bigl(\partial^0A^j(y)-\partial^jA^0(y)\bigr)
\rangle+{\cal O}(\alpha_s)
\nonumber\\ &&
~~~~~~~~~~~~~~~~~~~~~
=-i\frac{C_FC_A}{T_F}\int \frac{d^Dk}{(2\pi)^D}\,
\frac{e^{ik\cdot (x-y)}}{k^2+i0}\,
[k^ik^j-(k_0)^2\delta^{ij}]
+{\cal O}(\alpha_s)
,
\eea
where $d=D-1=3-2\epsilon$ represents the
dimensions of space.
Integrating over $k_0$, we obtain
\bea
&&
\langle\,\vec{E}^a(t)\cdot
\vec{E}^a(0)\,\rangle 
\approx
-i\frac{C_FC_A}{T_F}\int \frac{d^Dk}{(2\pi)^D}\,
\frac{e^{ik_0t}}{k^2+i0}\,
(\vec{k}^2-d\cdot k_0^2)
\nonumber\\ &&
~~~~~~~~~~~~~~~~~~~
=-\frac{C_FC_A}{T_F}
\Biggl[
\frac{1-d}{2}\int\! \frac{d^d\vec{k}}{(2\pi)^d}\,
|\vec{k}|\,e^{-i|\vec{k}|t}
-id\cdot\delta(t)
\int\! \frac{d^d\vec{k}}{(2\pi)^d}\,1
\Biggr]
.
\nonumber\\
\label{EEcorrelator}
\eea
In dimensional regularization the
second term vanishes, since it is proportional to
a scaleless integral.
Hence, we find
\bea
V_{\rm IR}(r)\approx
-4\pi\alpha_s\,\bar{\mu}^{2\epsilon}\,\frac{d-1}{2d}\,
C_F\, r^2\int\!\frac{d^d\vec{k}}{(2\pi)^d}\,
\frac{k}{k+\Delta V}
~;~~~k=|\vec{k}| ,
\label{PertNonLocalCond}
\eea
cf.~eqs.~(\ref{renormalization}) and (\ref{mubar}).
This has a form, which we anticipated in Sec.~\ref{sec:HistoricalBkg}.

\subsection{Matching to QCD}
\label{sec:MatchingToQCD}

Let us first examine a matching to full QCD in naive
expansion in $\alpha_s$.
Namely, we expand the integrand of eq.~(\ref{PertNonLocalCond})
in $\alpha_s$ before integrating over $\vec{k}$.
If we do so, the integral vanishes, since 
$
\Delta V\approx \frac{C_A}{2}\,\frac{\alpha_s}{r}$
[c.f., eq.~(\ref{Pert-DeltaV})], and
\bea
\int\!\frac{d^d\vec{k}}{(2\pi)^d}\,k^P=0
\eea
for an integer $P$.
(Scaleless integrals vanish in dimensional regularization.)
In fact, the second term of the matching relation
(\ref{MatchingRel}) [$V_{\rm IR}(r)$]
vanishes to all orders in $\alpha_s$,
if we expand it in $\alpha_s$ before integration,
since only scaleless integrals appear.
Consequently, we find that $V_S(r)$ coincides with
the naive expansion of $V_{\rm QCD}(r)$ in $\alpha_s$:
\bea
V_S(r) = V_{\rm QCD}(r)\Bigr|_\text{exp.\ in $\alpha_s$}
.
\label{V_SfromV_QCD}
\eea

Next we perform a matching consistent with the concept of
pNRQCD, which we explained in Sec.~\ref{sec:Concept-pNRQCD}.
There, we have specified the expansion parameters of the EFT.
In particular, since $E_g\simlt \Delta V$, we should not
expand the integrand of 
eq.~(\ref{PertNonLocalCond}) by $\Delta V$.
If we perform the integration of eq.~(\ref{PertNonLocalCond})
as it is,
it contains a UV divergence and is evaluated to be\footnote{
One can use a formula
\bea
\int\!\frac{d^d\vec{k}}{(2\pi)^d}\,
\frac{k^n}{(k+a)^\nu}
= 2^{1-d}\pi^{-\nu/2}\,\frac{\Gamma(n+d)\Gamma(\nu-n-d)}
{\Gamma(d/2)\,\Gamma(\nu)}\,a^{d+n-\nu}
\eea
to evaluate the integral.
Note that 
\bea
\Delta V = \frac{C_A}{2}\,\frac{\alpha_s}{r}\,
(\bar{\mu}r)^{2\epsilon}\,\frac{\Gamma(\frac{1}{2}-\epsilon)}{\pi^{\frac{1}{2}-\epsilon}}
+O(\alpha_s^2)
\eea
in dimensional regularization.
}
\bea
&&
V_{\rm IR}(r)\approx
\frac{C_F\alpha_s}{\pi}\,\frac{r^2}{3}\,\Delta V^3\,
\biggl[
\frac{1}{\epsilon}-\gamma_E-\log\biggl(
\frac{\Delta V^2}{\pi \bar{\mu}^2} \biggr)+\frac{5}{3}
\biggr]
\nonumber\\&&
~~~~~~~~
\approx
\frac{C_FC_A^3\alpha_s^4}{24\pi r}\biggl[
\frac{1}{\epsilon}+8\log(\mu r)-2\log(C_A\alpha_s)+
\frac{5}{3}+6\gamma_E
\biggr]
.
\label{VIRinDimReg}
\eea

At this stage we list some known facts concerning
this result and the matching relation (\ref{MatchingRel}).
\begin{itemize}
\item[(I)]
On the right-hand side of eq.~(\ref{MatchingRel}), the
following two contributions cancel:
(1) the IR divergence $\sim \frac{\alpha_s^4}{r}\times\frac{1}{\epsilon}$
contained in the first term $V_S(r)$ [= $\alpha_s$-expansion of
$V_{\rm QCD}(r)$] at three loop, and
(2) the UV divergence $\sim \alpha_s \, r^2 \Delta\! V^3\,\frac{1}{\epsilon}$
contained in the second term $V_{\rm IR}(r)$
[eq.~(\ref{VIRinDimReg})].

\item[(II)]
The right-hand side of eq.~(\ref{MatchingRel}) altogether is
finite as $\epsilon \to 0$ and includes
$$C_F\alpha_s r^2\Delta\! V^3\,\log(r^2\Delta\!V^2)
\sim \frac{C_FC_A^3\alpha_s^4}{r}\,\log(C_A^2\alpha_s^2).$$
[Compare the argument on collinear divergence and $\log (E_e/m_e)$
in Sec.~\ref{sec:HistoricalBkg}.]

\item[(III)]
The right-hand side of eq.~(\ref{MatchingRel}) altogether 
coincides with the resummation of ladder-type diagrams for 
$V_{\rm QCD}(r)$
shown below, which was suggested by Appelquist, Dine and Muzunich
to remedy apparent IR divergences of $V_{\rm QCD}(r)$.

\begin{center}
\includegraphics[width=7cm]{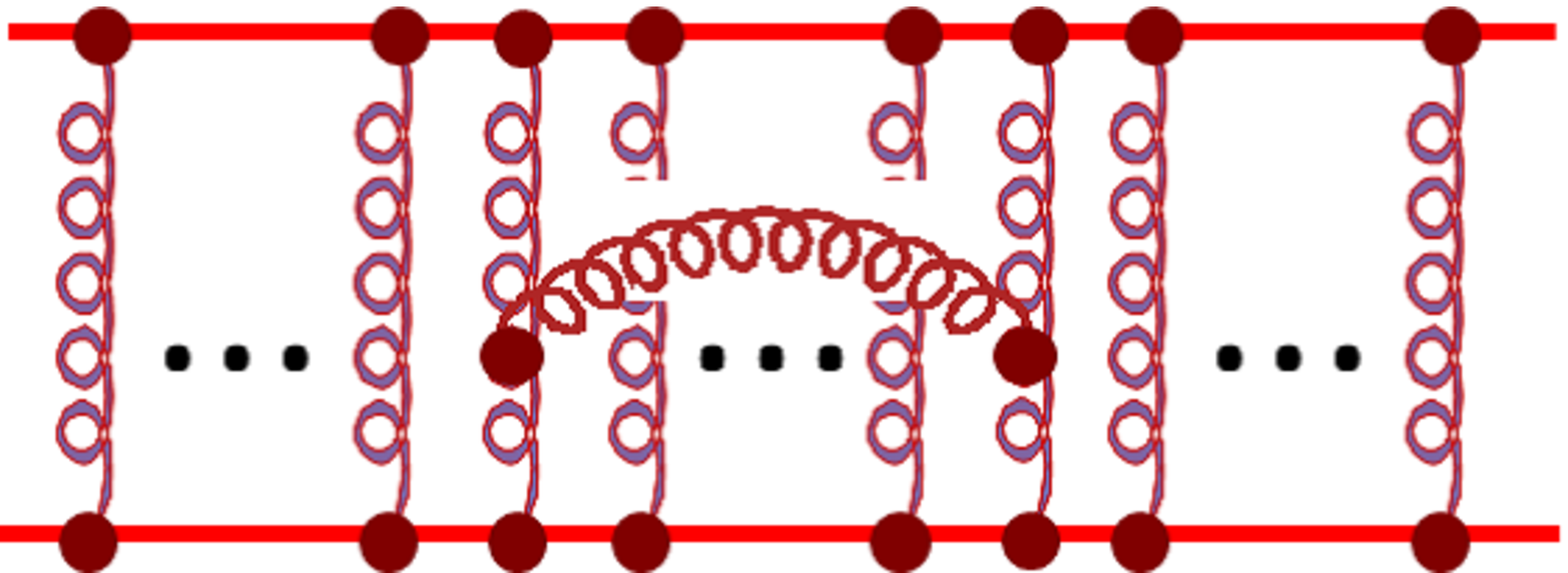}
\end{center}

\item[(IV)]
$V_{\rm QCD}(r)$ can be computed systematically in double
expansion in $\alpha_s$ and $\log \alpha_s$
using pNRQCD.

\end{itemize}

\subsection[Renormalization of Wilson coefficient and
$\mu_f$-independence of $V_{\rm QCD}(r)$]
{\boldmath Renormalization of Wilson coefficient and
$\mu_f$-independence of $V_{\rm QCD}(r)$}
\clfn

The IR divergence of the (bare) Wilson coefficient $V_S(r)$
and the UV divergence of $V_{\rm IR}(r)$ in pNRQCD
are considered as 
artifacts of dimensional regularization without
a cutoff 
\begin{wrapfigure}{l}{65mm}
\vspace*{-2mm}
\hspace*{3mm}
\includegraphics[width=6cm]{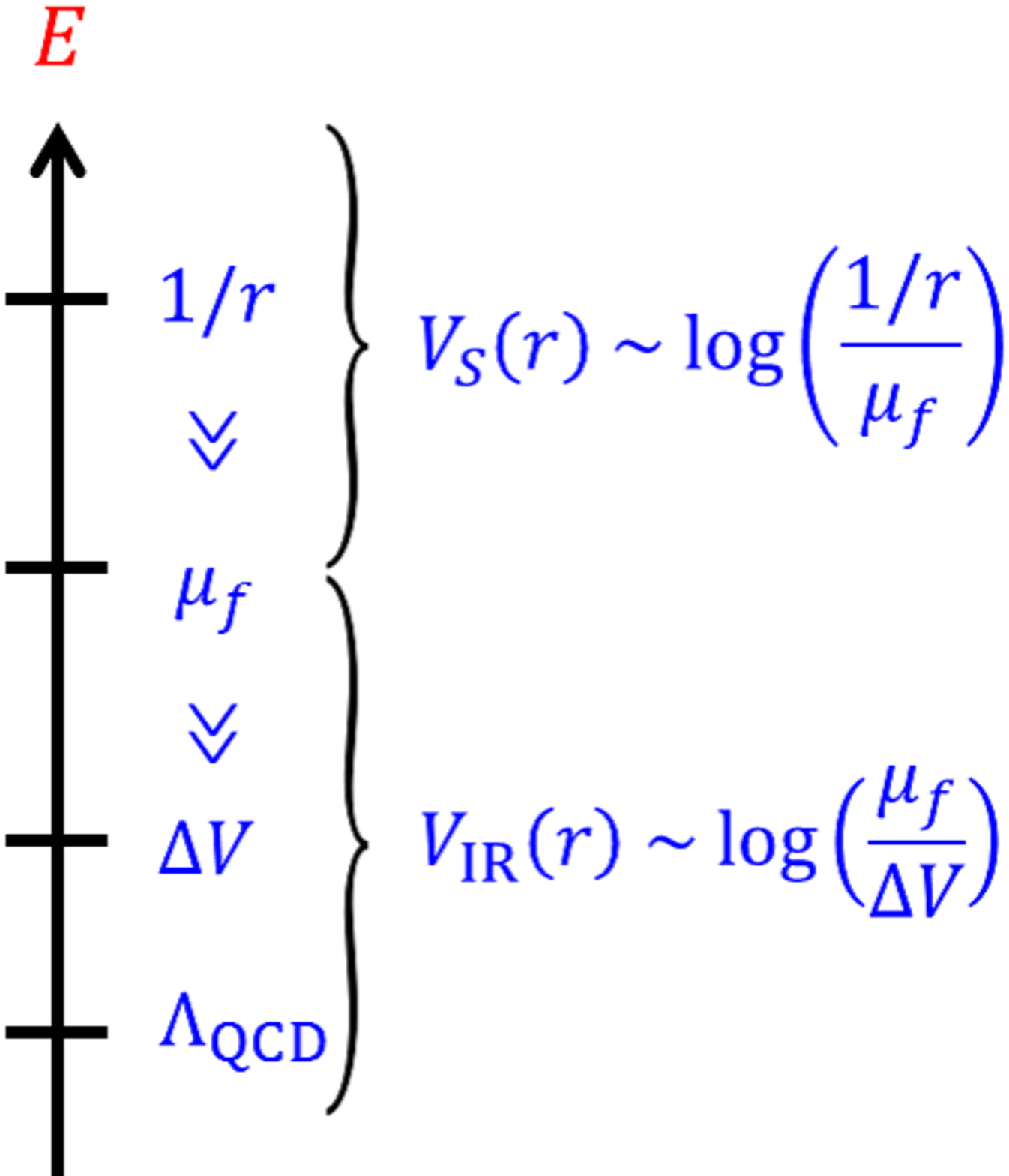}
\end{wrapfigure}
in momentum space.
Had we integrated out energy scales above the factorization
scale $\mu_f$ \`a la Wilson,
$V_S(r)$ would have contained $\log (r^{-1}/\mu_f)$ and
$V_{\rm IR}(r)$ would have contained $\log (\mu_f/\Delta V)$
in such a way that $\mu_f$ dependences cancel in the sum.
Physically we expect to replace the pole in $\epsilon$
corresponding to the IR divergence of $V_S(r)$ by $\log (r^{-1}/\mu_f)$,
while  we expect to replace the pole corresponding to the UV 
divergence in $V_{\rm IR}(r)$ by $\log (\mu_f/\Delta V)$.\footnote{
For instance,
one may compare with the chiral perturbation theory
formulated in dimensional regularization, in which
UV divergences represented by poles in $\epsilon$ 
are interpreted as logarithms of cut-off scale
$\sim\log (4\pi f_\pi)$.
}
This is the renormalization in pNRQCD.
Thus, the cancellation of $1/\epsilon$'s in
$V_S(r)$ and $V_{\rm IR}(r)$ implies cancellation of
$\mu_f$-dependences, which follows from the fact that 
$V_{\rm QCD}(r)$ should not depend on how to
factorize the energy scale.

The renormalization of $V_S(r)$ and $V_{\rm IR}(r)$ is not
important as long as we are interested in evaluating their
sum in double expansion in $\alpha_s$ and $\log\alpha_s$,
since the sum does not depend on the renormalization scheme
if expressed by the parameters of full QCD
(perturbative QCD).
The renormalization becomes important in the case that
we evaluate $V_{\rm IR}(r)$ non-perturbatively, such as by
using lattice computations, while $V_S(r)$ is evaluated
perturbatively.
In particular it becomes quite important to subtract
the IR renormalons contained in $V_S(r)$ in the case that
we use dimensional regularization.

Due to technical simplicity, 
$V_S(r)$ is often computed in dimensional regularization
using the relation (\ref{V_SfromV_QCD}).
Elimination of IR contributions from $V_S(r)$ is not
automatic in this case,
unlike the case where we explicitly introduce a cut-off
(corresponding to the factorization scale $\mu_f$)
in the computation.
Indeed the bare $V_S(r)$ in dimensional regularization
[= naive expansion of $V_{\rm QCD}(r)$ in $\alpha_s$]
contains IR renormalons just as we
explained in Sec.~\ref{sec:RenormalonsInV_QCD}.
The ${\cal O}(\LQ^3 r^2)$ renormalon should be subtracted from
$V_S(r)$.
It is absorbed in or replaced by a non-perturbative matrix element
in $V_{\rm IR}(r)$.\footnote{
Here and hereafter, 
we are mainly interested in the ${\cal O}(\LQ^3 r^2)$
renormalon and ignore the ${\cal O}(\LQ)$ renormalon
which can be canceled against that of $2m_{\rm pole}$ in the
total energy.
}
Otherwise one would inflate uncertainties in the
perturbative prediction of $V_S(r)$ as an artifact of
dimensional regularization.
We present specific renormalization schemes
of Wilson coefficients below.

\subsection[Renormalons in $V_{\rm IR}(r)$]{\boldmath Renormalons in $V_{\rm IR}(r)$}
\label{sec:RenormalonInV_IR}

As it is clear from its construction, pNRQCD EFT
reproduces correctly the dynamics of QCD in the IR region
relevant to the static $Q\bar{Q}$ system.
Therefore, the ${\cal O}(\LQ^3 r^2)$ IR renormalon 
of $V_{\rm QCD}(r)$, which we analyzed in
Sec.~\ref{sec:CancellationRenormalon}, should be reproduced
in $V_{\rm IR}(r)$.
Due to the relation (\ref{V_SfromV_QCD})
and the argument in the previous subsection, the
${\cal O}(\LQ^3 r^2)$ IR renormalon should arise
from the region $k\gg \Delta V$ in $V_{\rm IR}(r)$.
Namely, 
the region is IR compared to the scale $1/r$, but
we did not take account of the scale $\Delta V$ in
Sec.~\ref{sec:CancellationRenormalon}, which is invisible
in naive expansion in $\alpha_s$.

Let us check this statement
by introducing a cut-off in the gluon
momentum instead of dimensional regularization
in eq.~(\ref{EEcorrelator}).
In this case, the second term of 
eq.~(\ref{EEcorrelator}) cannot be dropped.
Integration of $\delta(t)$ over the range $t\geq 0$ has a
subtlety, and we find that the proper prescription
is to set this equal to 1/2.\footnote{
It is always possible to clarify the proper prescription, since
one can check equivalence of pNRQCD with full QCD in IR region
at every stage of deriving Feynman rules.
}
Hence, by comparing to eq.~(\ref{PertNonLocalCond}), 
one finds
\bea
&&
[{\cal O}(\LQ^3r^2)~\mbox{renormalon of}~V_{\rm IR}(r)]
\nonumber\\&&
~~~~~~~~~~
\sim
-4\pi\alpha_s\cdot\frac{2}{6}\cdot C_Fr^2
{\hbox to 18pt{
\hbox to -3pt{$\displaystyle \int$} 
\raise-18pt\hbox{$\scriptstyle k<\mu_f$} 
}}
\frac{d^3\vec{k}}{(2\pi)^3}\,
\biggl[
\frac{k}{k+\Delta V} - \frac{3}{2}
\biggr]_{k\gg\Delta V}
\nonumber\\&&
~~~~~~~~~~
\sim 
{\hbox to 18pt{
\hbox to -3pt{$\displaystyle \int$} 
\raise-18pt\hbox{$\scriptstyle k<\mu_f$} 
}}
\frac{d^3\vec{k}}{(2\pi)^3}\,
\biggl[-C_F\frac{4\pi\alpha_s}{k^2}\biggr]
\cdot\frac{1}{2}(i\vec{k}\!\cdot\!\vec{r})^2
,
\label{RenormalonVIR}
\eea
where we used the fact that the angular average of
$k^ik^j/k^2$ is equal to $\delta^{ij}/3$.
If we resum LLs, we can check that, in the region
$\Delta V\ll k<\mu_f$, 
the coupling constant $\alpha_s$ of
eq.~(\ref{RenormalonVIR}) is replaced by
the one-loop running coupling constant
$\alpha_{\rm 1L}(k)$.
Thus, indeed it has the same form as the ${\cal O}(\LQ^3 r^2)$
renormalon
which we analyzed in Sec.~\ref{sec:IRcancellation}.
Simultaneously this also shows that the $\mu_f$-dependences of
the $r^2$ terms cancel in eq.~(\ref{OPEofV_QCD}),
as we claimed.\footnote{
Note that, since $\mu_f\gg\LQ$, $\mu_f$-dependence
of $V_{\rm IR}$ can be estimated reliably
in expansion in $\alpha_s$, even in the
case that a dominant part of $V_{\rm IR}$ is
non-perturbative.
}

The above analysis suggests certain
renormalization schemes for $V_S(r)$ and $V_{\rm IR}(r)$,
which we discussed in the previous subsection.
We can convert IR contributions to $V_{\rm QCD}(r)$
from dimensional regularization to a cut-off regularization,
which matches the purpose of the
renormalization.
By taking the difference of $V_{\rm IR}(r)$ 
[eq.~(\ref{V_IR})] in dimensional regularization and
in a cut-off regularization, we obtain the counter term as
\bea
&&
[\delta_{c.t.}V_S(r)]_\text{cut-off}
= [{\rm eq}.~(\ref{VIRinDimReg})]-
\Biggl[
-\frac{4\pi\alpha_s}{3}\cdot C_Fr^2
{\hbox to 18pt{
\hbox to -3pt{$\displaystyle \int$} 
\raise-18pt\hbox{$\scriptstyle k<\mu_f$} 
}}
\frac{d^3\vec{k}}{(2\pi)^3}\,
\biggl(
\frac{k}{k+\Delta V} - \frac{3}{2}
\biggr)
\Biggr]
\nonumber\\&&
~~~~~~~~~~~~~~~~~~~
=
\frac{C_F\alpha_s}{9 \pi  r} 
\Biggl[
-(\mu_f r)^3
-\frac{3}{2}  C_A  \alpha_s\cdot
(\mu_f r)^2
+\frac{3}{2}   C_A^2  \alpha_s ^2\cdot
(\mu_f r)
\nonumber\\&&
~~~~~~~~~~~~~~~~~~~~~~~~~~~~~~~~
+C_A^3\alpha_s ^3 \left\{ \frac{3}{4} \log
   \left(\frac{\mu ^4 r^3}{2 \mu_f}\right)+\frac{3}{8 \epsilon
   }+\frac{9}{4} \gamma_E +\frac{5}{8}\right\}
\Biggr] ,
\eea
which should be added to
$V_S(r)$ in dimensional regularization [bare $V_S(r)$].
We see a strong dependence on the factorization scale
$\sim \mu_f^3 r^2$ reflecting the cubic divergence of the
$\vec{k}$ integration.

The above renormalization scheme may not be optimal,
since generally a cut-off in gluon momenta introduces
gauge dependences.
Another sensible renormalization scheme is to subtract
only the IR renormalon and IR divergences from the bare
$V_S(r)$, since they cause the main (known) problems of the
perturbative series and
can be extracted in a gauge-independent manner.
For instance, we can add the sum of the following two
counter terms, which are derived from
an estimate similar to eq.~(\ref{LOrenorm-Largebeta0})
for the ${\cal O}(\LQ^3r^2)$ renormalon and $\overline{\rm MS}$
renormalization of IR divergence, respectively:
\bea
&&
[\delta_{c.t.}V_S(r)]_\text{renormalons}
=-\frac{C_F\alpha_s}{9\pi}\, (\mu e^{5/6})^3r^2 \, \sum_{n=0}^\infty
\left(\frac{\beta_0\alpha_s}{4\pi}\cdot \frac{2}{3}\right)^n n!
,
\\&&
[\delta_{c.t.}\widetilde{V}_S(q)]_\text{IR-div.}
=
\frac{C_FC_A^3\alpha_s^4}{6 q^2}\,
\frac{1}{\epsilon}
+{\cal O}(\alpha_s^5)
,
\eea
\newpage\noindent
where the counter term in the $\overline{\rm MS}$ scheme is
added in momentum space.\footnote{
The relation between the potentials in coordinate space and
momentum space is given by
\bea
&&
V_{S}(r)
= \bar{\mu}^{2\epsilon}
\int \frac{d^d\vec{q}}{(2\pi)^d} \, e^{i \vec{q} \cdot \vec{r}}
\, \widetilde{V}_S(q) .
\eea
}

Using the result of Sec.~\ref{sec:UVcontrV_QCD},
we may define an alternative renormalization scheme as follows.
We define the renormalized $V_S(r)$ as
\bea
[V_S(r)]_\text{ren}=
V_C(r)+\sigma r,
\eea
where
\bea
&&
V_C(r)=\frac{A}{r}-
\frac{2C_F}{\pi}\,{\rm Im}\,\int_{C_1}
\!dq\, \frac{e^{iqr}}{qr}\, \alpha_{\rm 1L}(q) 
,
\\
&&
A
=-\frac{C_F}{\pi i} \int_{{C}_2}\!\!dq\,
\frac{\alpha_{\rm 1L}(q)}{qr} ,
~~~~~
\sigma=\frac{C_F}{2\pi i}
 \int_{C_2}\!\!dq\,
q\,{\alpha_{\rm 1L}(q)}
,
\eea
in the LL approximation.
This formula subtracts IR contrbutions
by resummation of logarithms and contour integral
surrounding the singularity at $q=\LQ$.
The formula can be extended naturally to include subleading
logarithms.
In this case the running coupling constant can
be determined from perturbative evaluation of
$V_S(r)$ in momentum space, $\widetilde{V}_S(q)$, after resummation of
logarithms by RG equation.
From NNNLL the bare $\widetilde{V}_S(q)$ includes 
IR divergences originating from the scale $\Delta V \ll k \ll 1/r$.
The contour deformation subtracts the ${\cal O}(\LQ^3r^2)$ 
renormalon but not the IR divergences which stem from 
deeper loop levels.
We subtract the IR divergences in the $\overline{\rm MS}$ 
scheme.

By making the scale $\Delta V$ explicit, 
an IR structure of $V_{\rm QCD}(r)$
concealed in the naive perturbative expansion
has become visible.
Eq.~(\ref{RenormalonVIR}) shows that the IR behavior of
$V_{\rm IR}(r)$ is different from the IR behavior of
$V_S(r)$.
From the behavior of $V_{\rm IR}(r)$ at $k\sim 0$,
one can estimate that it contains
an ${\cal O}(\LQ^4 r^3)$ IR renormalon. 
This means that, if $V_{\rm IR}(r)$ is examined in a
double expansion in $\alpha_s$ and $\log\alpha_s$,
it contains an ${\cal O}(\LQ^4 r^3)$ IR renormalon. 
This leads to an interesting consequence.
We found in Sec.~\ref{sec:CancellationRenormalon}
that the leading IR renormalon of
$E_{\rm tot}(r)$ is ${\cal O}(\LQ^3r^2)$.
In other words, the leading IR
renormalon of $2m_{\rm pole}+V_S(r)$  for the bare $V_S(r)$
is ${\cal O}(\LQ^3r^2)$.
In contrast,
the leading IR renormalon
of $2m_{\rm pole}+V_S(r)+V_{\rm IR}(r)$ is
${\cal O}(\LQ^4 r^3)$ if it
is examined in a
double expansion in $\alpha_s$ and $\log\alpha_s$.
Thus, the renormalon uncertainty of the
perturbative series for the total energy of
the $Q\bar{Q}$ system at small $r~(\ll \LQ^{-1})$
can be reduced step by step.
\pagebreak

\subsection[$V_{\rm QCD}(r)$ at very small $r$ and local gluon condensate]{\boldmath $V_{\rm QCD}(r)$ at very small $r$ and local gluon condensate\footnote{
The contents of this subsection are still premature
and many properties are yet to 
be tested quantitatively.
}
}
\begin{wrapfigure}{l}{45mm}
\hspace*{0mm}
\includegraphics[width=4cm]{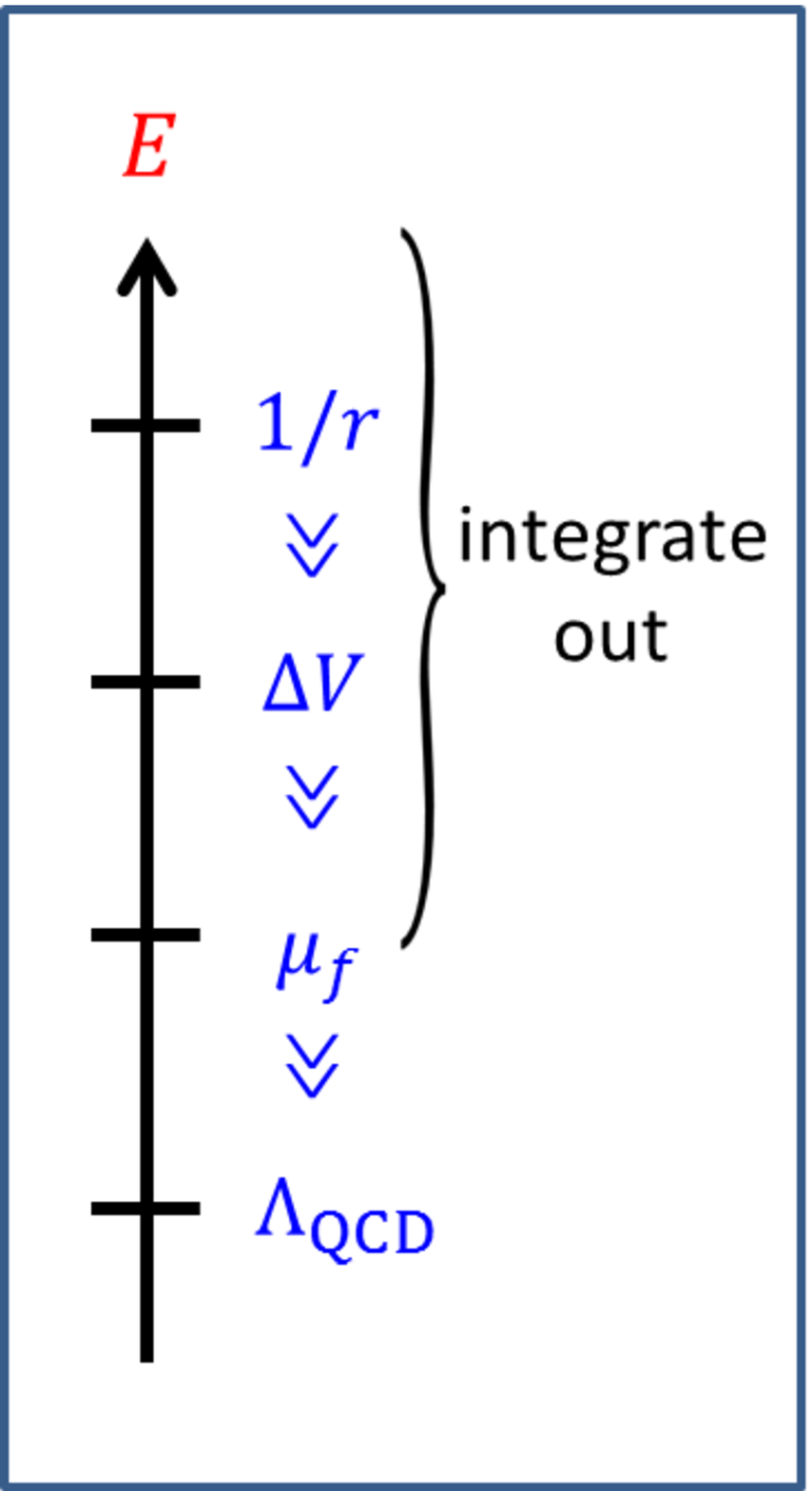}
\vspace*{-15mm}
\end{wrapfigure}
At very short distances $r\to 0$, we can assume
$\Delta V(r)\sim 1/|r\log(\LQ r)|\gg \LQ$.
Then it is possible to choose the factorization scale as
$\LQ \ll \mu_f \ll \Delta V$ and integrate out
all the scales above $\mu_f$.
In this case, only the scale $\LQ$ remains as
dynamical degrees of freedom.
Such an EFT describes dynamics of gluons with energies
$E_g\sim\LQ\ll\Delta V$, which interact with
the $Q\bar{Q}$ system.
The fields of the EFT are 
$S$, $O$ and $A_\mu$, 
the same as those of
pNRQCD.
In addition to the expansion parameters of
pNRQCD, we have $E_g/\Delta V\ll 1$.
Hence, we can regard $\partial_t$ operating on $A_\mu$
to be small.

The matching relation (\ref{MatchingRel}) changes to
\bea
&&
V_{\rm QCD}(r)=V_S(r) + V_{\rm IR}(r)
\nonumber\\&&
~~~~~~
-i g^2\,\frac{T_F}{N}\!\int_0^\infty \!\!\! dt
\, e^{-it\Delta V}
\Bigl[
\langle\,\vec{r}\!\cdot\! 
\vec{E}^a(0,\vec{0})\,
\vec{r}\!\cdot\! \vec{E}^a(0,\vec{0})
\,\rangle 
+{\cal O}(t) \Bigr]
\nonumber\\ &&
~~~~~~~~~~~~~~
+{\cal O}(r^3)
\label{MatchingRel2}
\eea
in this EFT.
We have expanded the matrix element in $t$;
if we transform the integral variable to $\tau=t\Delta V$,
it is easy to see that this expansion generates
essentially  an expansion
in $E_g/\Delta V$.
In this EFT, the energy region $E \sim\Delta V\gg \LQ$ is
included in Wilson coefficients.
Thus, perturbative evaluation of $V_{\rm IR}$, 
given by eq.~(\ref{VIRinDimReg}), is treated as a UV contribution.
Note that the above matching relation is consistent
both in naive expansion in $\alpha_s$ ($V_{\rm IR}=0$ in this
case) and in double expansion in $\alpha_s$ and $\log\alpha_s$
in dimensional regularization;
in both cases the third term evaluates
to zero, since it is given by scaleless integrals.

Noting that
\bea
\bra{0}G_{\mu\nu}^a(0)G_{\rho\sigma}^a(0)\ket{0}\propto
g_{\mu\rho}g_{\nu\sigma}-g_{\nu\rho}g_{\mu\sigma}
\eea
by antisymmetry of $G^a_{\mu\nu}$ and Lorentz invariance
of the vacuum,
it follows that
\bea
\bra{0}G^{\mu\nu a}(0)G_{\mu\nu}^a(0)\ket{0}=
-D \bra{0}\vec{E}^{a}(0)\!\cdot\!\vec{E}^{a}(0)\ket{0} .
\eea
Hence, the leading non-perturbative contribution in eq.~(\ref{MatchingRel2})
can be expressed by the local gluon condensate (Voloshin, Leutweyler):
\bea
\frac{T_F}{N D(D-1)}\,\frac{r^2}{\Delta V}
\bra{0}g^2G^{\mu\nu a}(0)G_{\mu\nu}^a(0)\ket{0}
.
\label{NonPT-VL}
\eea
This shows that the non-perturbative contribution is given by
$\sim \LQ^4r^3$ at very short distances.
In fact, it follows from a
purely dimensional analysis, since there is no other scale than
$\LQ$ and
the lowest-dimensional
local operator of the gauge field is the local gluon condensate
$\sim \LQ^4$.
We also note that this matches the IR renormalon of $V_{\rm IR}$
discussed in the previous section.

The bare $V_S(r)+V_{\rm IR}(r)$ 
(which is evaluated in double expansion in $\alpha_s$ and $\log\alpha_s$
in dimensional regularization)
is free from IR divergences, as stated 
at the end of Sec.~\ref{sec:MatchingToQCD}.
Nevertheless, it contains IR renormalons, and in general
all the IR contributions (including the renormalons) should
be subtracted by renormalization.
Noting that the non-perturbative contribution 
(\ref{NonPT-VL}) is zero 
in double expansion in $\alpha_s$ and $\log\alpha_s$,
and according to the interpretation
which we explained below eq.~(\ref{OPEofV_QCD}),
it is legitimate to define the renormalized
$V_S(r)+V_{\rm IR}(r)$ in the following way:
\bea
[V_S(r)+V_{\rm IR}(r)]_\text{ren}=
V_C(r)+\sigma r,
\eea
where, in this case, 
the running coupling constant 
is determined from perturbative evaluation of
$V_S(r)+V_{\rm IR}(r)$ in momentum space, after resummation of
logarithms by RG equation.
It is free of IR divergences and includes contributions from
scales $E\sim \Delta V$ and higher.
By contour deformation
method, contrbutions only from the
scale $\LQ$ are subtracted from it.
This definition is consistent with the
fact that the residual non-perturbative contributions include only
the scale $\LQ$.
Note that, up to NNLL, the perturbative evaluation of $V_{\rm IR}$ 
in dimensional regularization vanishes, 
so that there is no distinction whether the
scale $\Delta V$ is included in $V_C(r)+\sigma r$ or not
up to this order.

\medbreak

To end Sec.~\ref{Sec:pNRQCD}, let us make some remarks.
As we have seen, pNRQCD (in the static limit) provides a powerful
tool to analyze IR dynamics of the heavy $Q\bar{Q}$ states
systematically using multipole expansion.
It clarifies how to factorize IR contributions.
In this way, perturbative predictions of UV contributions 
can be made accurate, free from
IR divergences and IR renormalons,
by replacing IR contributions by non-perturbative
matrix elements.

Nevertheless, at the present status, we find that
the following question needs to be understood better.
In principle,
IR contributions in perturbative computations 
can be defined unambiguously
if we introduce a cut-off in gluon momenta, which is
close to the original idea of Wilson.
This introduces, however, gauge dependence and
perhaps artificially strong dependences
on the cut-off $\mu_f$.
Hence, it is desirable to find more sophisticated ways to subtract
IR contributions from Wilson coefficients.
We have suggested two ways in this lecture:
(i) to subtract IR divergences and estimates of IR renormalons,
and (ii) to subtract IR contribution as a contour integral surrounding
a singularity after resummations of logarithms.
In either method, one needs to understand the structure of
IR renormalons in advance.
In detailed analyses which go beyond the LL level, one finds different
types of IR renormalons
by inspection of certain series of higher-order terms of perturbative
expansion.
It is quite challenging to 
identify them exhaustively or organize them systematically.
We would like to study further to which extent this is possible.
The correspondence between the renormalons and non-perturbative
matrix elements of EFT would play a key role in
such an analysis.

\section{References for Further Studies}

In this section we list some references
which are useful for further studies.

\subsection*{Sec.~2}
For reviews of semi-quantitative descriptions
of the dynamics of chiral symmetry breaking in QCD using
Nambu-Jona-Lasino model and Schwinger-Dyson equation, see:
\begin{itemize}
\item[2-1.]
  K.~Higashijima,
  ``Dynamical Chiral Symmetry Breaking,''
  Phys.\ Rev.\ D {\bf 29} (1984) 1228;
  K.~Higashijima,
  ``Theory of dynamical symmetry breaking,''
  Prog.\ Theor.\ Phys.\ Suppl.\  {\bf 104} (1991) 1.

\item[2-2.]
  K.~I.~Aoki, M.~Bando, T.~Kugo, M.~G.~Mitchard and H.~Nakatani,
  ``Calculating the decay constant $F_\pi$,''
  Prog.\ Theor.\ Phys.\  {\bf 84} (1990) 683.
\end{itemize}

\subsection*{Sec.~3}

There are many good textbooks and reviews on 
renormalization and renormalization-group equation
in QCD.
See, for example:
\begin{itemize}
\item[3-1.]
  J.~C.~Collins,
  ``Renormalization --- {\it
An introduction to renormalization, the renormalization group, and the 
operator-product expansion},''
  Cambridge University Press, (1986),
ISBN: 9780521311779.
\item[3-2.]
D. J. Gross, 
``Applications of the Renormalization Group to High-Energy Physics,'' 
in {\it Les Houches 1975, Proceedings, Methods
In Field Theory}, Amsterdam 1976, p.141--250.
\end{itemize}

\subsection*{Sec.~4}

For a connection between the static QCD potential and the
heavy quark effective theory, see, for example:
\begin{itemize}
\item[4-1.]
  M.~Peter,
  ``The Static potential in QCD: A Full two loop calculation,''
  Nucl.\ Phys.\ B {\bf 501} (1997) 471
  [hep-ph/9702245].

\end{itemize}
A review on the lattice computations of the static QCD potential
in maximally abelian gauge and connection with dual-Meissner effects
can be found in:
\begin{itemize}
\item[4-2.]
  H.~Ichie,
  ``Dual Higgs theory for color confinement in quantum chromodynamics,''
  hep-lat/9906005.

\end{itemize}

\subsection*{Sec.~5}

Reviews on large-order behaviors of perturbative series
and asymptotic series
can be found in:
\begin{itemize}
\item[5-1.]
  J.~C.~Le Guillou and J.~Zinn-Justin,
  ``Large order behavior of perturbation theory,''
  Amsterdam, Netherlands: North-Holland (1990),
  ISBN: 978-0-444-88597-5.
\item[5-2.]
  J.~Zinn-Justin,
  ``Perturbation Series at Large Orders in Quantum Mechanics and Field Theories: Application to the Problem of Resummation,''
  Phys.\ Rept.\  {\bf 70} (1981) 109.
\item[5-3.]
R. B. Dingle, 
``Asymptotic Expansions: Their Derivation and Interpretation,'' 
Academic Press, 1973,
ISBN-10: 0122165500.
\end{itemize}
For a review on renormalons, see:
\begin{itemize}
\item[5-4.]
  M.~Beneke,
  ``Renormalons,''
  Phys.\ Rept.\  {\bf 317} (1999) 1
  [hep-ph/9807443].
\end{itemize}
Renormalons in the static potential were first discussed in:
\begin{itemize}
\item[5-5.]
  U.~Aglietti and Z.~Ligeti,
  ``Renormalons and confinement,''
  Phys.\ Lett.\ B {\bf 364} (1995) 75
  [hep-ph/9503209].
\end{itemize}

\subsection*{Sec.~6}

The cancellation of renromalons in the total energy was
discovered by:
\begin{itemize}
\item[6-1.]
A.~Pineda, 
``Heavy Quarkonium and Nonrelativistic Effective Field Theories,''
Ph.D. Thesis,
~\texttt{
http:\,//www.slac.stanford.edu/spires/find/hep/www?irn=}
\texttt{5399084};
A.~H.~Hoang, M.~C.~Smith, T.~Stelzer and S.~Willenbrock,
``Quarkonia and the pole mass,''
Phys.\ Rev.\ D {\bf 59}, 114014 (1999)
[arXiv:hep-ph/9804227];
M.~Beneke,
``A quark mass definition adequate for threshold problems,''
Phys.\ Lett.\ B {\bf 434}, 115 (1998)
[arXiv:hep-ph/9804241].
\end{itemize}
Details of the
numerical results of the perturbative series in Sec.~6.3
are given in:
\begin{itemize}
\item[6-2.]
Y.~Sumino,
``A connection between the perturbative QCD potential and  phenomenological
potentials,''
Phys.\ Rev.\ D {\bf 65}, 054003 (2002)
[arXiv:hep-ph/0104259];
S.~Recksiegel and Y.~Sumino,
``Perturbative QCD potential, renormalon cancellation and  phenomenological
potentials,''
Phys.\ Rev.\ D {\bf 65}, 054018 (2002)
[arXiv:hep-ph/0109122].
\end{itemize}
The three-loop [${\cal O}(\alpha_s^4)$] corrections to the QCD potential
were computed in:
\begin{itemize}
\item[6-3.]
  A.~V.~Smirnov, V.~A.~Smirnov and M.~Steinhauser,
  ``Fermionic contributions to the three-loop static potential,''
  Phys.\ Lett.\  B {\bf 668}, 293 (2008)
  [arXiv:0809.1927 [hep-ph]].
\item[6-4.]
  C.~Anzai, Y.~Kiyo and Y.~Sumino,
  ``Static QCD potential at three-loop order,''
  Phys.\ Rev.\ Lett.\  {\bf 104}, 112003 (2010)
  [arXiv:0911.4335 [hep-ph]];
  A.~V.~Smirnov, V.~A.~Smirnov and M.~Steinhauser,
  ``Three-loop static potential,''
  Phys.\ Rev.\ Lett.\  {\bf 104}, 112002 (2010)
  [arXiv:0911.4742 [hep-ph]].
\end{itemize}
The relation between the pole and $\overline{\rm MS}$ masses in the
large-$\beta_0$ approximation is computed in:
\begin{itemize}
\item[6-5.]
  M.~Beneke and V.~M.~Braun,
  ``Naive non-abelianization and resummation of fermion bubble chains,''
  Phys.\ Lett.\ B {\bf 348} (1995) 513
  [hep-ph/9411229].
\end{itemize}
A compelling evidence of IR renormalons is presented in:
\begin{itemize}
\item[6-6.]
  C.~Bauer, G.~S.~Bali and A.~Pineda,
  ``Compelling Evidence of Renormalons in QCD from High Order Perturbative Expansions,''
  Phys.\ Rev.\ Lett.\  {\bf 108} (2012) 242002
  [arXiv:1111.3946 [hep-ph]].
\end{itemize}

\subsection*{Sec.~7}

The idea of Wilsonian low-energy EFT is explained
fully in:
\begin{itemize}
\item[7-1.]
  K.~G.~Wilson and J.~B.~Kogut,
  ``The Renormalization group and the epsilon expansion,''
  Phys.\ Rept.\  {\bf 12} (1974) 75.
\end{itemize}
A short-distance expansion of UV contributions to 
$V_{\rm QCD}(r)$
and their analytic evaluation can be found in:
\begin{itemize}
\item[7-2.]
  Y.~Sumino,
  ``Static QCD Potential at 
$r < \LQ^{-1}$: perturbative expansion  and
  operator-product expansion,''
  Phys.\ Rev.\  D {\bf 76}, 114009 (2007)
  [arXiv:hep-ph/0505034].
\end{itemize}
The method of ``integration by regions'' 
to calculate asymptotic expansion of Feynman diagrams
is explained in:
\begin{itemize}
\item[7-3.]
  V.~A.~Smirnov,
  ``Applied asymptotic expansions in momenta and masses,''
  Springer Tracts Mod.\ Phys.\  {\bf 177} (2002),
  ISBN-10: 3540423346.
\item[7-4.]
  M.~Beneke and V.~A.~Smirnov,
  ``Asymptotic expansion of Feynman integrals near threshold,''
  Nucl.\ Phys.\ B {\bf 522} (1998) 321
  [hep-ph/9711391].
\end{itemize}
A justification of this method is given in:
\begin{itemize}
\item[7-5.]
  B.~Jantzen,
  ``Foundation and generalization of the expansion by regions,''
  JHEP {\bf 1112} (2011) 076
  [arXiv:1111.2589 [hep-ph]].
\end{itemize}

\subsection*{Sec.~8}

Micrscopic composition of the energy
inside the bottomonium states using their wave functions
is discussed in:
\begin{itemize}
\item[8-1.]
  N.~Brambilla, Y.~Sumino and A.~Vairo,
  ``Quarkonium spectroscopy and perturbative QCD: A New perspective,''
  Phys.\ Lett.\ B {\bf 513} (2001) 381.
\item[8-2.]
  S.~Recksiegel and Y.~Sumino,
  ``Improved perturbative QCD prediction of the bottomonium spectrum,''
  Phys.\ Rev.\ D {\bf 67} (2003) 014004
  [hep-ph/0207005].
\end{itemize}

\subsection*{Sec.~9}

Potential-NRQCD EFT in the static quark limit is discussed
extensively in: 
\begin{itemize}
\item[9-1.]
  N.~Brambilla, A.~Pineda, J.~Soto and A.~Vairo,
  ``Potential NRQCD: An Effective theory for heavy quarkonium,''
  Nucl.\ Phys.\ B {\bf 566} (2000) 275
  [hep-ph/9907240].
\end{itemize}
The IR divergence of $V_{\rm QCD}(r)$ from three loops
was found by:
\begin{itemize}
\item[9-2.]
T.~Appelquist, M.~Dine and I.~J.~Muzinich,
``The Static Limit Of Quantum Chromodynamics,''
Phys.\ Rev.\ D {\bf 17}, 2074 (1978).
\end{itemize}
Its full identification within pNRQCD
was given in Ref.[9-1] and
\begin{itemize}
\item[9-3.]
B.~A.~Kniehl and A.~A.~Penin,
``Ultrasoft effects in heavy quarkonium physics,''
Nucl.\ Phys.\ B {\bf 563}, 200 (1999).
\end{itemize}
The facts (I)--(IV) listed at the end of Sec.~9.5 are
verified in:
\begin{itemize}
\item[9-4.]
  C.~Anzai, Y.~Kiyo and Y.~Sumino,
  ``Violation of Casimir Scaling for Static QCD Potential at Three-loop Order,''
  Nucl.\ Phys.\ B {\bf 838} (2010) 28
  [arXiv:1004.1562 [hep-ph]].
\end{itemize}
Renormalization of Wilson coefficients in OPE
of $V_{\rm QCD}(r)$ is discussed in: 
\begin{itemize}
\item[9-5.]
  A.~Pineda,
  ``The Static potential: Lattice versus perturbation theory in a renormalon based approach,''
  J.\ Phys.\ G {\bf 29} (2003) 371
  [hep-ph/0208031],
\end{itemize}
and in Ref.[7-2].
\\
\\
The mechanism how a double expansion in $\alpha_s$ and $\log\alpha_s$
can modify the structure of renormalon from ${\cal O}(\LQ^3)$
to ${\cal O}(\LQ^4)$ is discussed in:
\begin{itemize}
\item[9-6.]
  Y.~Kiyo and Y.~Sumino,
  ``Off-shell suppression of renormalons in nonrelativistic QCD bound states,''
  Phys.\ Lett.\ B {\bf 535} (2002) 145
  [hep-ph/0110277].
\end{itemize}
An analysis of renormalons in $V_{\rm IR}(r)$ in
the large-$\beta_0$ approximation and using
its Borel transform
is given in:
\begin{itemize}
\item[9-7.]
  Y.~Sumino,
  ``'Coulomb + linear' form of the static QCD potential in operator product expansion,''
  Phys.\ Lett.\ B {\bf 595} (2004) 387
  [hep-ph/0403242].
\end{itemize}
The non-perturbative correction in terms of local gluon condensate
was derived by
\begin{itemize}
\item[9-8.]
  M.~B.~Voloshin,
  ``On Dynamics of Heavy Quarks in Nonperturbative QCD Vacuum,''
  Nucl.\ Phys.\ B {\bf 154} (1979) 365;
  H.~Leutwyler,
  ``How to Use Heavy Quarks to Probe the QCD Vacuum,''
  Phys.\ Lett.\ B {\bf 98} (1981) 447,
\end{itemize}
and in Ref.[9-1] in the framework of pNRQCD.

\subsection*{Applications}

To compute observables of heavy quarkonium
states using pNRQCD, it is necessary to go beyond the
static limit and include corrections in expansion
in $1/m_Q$.
For the study in this direction we refer to a review:
\begin{itemize}
\item[A-1.]
  N.~Brambilla, A.~Pineda, J.~Soto and A.~Vairo,
  ``Effective field theories for heavy quarkonium,''
  Rev.\ Mod.\ Phys.\  {\bf 77} (2005) 1423
  [arXiv:hep-ph/0410047].
\end{itemize}
In particular, the full NNNLO Hamiltonian for the
heavy quarkonium was computed in:
\begin{itemize}
\item[A-2.]
  B.~A.~Kniehl, A.~A.~Penin, V.~A.~Smirnov and M.~Steinhauser,
  ``Potential NRQCD and heavy quarkonium spectrum at next-to-next-to-next-to-leading order,''
  Nucl.\ Phys.\ B {\bf 635} (2002) 357,
\end{itemize}
(completed with the three-loop static potential in Refs.[6-3] and [6-4]).
\\ \\
Various applications and related subjects
at the frontiers are covered by 
comprehensive reviews by Quarkonium Working
Group:
\begin{itemize}
\item[A-3.]
  N.~Brambilla {\it et al.}  [Quarkonium Working Group Collaboration],
  ``Heavy quarkonium physics,''
  hep-ph/0412158.

\item[A-4.]
  N.~Brambilla, S.~Eidelman, B.~K.~Heltsley, R.~Vogt, G.~T.~Bodwin, E.~Eichten, A.~D.~Frawley and A.~B.~Meyer {\it et al.},
  ``Heavy quarkonium: progress, puzzles, and opportunities,''
  Eur.\ Phys.\ J.\ C {\bf 71} (2011) 1534
  [arXiv:1010.5827 [hep-ph]].

\end{itemize}

\section*{Acknowledgments}

This lecture note is
based on the lecture courses given at Rikkyo University, Kyoto University,
Karlsruhe University and 
Nagoya University, during the years 2012--2014.
The author is grateful to the members of the institutes of
elementary particle and nuclear physics theory groups
at these universities,
for invitation and kind hospitality during
the courses.
In particular, the author would like to express his
gratitude to
Y.~Takada, H.~Suganuma, M.~Steinhauser, J.H.~K\"uhn and J.~Hisano.
This work was supported in part by Grant-in-Aid for
scientific research No.\ 26400238 from
MEXT, Japan.

\newpage
\appendix
\clfn

\section{\boldmath Formulas for Perturbative Series of $E_{\rm tot}(r)$}
\label{AppA}

In this appendix we collect some formulas necessary to
compute the perturbative series of $E_{\rm tot}(r)$,
to facilitate the reading in Sec.~6.
The formulas are given only for the case in which
the masses of the quarks in internal loops are neglected,
for the sake of simplicity.

We set the number of quark flavors to be $n_f$.
The total energy is given by
\bea
E_{{\rm tot}}(r) = 
2 m_{{\rm pole}} + V_{{\rm QCD}}(r) .
\eea
The relation between the pole mass and the $\overline{\rm MS}$ mass
has been computed up to three loops in a full theory,
which contains $n_h$ heavy flavors (with equal masses) 
and $n_l$ massless flavors in
general.
Setting $n_h=1$ and rewriting the relation in terms of the coupling
constant of the
theory with $n_f$ massless flavors only, we obtain
\bea
m_{{\rm pole}} = \overline{m}
\left\{ 1 + 
{\alpha_s(\overline{m})\over \pi}\, d_0
+   \left({\alpha_s(\overline{m})\over \pi}\right)^{\! 2} d_1 
+  \left({\alpha_s(\overline{m})\over \pi}\right)^{\! 3} d_2
+  \left({\alpha_s(\overline{m})\over \pi}\right)^{\! 4} d_3 \right\}
\label{massrel},
\eea
where
$\overline{m} \equiv m_{\overline{\rm MS}}(m_{\overline{\rm MS}})$
denotes the $\overline{\rm MS}$ mass renormalized at the 
$\overline{\rm MS}$-mass scale.
The first three coefficients  are given by
\bea
d_0&=&\frac{4}{3} ,
\\
d_1 &=&
{\frac{307}{32}} + {\frac{{{\pi }^2}}{3}} 
      + {\frac{{{\pi }^2}\,\log 2}{9}} - {\frac{\zeta_3}{6}}
+ 
  n_f\,\left( -{\frac{71}{144}} - {\frac{{{\pi }^2}}{18}} \right)
\nonumber \\ &\simeq&
13.4434-1.04137 \, n_f ,
\\ ~~~ \nonumber \\
d_2 &=&
{\frac{8462917}{93312}} + {\frac{652841\,{{\pi }^2}}{38880}} - 
  {\frac{695\,{{\pi }^4}}{7776}} - {\frac{575\,{{\pi }^2}\,\log 2}{162}} 
\nonumber \\ &&
- 
  {\frac{22\,{{\pi }^2}\,{{\log^2 2}}}{81}} - 
  {\frac{55\,{{\log^4 2}}}{162}} 
- 
  {\frac{220\,{\rm Li}_4(\frac{1}{2})}{27}} 
+ {\frac{58\,\zeta_3}{27}} - 
  {\frac{1439\,{{\pi }^2}\,\zeta_3}{432}} + 
  {\frac{1975\,\zeta_5}{216}}
\nonumber \\ &&
+ 
  n_f\,\left( -{\frac{231847}{23328}} - 
     {\frac{991\,{{\pi }^2}}{648}} + {\frac{61\,{{\pi }^4}}{1944}} - 
     {\frac{11\,{{\pi }^2}\,\log 2}{81}} + 
     {\frac{2\,{{\pi }^2}\,{{\log^2 2}}}{81}} + 
     {\frac{{{\log^4 2}}}{81}} 
\right.
\nonumber \\ &&
\left.
+ 
     {\frac{8\,{\rm Li}_4(\frac{1}{2})}{27}} - 
     {\frac{241\,\zeta_3}{72}} \right)  + 
  {n_f^2}\,\left( {\frac{2353}{23328}} + 
     {\frac{13\,{{\pi }^2}}{324}} + {\frac{7\,\zeta_3}{54}}
     \right)  
\nonumber \\ &\simeq &
190.391 - 26.6551\, n_f +  0.652691\, n_f^2  \, ,
\eea
where
$\zeta(z)=\sum_{n=1}^\infty {1}/{n^z}$ 
denotes the Riemann zeta function, 
and $\zeta_3=\zeta(3)=1.2020...$,
$\zeta_5=\zeta(5)=1.0369...$;
${\rm Li}_n(x)=\sum_{k=1}^\infty\frac{x^k}{k^n}$ denotes 
the polylogarithm, and
${\rm Li}_4(\frac{1}{2})=0.517479\cdots $. 

The fourth coefficient $d_3$ is not known exactly yet.
Its value in the large-$\beta_0$ approximation 
is given by
\bea
d_3(\mbox{large-}\beta_0) &=&
\frac{ \beta_0^3}{64}\,
\left( {\frac{42979}{5184}} + {\frac{89\,{{\pi }^2}}{18}} + 
       {\frac{71\,{{\pi }^4}}{120}} + 
       {\frac{317\, \zeta_3}{12}}\
\right)
\nonumber \\ &\simeq & 
3046.29 - 553.872\,{n_f} +   33.568\,{n_f^2} - 0.678141\,{n_f^3}.
\eea

The QCD potential of the theory with $n_f$ massless flavors only
is given, up to ${\cal O}(\alpha_s^4)$ and
${\cal O}(\alpha_s^4\log\alpha_s)$, by 
\bea
&&
V_{{\rm QCD}}(r)
=-C_F\frac{\alpha_s(\mu)}{r}
\, \sum_{n=0}^{3} P_n(L_r ) \,
\biggl( \frac{\alpha_s(\mu)}{4\pi} \biggr)^n \,,
\eea
where
\bea
{L}_r=\log(\mu^2 r^2) + 2{\gamma_E}\,,
\eea
and
\bea
&&
P_0 = a_0 ,~~~
P_1 = a_1 + a_0\beta_0  L_r ,
~~~
P_2 = a_2 + (2a_1\beta_0 + a_0\beta_1 )L_r +
a_0{\beta_0}^2 \biggl(L_r^{\,2}+\frac{\pi^2}{3}\biggr) ,
\nonumber \\ &&
P_3 =
\bar{a}_3 +\delta a_3^{\rm US} + 
( 3a_2\beta_0+   2 a_1 \beta_1 +
    a_0 \beta_2) L_r
\nonumber\\ &&
~~~~~~~~
+
   \biggl(3 a_1 {\beta_0}^2
+ \frac{5}{2}  a_0 \beta_0 \beta_1\biggr) \biggl(L_r^{\,2}+\frac{\pi^2}{3}\biggr)
+  a_0 {\beta_0}^3 (L_r^{\,3}+\pi^2 L_r+16\zeta_3)
\,,
\eea
with
\bea
&&
\delta a_3^{\rm US} = \frac{16}{3}\pi^2C_A^3
\biggl[ \log \Bigl(C_A\alpha_s(\mu)\Bigr) + \gamma_E -\frac{5}{6}
\biggr] \,;
~~~~~C_A=3 .
\eea
The coefficients of the beta function $\beta_n$ are given by 
\bea
&&
\beta_0 = 11 - \frac{2}{3} n_f ,
~~~~~~
\beta_1 = 102 - \frac{38}{3} n_f,
\\&&
\beta_2 = \frac{2857}{2} - \frac{5033}{18}n_f
+ \frac{325}{54} n_f^2 .
\eea
The constants $a_n$ of the potential, not determined by
the RG equation, are given by
\bea
&&
a_0=1,
~~~~~~~
a_1 = \frac{31}{3} - \frac{10}{9} n_f 
\\ &&
a_2 = {\frac{4343}{18}} + 36\,{{\pi }^2} +   66\,{\zeta_3} - 
  {\frac{9\,{{\pi }^4}}{4}} - 
   \left( {\frac{1229}{27}} + {\frac{52\,{\zeta_3}}{3}} \right)  \,{n_f}
+ {\frac{100}{81}} \,{n_f^2} 
\\ &&
~~~
\simeq 456.749 -66.3542 \,n_f +
1.23457 \,n_f^2
,
\\&&
\bar{a}_3\simeq 13431.7-3289.91 \,{n_f}+185.99\, n_f^2
-1.37174 \,n_f^3
,
\label{a2}
\eea
where presently $\bar{a}_3$ is known only numerically.

In order to achieve the renormalon cancellation between $2\, m_{\rm pole}$ and 
$V_{\rm QCD}(r)$ order by order in $\alpha_s$ expansion, 
we must use the same coupling constant $\alpha_s(\mu)$ in the series
expansions of 
$2m_{\rm pole}$ and $V_{\rm QCD}(r)$. 
Therefore, $\alpha_s(\overline{m})$ is re-expressed in 
terms of $\alpha_s(\mu)$ as 
\begin{eqnarray}
&&
\alpha_s(\overline{m})
=
\alpha_s(\mu)
\left\{ 1 
+ \frac{\beta_0  \log \left(\frac{\mu}{\overline{m}}\right) }{2} \, 
  \left( \frac{  \alpha_s(\mu)}{\pi} \right)
+ \left( \frac{\beta_1  \log\left(\frac{\mu}{\overline{m}}\right)}{8}
        +\frac{\beta_0^2  \log^2 \left(\frac{\mu}{\overline{m}}\right)}{4} \right)
  \left(  \frac{ \alpha_s(\mu)}{\pi} \right)^2
\right.
\nonumber \\
&& ~~~~~~~~~~
\left.
+ \left( \frac{\beta_2  \log \left(\frac{\mu}{\overline{m}}\right)}{32}
       + \frac{5\beta_0  \beta_1 \log^2 \left(\frac{\mu}{\overline{m}}\right)}{32}
       +\frac{\beta_0^3  \log^3 \left(\frac{\mu}{\overline{m}}\right)}{8}  
 \right)   \left(  \frac{ \alpha_s(\mu)}{\pi} \right)^3
+{\cal O}(\alpha_s^4)
\right\} \, ,
\label{a-m}
\end{eqnarray}
which follows from the RG equation
\bea
\mu^2 \, \frac{d}{d\mu^2} \, \alpha_s(\mu) =
- \alpha_s(\mu) \sum_{n=0}^{\infty} \beta_n
\biggl( \frac{\alpha_s(\mu)}{4\pi} \biggr)^{n+1}.
\label{RGeq}
\eea
Using Eqs. (\ref{massrel}) and (\ref{a-m}), we obtain the expansion 
of $m_{\rm pole}$ in terms of $\alpha_s(\mu)$, 
\bea
m_{\rm pole}
&=&
\overline{m} 
\times
\left( 1 + \sum_{n=1}^{4}\, \widetilde{d}_{n-1}(l_\mu) \, 
           \left(\frac{\alpha_s(\mu)}{\pi}\right)^{n} 
\right)
+{\cal O}(\alpha_s^5) \, ,
\label{m-pole-mu}
\end{eqnarray}
where the coefficients $\widetilde{d}_{n}(l_\mu)$ are 
functions of $l_\mu=\log(\mu/\overline{m})$.

\section{\boldmath Computation of $V_C(r)$}
\label{AppB}

We show how to compute $V_C(r)$, given by
eq.~(\ref{Vc(r)}).
We rotate the integral contour to imaginary axis in the
complex $q$-plane.
Setting $q=it/r$, we may rewrite
\bea
&&
- \frac{2C_F}{\pi} \, {\rm Im}
\int_{C_1}\! dq \, \frac{e^{iqr}}{qr} \, \alpha_{\rm 1L}(q) 
= -\frac{4C_F}{\beta_0r}
\, {\rm Im}
\int_0^\infty\! dt \, \frac{e^{-t}}{t\,\log\bigl({it}/{\LQ r}\bigr)}
\nonumber\\&&
~~~~~~~~~~~~
= -\frac{4C_F}{\beta_0r}
\, 
\left[ -\pi + {\rm Im}
\int_0^\infty\! dt \, {e^{-t}}\,
{\log\Bigl[\log\bigl({it}/{\LQ r}\bigr)}
\Bigr]
\right]
\nonumber\\&&
~~~~~~~~~~~~
= -\frac{4C_F}{\beta_0r}
\left[ -\pi +
\int_0^\infty\!\! dt \, {e^{-t}}\, {\rm Im}
\,\Bigl[
{\log\Bigl\{\log t-\log({\LQ r})+i\mbox{$\frac{\pi}{2}$} }
\Bigr\}\Bigr] 
\right],
\eea
where we used integration by parts.
Combining with $A/r=-4\pi C_F/(\beta_0 r)$, we obtain
an expression of
$V_C(r)$ given by an integral over $t$:
\bea
&&
V_C(r)=-\frac{4C_F}{\beta_0r}
\int_0^\infty\!\! dt \, {e^{-t}}\, {\rm Im}
\,\Bigl[
{\log\Bigl\{\log t-\log({\LQ r})+i\mbox{$\frac{\pi}{2}$} }
\Bigr\}\Bigr] 
.
\eea
The integral is easily evaluated numerically for a given $r$.

To obtain asymptotic behaviors analytically we can expand
${\rm Im}\,\log\Bigl[\log t-\log({\LQ r})+i\mbox{$\frac{\pi}{2}$}\Bigr]$
in $\frac{1}{|\log({\LQ r})|}$
before integration over $t$, which reduces to
$\pi -\frac{\pi}{2|\log({\LQ r})|}$ as $r\to \infty$, and
to $\frac{\pi}{2|\log({\LQ r})|}$ as $r\to 0$.

\section{Integration-by-regions Method and Relation to EFT}
\label{AppC}
\clfn

In this appendix, we explain the technique called asymptotic
expansion of a diagram or integration by regions.
This can be used to identify operators ${\cal O}_i$
(effective interactions)
in the Lagrangian of a Wilsonian EFT, eq.~(\ref{EFT-Lagrangian}).
At the same time the technique provides an efficient method for
perturbative computations of Wilson coefficients $g_i(\mu_f)$.

Let us first explain the idea of the asymptotic expansion
in a simplified example.
We consider an integral
\bea
&&
I(m;\epsilon)=\int_0^\infty dp\, \frac{p^\epsilon}{(p+m)(p+1)} .
\label{defI}
\eea
It is a toy model imitating an
integral in dimensional regularization.
In fact, it is a one-parameter integral,
imitating integral over the radial direction of
a dimensionally-regulated integral, with
$p^\epsilon dp$ representing a volume element;
furthermore each propagator denominator is a linear
function of $p$ rather than a quadratic function.
Suppose $m \ll 1$ is a small parameter and consider
expanding $I$ in $m$.
Let us presume as if the integral region is
divided into two regions
$p< 1$ and ${p> 1}$, and
expand the integrand in each region in a small
parameter.
Nevertheless, we restore the original integral region in
each integral, as follows.
\bea
&&
I=
\int_0^\infty dp\, 
\underline{
\frac{p^\epsilon}{p+m}
(1-p+p^2+\cdots)
} ~+~
\int_0^\infty dp\, 
\underline{
\frac{p^\epsilon}{p+1}\,\frac{1}{p}
\left(1-\frac{m}{p}+\cdots\right) 
}.
\label{ExpansionOfI}
\\ &&
~~~~~~~~~~~
~~~~~~~~~~~
{p< 1}
~~~~~~~~~~~
~~~~~~~~~~~
~~~~~~~~~~~
~~~~~
{p> 1\gg m}
\nonumber
\eea
At a first glance, this seems to give a wrong result,
since firstly we have extended each integral region to a
region where the expansion is not justified, and secondly
there would be a problem of double counting of region.
Surprisingly, however, 
if we evaluate the individual terms of the above
integrals and take their sum, it gives 
the correct expansion in $m$ of the original integral $I$.

\begin{figure}
\begin{center}
\vspace*{-0.5cm}
\includegraphics[width=8.5cm]{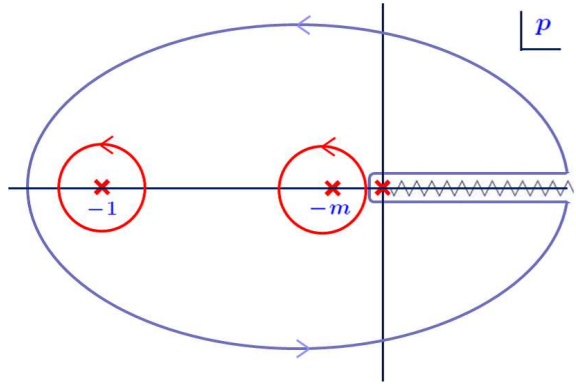}
\end{center}
\vspace*{-.5cm}
\caption{\footnotesize
Analyticity of
the integrand of eq.~(\ref{defI}) in the
complex $p$-plane.
The blue contour can be
deformed to the sum of the red contours.
\label{ContourAsymptExp}}
\end{figure}
The reason can be understood as follows.
Fig.~\ref{ContourAsymptExp} shows the analyticity of
the integrand of eq.~(\ref{defI}) in the
complex $p$-plane:
there are poles at $p=-1$ and $p=-m$;
the origin is a branch point due to
$p^\epsilon$ and the
branch cut lies along the positive $p$-axis.
The integral of $p$ along the positive $p$-axis
is equal to, up to a proportionality factor, 
an integral along the
contour wrapping the branch cut.
We may close the contour at negative infinity and deform the
contour into the sum of 
two closed contours surrounding the
two poles.
(See Fig.~\ref{ContourAsymptExp}:
we deform the blue contour
to the sum of the red contours.)
Along the contour surrounding the pole at $-m$, it is justified to
expand the integrand using the fact $|p|\approx m \ll 1$;
this gives the integrand of the first term of eq.~(\ref{ExpansionOfI}).
After the expansion, the contour of the integral of each term
of the expansion
can be brought back to the original contour wrapping the
branch cut along the positive $p$-axis.
Similarly, along the contour surrounding the pole at $-1$,
we may expand the integrand using $|p|\approx 1\gg m$,
which gives the second term of eq.~(\ref{ExpansionOfI}).
Again, after the expansion, the integral contour can be
brought back to the one surrounding the branch cut.\footnote{
In these manipulations, the value of $\epsilon$ in each term needs to
be varied appropriately by
analytical continuation into the domain where each integral
is well defined.
}
In this way, we obtain eq.~(\ref{ExpansionOfI}).

Thus, for an integral that imitates a dimensionally-regulated one,
we can expand the integral in a small parameter, 
without introducing a cut-off in the integral region.
The important point in the above example is that the contribution from
each of the scales $|p|\sim 1$ and $|p|\sim m$ is expressed
by a contour integral surrounding the corresponding pole
in the integrand (i.e., by the residue of each pole).

The method for the asymptotic expansion of a loop integral
in dimensional regularization is the same:
we divide the integral region into separate regions according to
the scales contained in the integrand and
expand the integrand in appropriate small parameters in
respective regions;
we nevertheless integrate individual terms of the expansions over
the original integral region, namely, over the entire $D$-dimensional
phase space for each loop integral.\footnote{
At the moment, 
the proof of this method using contour deformation 
as in the above toy model is missing,
for general loop integrals in dimensional regularization.
While it is likely that such an interpretation is possible generally,
presently this type of proof is valid only in some selective
cases.
There exists a general proof based on different reasonings.
}

\begin{figure}[h]
\vspace*{-.5cm}
\begin{center}
\includegraphics[width=14cm]{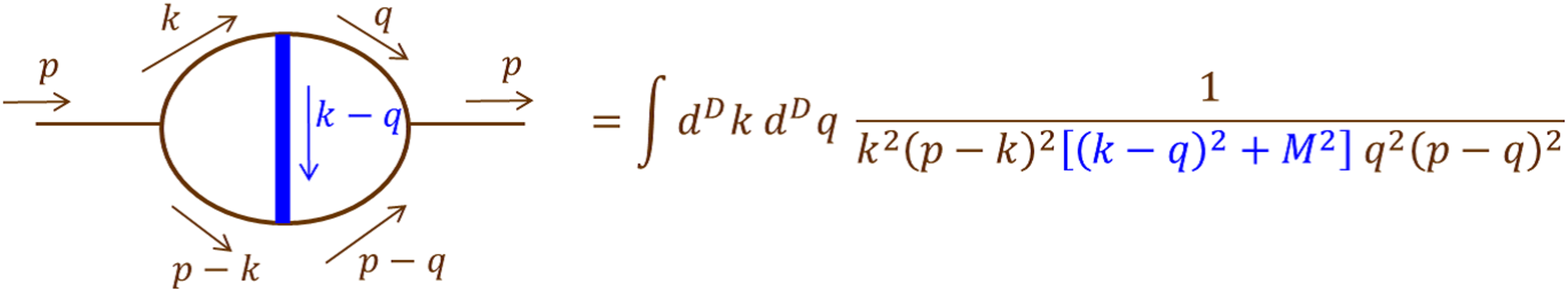}
\end{center}
\vspace*{-.5cm}
\caption{\footnotesize
Two-loop diagram used to illustrate asymptotic expansion
in $p^2/M^2$.
The thick blue line represents a propagator
with mass $M$, while all other lines represent massless
propagators.
\label{2loopDiag-AsymptExp}}
\vspace*{-.5cm}
\begin{center}
\includegraphics[width=16cm]{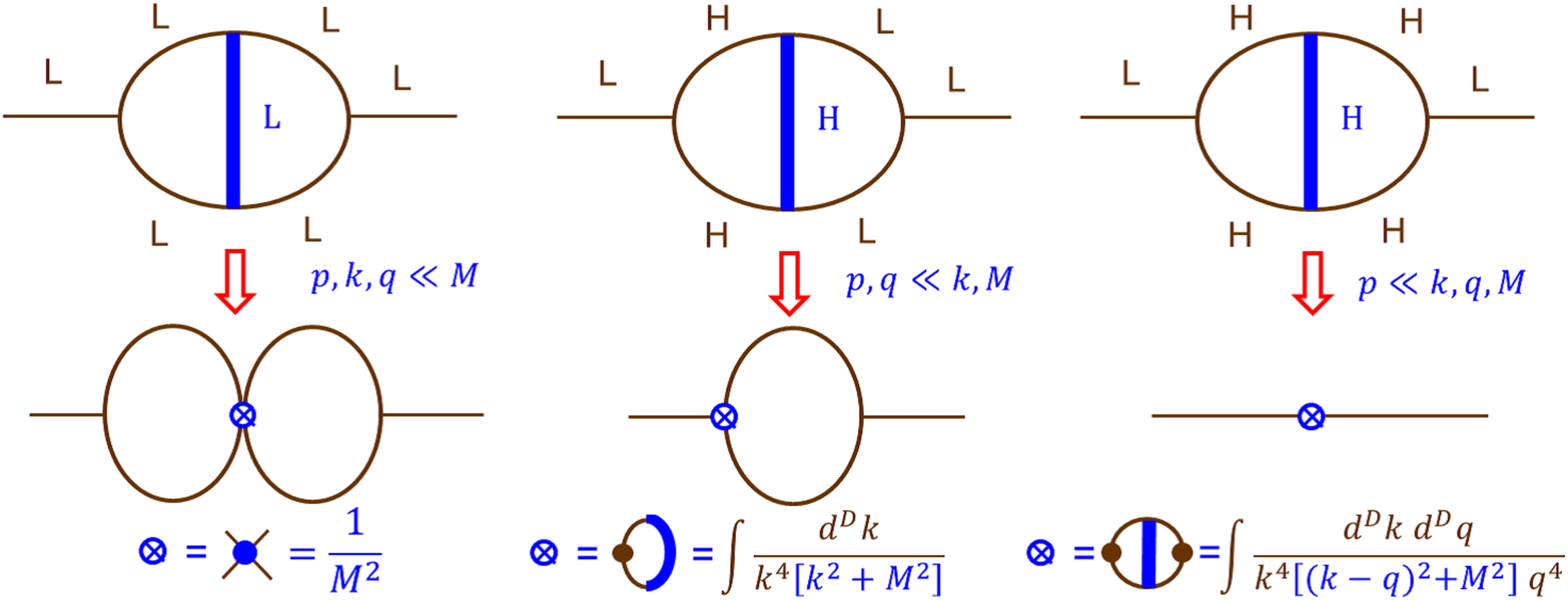}
\end{center}
\vspace*{-.5cm}
\caption{\footnotesize
Diagrams showing procedure of the asymptotic expansion.
The bottom line represent the Wilson coefficients
of the leading-order effective vertices in
respective regions.
\label{AsymptExpOf2loopDiag}}
\end{figure}

For illustration we consider the following two-loop integral
in the case $p^2\ll M^2$:
\bea
J(p^2,M^2)=
\int d^D\!k\,d^D\!q \, \frac{1}{k^2(p-k)^2[(k-q)^2+M^2]q^2(p-q)^2} .
\eea
The corresponding
diagram is shown in Fig.~\ref{2loopDiag-AsymptExp}, where the
thick blue line represents a heavy particle with
mass $M$ and all other lines
represent massless particles.
We expand $J$ in $p^2/M^2$.
The integral region of each loop integral is divided into two
regions: high momentum region (H), $|k|>M$ or $|q|>M$, and 
low momentum region (L), $|k|<M$ or $|q|<M$.
Hence, the whole integral region is divided into four regions:
(H,H),(H,L),(L,H),(L,L).
Of these (H,L) and (L,H) are the same due to the exchange symmetry
between $k$ and $q$.
Fig.~\ref{AsymptExpOf2loopDiag} shows how to perform the asymptotic
expansion in each of these regions.

In the region (L,L) we expand the massive propagator
$1/[(k-q)^2+M^2]$ in $k$ and $q$.
Each term represents an effective four-point vertex, where
the leading vertex is given by a constant coupling $1/M^2$.
This is depicted in the left-most part of the figure.
Higher-order vertices are associated with 
powers of the factor $(k-q)^2/M^2$, which correspond to
four-point interactions given by higher derivative operators.

In the region (H,L) we expand the propagator $1/(p-k)^2$
in $p$ and the propagator $1/[(k-q)^2+M^2]$ in $q$.
In each term of the expansion, integral over $k$ can be factorized, 
since $p,q$ enter only the numerator of the integrand and can be
pulled outside of the integral.
This produces effective three-point vertices, which
correspond to three-point interactions given by
local operators.
The leading term of this expansion is depicted in the
middle part of the figure.
Since high momenta flow through the $k$-loop,
it is natural to expect that the loop 
effectively shrinks to a point.

In the region (H,H) we expand $1/(p-k)^2$ and $1/(p-q)^2$
in $p$.
In this case, the whole integral over $k$ and $p$ can be
factorized at each order of the expansion.
Thus, each term can be regarded as an effective two-point
interaction corresponding to a local operator.
See the right-most part of the figure.

We may compute the same process in a low-energy EFT in which the massive
particle has been integrated out.
The asymptotic expansion of the diagram in the full theory obtained above
can be interpreted as the computation in the EFT.
The bottom-left diagram in Fig.~\ref{AsymptExpOf2loopDiag} represents
a two-loop computation of this process in the EFT with an insertion of
a four-point vertex, which is generated at tree-level of the full
theory.
The factor $1/M^2$ below the diagram represents the Wilson coefficient
of the leading-order vertex in expansion in $1/M^2$.
The bottom-middle diagram represents a one-loop computation of this
process in the EFT
with an insertion of a three-point vertex, which is generated at
one-loop level in the full theory.
The one-loop integral shown below the diagram represents the
Wilson coefficient of the leading-order vertex in expansion in $1/M^2$.
The bottom-right diagram represents a tree-level computation of this
process in the EFT with an insertion of a two-point vertex, which
is generated at two-loop level in the full theory.
The corresponding leading-order Wilson coefficient is shown
as a two-loop integral.
Thus, the relevant operators and Wilson coefficients of EFT can be
identified.

The Wilson coefficients, given by loop integrals in 
dimensional regularization, are 
particularly convenient in practical computations.
They are homogeneous in a single dimensionful parameter $M$, 
which can be computed relatively easily.
In contrast, if we adopt a cut-off regularization,
usually it becomes much more difficult to evaluate the 
corresponding integrals
(especially at higher loops),
since more scales are involved.

\end{document}